\documentclass{elsart}

\usepackage{amsmath}
\usepackage{amssymb}

\makeatletter
\@addtoreset{equation}{section}
\renewcommand{\theequation}{\thesection.\@arabic\c@equation}
\makeatother

\newcommand{\ie}{\textit{i.e.}}

\begin{document}

\begin{frontmatter}

\title{Dynamic aspect of the chiral phase transition\\
 in the mode coupling theory}

\author[Tokyo]{K.~Ohnishi\thanksref{Yukawa}},
\ead{konishi@nt1.c.u-tokyo.ac.jp}
\author[MIT,Hongo]{K.~Fukushima} and
\ead{kenji@lns.mit.edu}
\author[Tokyo]{K.~Ohta}
\ead{ohta@nt1.c.u-tokyo.ac.jp}

\address[Tokyo]{Institute of Physics, University of Tokyo,\\
                3-8-1 Komaba, Meguro-ku, Tokyo 153-8902, Japan}
\address[MIT]{Center for Theoretical Physics, Massachusetts
              Institute of Technology,\\
              77 Massachusetts Avenue, Cambridge, MA 02139, U.S.A.}
\address[Hongo]{Department of Physics, University of Tokyo,\\
                7-3-1 Hongo, Bunkyo-ku, Tokyo 113-0033, Japan}
\thanks[Yukawa]{\textit{Present address}:
 Yukawa Institute for Theoretical Physics, Kyoto University,
 Kyoto 606-8502, Japan}

\begin{abstract}
We analyze the dynamic aspect of the chiral phase transition. We apply
the mode coupling theory to the linear sigma model and derive the
kinetic equation for the chiral phase transition. We challenge
Hohenberg and Halperin's classification scheme of dynamic critical
phenomena in which the dynamic universality class of the chiral phase
transition has been identified with that of the antiferromagnet. We
point out a crucial difference between the chiral dynamics and the
antiferromagnet system. We also calculate the dynamic critical
exponent for the chiral phase transition. Our result is
$z=1-\eta/2\cong 0.98$ which is contrasted with $z=d/2=1.5$ of the
antiferromagnet.
\end{abstract}

\end{frontmatter}

\newpage

\section{Introduction}
\label{intro}
Among the central issues in the study of QCD (Quantum ChromoDynamics),
there exists the problem of the spontaneous chiral symmetry breaking
and its restoration,
\ie, the chiral phase transition. It is believed that the chiral phase
transition occurs under an extreme condition like high temperature or
high density. The critical dynamics of the second order chiral phase
transition for the two flavor massless QCD is not only of theoretical
interest but also of importance in understanding the relativistic
heavy ion collision experiments. The critical phenomena significantly
affect how hot matter cools down. There are a lot of works aimed to
look into the \textit{static} (equilibrium) nature of the chiral phase
transition, including the lattice QCD simulation
\cite{Pisarski:ms,Karsch:1994hm,Aoki:1998wg,Bernard:1999fv,Laermann:2003cv}.
Our objective in the present paper is to investigate the
\textit{dynamic} (non-equilibrium or real-time) property of the
second order chiral phase transition
\cite{Hatsuda:1985eb,Rajagopal:1992qz,Son:2001ff,Boyanovsky:2001pa,Koide:2003ax,Blaizot:2001nr,Berges:2003pc,Ikeda:2004in}.
Knowledge of the dynamic critical behavior would
be indispensable to understand the relativistic heavy ion collision
experiments because the system evolves essentially through
non-equilibrium states. Anomalous dynamic critical phenomena such as
the \textit{critical slowing down} and the \textit{softening of
propagating modes} have significant relevance to the matter going
toward or coming from a quark-gluon plasma. Throughout this paper, we
will restrict ourselves to the chiral phase transition at zero baryon
number density.

In general, whenever we consider a critical point of second order, it
is useful to classify the critical point according to the universality
class. The critical points in the same universality class are to show
exactly the same critical behavior unless there appears another fixed
point or the interaction is infinitely long-ranged. For the static
case, as is well known, the universality class is determined solely by
the global symmetry of the system. The systems having the same
symmetry belong to the same static universality class. On the other
hand, the dynamic universality class is more complicated to classify.
A classification method, which we will briefly review in the next
section, has been proposed and established by Hohenberg and Halperin.
Actually, all the observed critical points, such as the liquid-gas
transition, transitions in a ferromagnet, in an antiferromagnet and in
a superfluid, have been classified systematically and comprehensively
in the review article by Hohenberg and Halperin
\cite{Hohenberg:1977ym}.

Consideration on the universality class of the chiral phase transition
has already been given. The static universality class of the
two-flavor chiral phase transition is identified with that of the
ferromagnet and antiferromagnet because they all have the rotational
O(4) symmetry \cite{Pisarski:ms}. The dynamic universality class of
the chiral phase transition has also been discussed by Rajagopal and
Wilczek \cite{Rajagopal:1992qz}. They argue that the chiral phase
transition belongs to the same dynamic universality class as that of
the antiferromagnet. Certainly it is concluded that the dynamic
universality classes of the two systems would be identical according
to the classification method of Hohenberg and Halperin. Since the
nature of the antiferromagnet has been analyzed and understood enough,
the identification of the dynamic universality class means that the
nature of the chiral phase transition is understood as well. There
does exist, however, a crucial difference between the two systems in
the appearance of physical modes, as pointed out in
Ref.~\cite{Koide:2003ax}. The difference is so crucial that the
dynamic universality classes of the two systems cannot be regarded as
identical any longer and Hohenberg and Halperin's classification
method itself must be reexamined carefully.
In the next section, we will take a close look at the two systems to
reveal the difference.

Once the two systems are realized to be distinctive, it is necessary
to reanalyze the dynamic aspect of the chiral phase transition without
resorting to Hohenberg and Halperin's classification method. We will
adopt the mode coupling theory
\cite{kawaprogress,kawaannals,kawareview,kawahiheiko} in order to
disclose the physical modes and write down the kinetic equations in a
chiral effective model.

The plan of this paper is as follows. In Sec.~\ref{univofchiral}, we
will give a general review on the dynamic critical phenomena and point
out the difference between the antiferromagnet and the chiral phase
transition. In Sec.~\ref{review}, we will review the mode coupling
theory briefly. In Sec.~\ref{O(2)}, we will apply the mode coupling
theory to the O(2) linear sigma model in order to see what kinds of
slow modes appear near the phase transition. In Sec.~\ref{O(4)} we
will derive the kinetic equation in the O(4) linear sigma model and we
will calculate the dynamic critical exponents in Sec.~\ref{exponent}.
The exponent will be found to be $z=1-\eta/2\cong 0.98$, which is
contrasted with $z=d/2$ of the antiferromagnet. Finally, we will give
the summary in Sec.~\ref{summary}.

\section{Dynamic universality class of the chiral phase transition}
\label{univofchiral}

In this section, we will discuss the dynamic universality class of the
chiral phase transition with emphasis on the difference from the
antiferromagnet. For that purpose, it is inevitably necessary to
explain the notion of the slow mode. In Subsec.~\ref{generalreview},
we will give a general review on the dynamic critical phenomena, in
which we will explain the slow variable, the slow mode and the
classification method of the dynamic universality class by Hohenberg
and Halperin. In Subsec.~\ref{difference}, we will discuss the chiral
phase transition.

\subsection{Dynamic critical phenomena}
\label{generalreview}

One of the prominent ingredients of the dynamic critical phenomena
would be the critical slowing down, that is, the long (and infinite)
relaxation time near (and at) a critical point. The critical slowing
down is attributed to the slow motion of the long wavelength
fluctuations of the slow variables. The slow variables near a critical
point consist of the \textit{order parameter} and the
\textit{conserved quantities} involved in the system. The slowness of
the order parameter fluctuations can be understood by considering a
flat thermodynamic potential with respect to the order parameter near
the critical point, which indicates a weak restoring force on the
order parameter fluctuations. Alternatively we can imagine the
correlated domains of the order parameter to understand the critical
slowing down. Near the critical point, the size of the domains, \ie,
the correlation length $\xi$, becomes so large that the motion of the
domains gets very dull and slow. Since $\xi$ diverges just at the
critical point, the slowness of the motion becomes infinite, leading
to the divergent relaxation time.

The motion of the long wavelength fluctuation of a conserved quantity
is kinematically slow even away from the critical point. In fact, the
fluctuation of a conserved quantity with zero wavenumber, $\vec{q}=0$,
has an infinite relaxation time because of conservation of the
quantity. [Note that $\partial_\mu j^\mu=0$ means $j^0$ is time
independent when $\vec{q}=0$.] The fluctuation with a small but
finite wavenumber has a long but finite relaxation time. Near the
critical point, it is known that the coupling between the order
parameter and the conserved quantities becomes very strong. Through
the coupling, the fluctuations of the conserved quantity even with
a finite wavenumber come to have an infinite relaxation time at the
critical point.

We have seen the slowness of the long wavelength fluctuation of the
slow variables. The relaxation time for those fluctuations is so long
and amounts to the macroscopic scale belonging to the kinetic regime.
On the other hand, for the degrees of freedom other than the long
wavelength fluctuation of the slow variables, namely the short
wavelength fluctuations of the slow variables and the fluctuations of
the coordinates other than the slow variables, the motions remain
rapid near the critical point. Consider a non-equilibrium system near
the critical point after a macroscopic time has elapsed. The
non-equilibrium fluctuations of the rapid degrees of freedom will have
already relaxed to constitute the heat bath, while the long wavelength
fluctuations of the slow variables will still remain as
non-equilibrium fluctuations. Thus, by tracing the time evolution of
the long wavelength fluctuations of the slow variables, we can make
the sufficient and even complete description of non-equilibrium
processes in the kinetic regime. The mode coupling theory developed by
Kawasaki
\cite{kawaprogress,kawaannals,kawareview,kawahiheiko} is one of the
schemes which provides us with the equation of motion for the critical
fluctuations, that is, the kinetic equation near the critical point.

The long wavelength fluctuations of the slow variables or their linear
combinations constitute what is called the slow modes. The slow modes
appear as narrow peaks in the spectral functions $N(\omega,\vec{q})$
associated with the slow variables. There are two kinds of slow modes;
a \textit{diffusive mode} and a \textit{propagating mode}. The
diffusive mode is a mode that is purely dissipative, while the
propagating mode involves an oscillation besides a
dissipation. Examples for the propagating mode are the sound wave, the
spin wave and the ``particle mode.'' For definiteness, consider the
critical fluctuation with a fixed wavenumber $\vec{q} \ll 1$
[$\vec{q}$ not necessarily vanishing] in the following explanation.
The diffusive mode corresponds to the pole of the spectral function in
the complex plane of the frequency $\omega$ having only an imaginary
part $\Gamma$. In the spectrum as a function of $\omega$, the
diffusive mode is seen as a narrow peak with the width $\Gamma$, the
peak position of which is located at $\omega=0$ (so called the central
peak). The width $\Gamma$ represents the decay rate of the
fluctuation, that is, the inverse relaxation time. As the critical
point is approached, $\Gamma$ goes to zero, implementing the critical
slowing down.

As for the propagating mode, the spectral function has the form that
the pole with respect to complex $\omega$ has the real part $\Omega$
as well as the imaginary part $\Gamma$. The peak for the propagating
mode thus stands at $\omega=\Omega$ with the width $\Gamma$ in the
spectrum. The $\Omega$ represents the frequency of the collective
oscillation. For the sound wave, for example,
$\Omega=v_{\text s}q$ where $v_{\text s}$ is the sound velocity. We
note that because of the symmetry of the spectral function requested
by general arguments, a propagating mode with the frequency $\Omega$
is always accompanied by a propagating mode with the frequency
$-\Omega$, which propagates in the opposite direction in space. The
emergence of a pair of propagating modes means that two slow variables
must be involved. For the ``particle mode,'' these two modes standing
at $\omega=\pm\Omega$ in the spectrum of course correspond to the
particle and anti-particle modes. As we approach the critical point,
the width $\Gamma$ gets small as a consequence of the critical slowing
down. At the same time, the frequency $\Omega$ also goes to zero. This
phenomenon is called the softening of a propagating mode. For the
sound wave, the softening implies the decrease of the  sound velocity.
Noting that the sound velocity measures the stiffness of the system
explains the meaning of `softening.'

It may be helpful to express the above statements in terms of the pole
position in the complex $\omega$ plane. The diffusive mode is a mode
that is located on the imaginary axis. Near the critical point, the
pole moves on the imaginary axis toward the origin. The propagating
mode is given by a pole located at an arbitrary point in the plane
except on the imaginary axis, which shifts toward the origin as the
critical point is approached.

Thus, both the width $\Gamma$ of the slow (either diffusive or
propagating) mode and the frequency $\Omega$ of the propagating mode
become small near the critical point, representing the critical
slowing down and the softening, respectively. Concerning how $\Gamma$
and $\Omega$ go to zero, Hohenberg and Halperin have proposed the
dynamic scaling hypothesis \cite{hhhypothesis}: Let $\Gamma_q$ be a
typical inverse time scale for each slow mode. We can take the width
for $\Gamma_q$ for the diffusive mode. For the propagating mode, the
frequency or width plays the role of $\Gamma_q$. The dynamic scaling
hypothesis states that $\Gamma_q$ scales as a power law of the
correlation length $\xi$,
\begin{equation}
 \Gamma_{q}(\xi)=\xi^{-z}f(q\xi),
\end{equation}
where $f$ is some function. The index $z$ is called the dynamic
critical exponent and characterizes how the motion of the critical
fluctuation gets slow or how the propagating mode is softened. The
exponent $z$ may differ from mode to mode at one critical point. The
original dynamic scaling hypothesis insists that the exponents should
be identical for all the modes in the system. This seems to be a
reasonable assumption because the couplings among the modes are so
strong that all the modes are in the same motion. However it is known
that, in many systems a single value of $z$ is not always shared among
all the slow modes.

As is already described, the non-equilibrium processes near a critical
point are dominated by the long wavelength fluctuations of the slow
variables. It is reasonable to expect that the idea of universality,
which means that the dynamics is irrelevant to the microscopic
interactions in the system, is still meaningful for the dynamic case
just as for the static case. As is well known, the static universality
class is determined solely by the symmetry of the system. The dynamic
universality class, on the other, is dependent on what kinds of slow
variables are contained, besides on the symmetry. Hohenberg and
Halperin have defined their dynamic universality class
\cite{Hohenberg:1977ym}. The rules for classification are; (i) whether
or not the order parameter is conserved, (ii) what kinds of conserved
quantities are possessed by the system. According to the rules, the
ferromagnet and the antiferromagnet, for instance, belong to
distinctive dynamic universality classes because the slow variables in
the two systems are quite different, although their static
universality classes are the same.

\subsection{Dynamic universality class of the chiral phase transition}
\label{difference}

Now let us consider the chiral phase transition which is of our major
interest. The static universality class of the two flavor chiral phase
transition is, as is well known, the same as that of the O(4)
Heisenberg ferro- and antiferromagnet
\cite{Pisarski:ms,Rajagopal:1992qz}. In Ref.~\cite{Rajagopal:1992qz},
Rajagopal and Wilczek have discussed the dynamic universality class of
the chiral phase transition. They argued that the chiral phase
transition belongs to the same dynamic universality class as the
Heisenberg antiferromagnet, by comparing the slow variables of the
both systems. The slow variables of the Heisenberg antiferromagnet are
composed of the staggered magnetization $N$ which is the order
parameter and not conserved, the magnetization $M$ which is the
generator of rotation, and the energy $E$. In the chiral phase
transition, on the other hand, we have the four component meson field
$\phi_\alpha$ which plays the role of the non-conserved order
parameter, the six generators of the chiral rotation $Q_\alpha$, the
energy $E$ and the momentum $P^i$. We note that the momentum is
lacking in the Heisenberg antiferromagnet in which the spin is sticked
on the lattice sites. The difference, however, is not so crucial in
the following discussions. We have only to consider the itinerant
antiferromagnet rather than the Heisenberg antiferromagnet for
complete correspondence. Thus, according to Hohenberg and Halperin,
the two systems belong to the common dynamic universality class and
we could expect the same dynamic critical behavior in both systems.
Since the antiferromagnet had already been well studied \cite{hhma},
the analysis went on in the same way. Rajagopal and Wilczek employed
the kinetic equation provided by the time dependent Ginzburg Landau
theory. By applying the renormalization group method to the kinetic
equation, the critical exponent $z=d/2=3/2$ was obtained, where $d$ is
the spatial dimensionality.

It is at this point where we should find the crucial difference
between the antiferromagnet and the chiral phase transition by
thinking of the \textit{slow modes} rather than the \textit{slow
variables} in each system. Consider the disordered phase of the two
systems. For the antiferromagnet, the slow mode associated with the
order parameter is known to be diffusive. On the other, the slow mode
for the chiral order parameter is apparently the meson mode or a
``particle mode'' \cite{Hatsuda:1985eb,Son:2001ff,Koide:2003ax}, which
is a propagating mode. This difference should be taken seriously.

As is already mentioned, in general, we need two slow variables in
order to describe a propagating mode. This is a reflection of the fact
that a propagating mode always accompanies its counter partner in the
spectral function. The counter partner for a ``particle mode'' is just
an anti-particle mode. What are the two slow variables for the
description of the meson mode? The answer is that the two slow
variables for the meson mode are the meson field $\phi_\alpha$, which
is the order parameter, and the canonical momentum $\pi_\alpha$
conjugate to $\phi_\alpha$. This statement is
just what we would like to propose and emphasize in this paper. Note
that the canonical momentum is neither an order parameter nor a
conserved quantity. Since the slow modes in the antiferromagnet and
in the chiral phase transition are different, and accordingly, since
the slow variables in the two systems are different, it is not
possible any longer to identify the dynamic universality class of the
chiral phase transition with that of the antiferromagnet. In fact, as
we will calculate later, the dynamic critical exponent for the chiral
phase transition will be found to be $z=1-\eta/2\cong0.98$ within
our framework, which is apparently not the same as $z=d/2=1.5$ of the
antiferromagnet. The difference of the dynamic critical exponents
explicitly shows the different dynamic behavior in the two systems.

We have found that the dynamic universality class of the chiral phase
transition is not identical with that of the antiferromagnet. This
means that the classification method of the dynamic universality class
due to Hohenberg and Halperin is not applicable at least to the chiral
phase transition. What is the problem? The classification method of
Hohenberg and Halperin is essentially to compare the slow variables in
the system. The usual prescription to determine the slow variable,
which has been accepted without any doubt, is to gather the order
parameter and the conserved quantities of the system. As we have
noted, however, the canonical momentum $\pi_\alpha$ which is necessary
for a proper description of the meson mode is neither the order
parameter nor a conserved quantity. The chiral phase transiton
presents us with the situation in which the order parameter and the
conserved quantities do not exhaust all the slow variables
necessary for a description of the slow modes. The prescription of
just gleaning the order parameter and the conserved quantities for
determination of the slow variables has thus broken down. Accordingly
the classification method of Hohenberg and Halperin turns out
insufficient because it involves just comparing the order parameter
and the conserved quantities. If we consider all the slow modes or the
slow variables necessary for their description, it can happen that the
dynamic universality class given by Hohenberg and Halperin's method
splits into finer classes.

Now what we have to do is to reanalyze the chiral phase transition,
the dynamic universality class of which is not the same as that of the
antiferromagnet nor that of any other critical point that has ever
been considered. For that purpose, we will employ the mode coupling
theory. In the analysis, we will find that a propagating mode for the
meson mode appears appropriately from the meson field and the
canonical momentum conjugate to it. In the next section, we will
briefly review the mode coupling theory.

\section{Mode coupling theory}
\label{review}

The mode coupling theory is a phenomenological theory which provides
us with the kinetic equation for the critical fluctuations. It was
developed and established by Kawasaki
\cite{kawaprogress,kawaannals,kawareview,kawahiheiko}, and played an
important role in the extensive studies of the dynamic critical
phenomena in 1960-70's. Although the theory does not go beyond
phenomenology, it gives us the useful quantitative predictions.

In Subsec.~\ref{generalkinetic}, we will explain the general form of
the kinetic equation in the mode coupling theory. In
Subsec.~\ref{antiferromagnet}, as a demonstration, we will give
the kinetic equation for the Heisenberg antiferromagnet in the
disordered phase.

\subsection{Kinetic equation in the mode coupling theory}
\label{generalkinetic}

Our kinetic equation is based on the Mori equation \cite{mori}. In
general, it is necessary to separate the macroscopic degrees of
freedom (slow motions) from the microscopic ones (rapid motions) in
order to obtain a kinetic equation. This can be implemented by the
projection operator method \cite{zwanzig}. The Mori equation is
derived by applying the projection operator method to the microscopic
equation of motion, \ie, the Heisenberg equation. The Mori equation
deals with only the slow motions directly. In a sense, the Mori
equation is just the Heisenberg equation after the rapid fluctuations
are integrated out. Thus we can state that the kinetic equation in the
mode coupling theory originates from the microscopic Heisenberg
equation.

The explicit form of the kinetic equation in the mode coupling theory
is given by
\begin{eqnarray}
\frac{\text d}{\text d t}a_{\vec{q}}^j(t)
&=&\sum_l\left(
\text i\omega_{\vec{q}}^{jl}-\frac{k_{\text B}}{\chi_{\vec{q}}^l}
L_{jl}^0\right)
a_{\vec{q}}^l(t)\nonumber\\
&&+ \frac{\text i}{2}\sum_{\vec{k}}\sum_{lm}\Omega_{j;lm}
\left(\chi_{\vec{k}}^l\chi_{\vec{q}-\vec{k}}^m\right)^{-\frac{1}{2}}
\left(
a_{\vec{k}}^la_{\vec{q}-\vec{k}}^m
-\left\langle a_{\vec{k}}^la_{\vec{q}-\vec{k}}^m\right\rangle
_{\text{eq}}
\right)
+f_{\vec{q}}^j \:,
\label{finalkinetic}
\end{eqnarray}
where
\begin{align}
\omega_{\vec{q}}^{jl}
=&
-k_{\text B}T
\left(\chi_{\vec{q}}^l\right)^{-1}
\left\langle
\left[a_{\vec{q}}^j,a_{\vec{q}}^{l\dagger}\right]
\right\rangle_{\text{eq}},\label{frequency}\\
\Omega_{j;lm}
=&
-k_{\text B}T
\left(\chi_{\vec{k}}^l\chi_{\vec{q}-\vec{k}}^m\right)^{-\frac{1}{2}}
\biggl\{
\left\langle
\left[a_{\vec{q}}^j,a_{\vec{k}}^{l\dagger}a_{\vec{q}-{k}}^{m\dagger}\right]
\right\rangle_{\text{eq}} \notag\\
&\qquad\qquad\qquad\qquad\qquad
-\sum_p
\left\langle
\left[a_{\vec{q}}^j,a_{\vec{q}}^{p\dagger}\right]
\right\rangle_{\text{eq}}
{\chi_{\vec{q}}^p}^{-1}
\left(a_{\vec{q}}^p,a_{\vec{k}}^{l\dagger}a_{\vec{q}-\vec{k}}^{m\dagger}\right)
\biggr\}.
\label{coupling}
\end{align}

The $a_{\vec{q}}^j$ is the microscopic expression for the Fourier
transformed fluctuation of the $j$th slow variable. The
$\chi_{\vec q}^j$ is the static susceptibility given by
\begin{equation}
\chi_{\vec q}^j \;\delta_{jl}=
\left(a_{\vec q}^j, a_{\vec q}^l\right)\cong
\langle a_{\vec q}^j \, a_{\vec q}^l\rangle_{\text{eq}} \:,
\end{equation}
where the inner product $(F,G)$ is defined by
\begin{equation}
(F,G)\equiv
\frac{1}{\beta}
\int_0^\beta \text{d}\lambda
\left\langle e^{\lambda H}Fe^{-\lambda H}G
\right\rangle_{\text{eq}}
\end{equation}
with the inverse temperature $\beta=1/k_{\text B}T$ and the
Hamiltonian $H$. In the above expressions,
$\langle\cdots\rangle_{\rm eq}$ denotes the statistical average in the
equilibrium ensemble. In the following, we will omit the index `eq'
except for emphasizing it.

There are three terms in the right-hand side of
Eq.~(\ref{finalkinetic}), which are the linear term, the nonlinear
term and the noise term $f_{\vec q}^j$. The equation without the
nonlinear term describes the free motion of the slow modes. The
nonlinear term corresponds to the interactions among them.

The linear term consists of the frequency term and the dissipative
term, which give the peak position and its width in the spectral
function of the slow variables, respectively. The frequency matrix
$\omega_{jl}$ is given by the statistical average of the commutator of
the slow variables in the equilibrium ensemble, as is shown in
Eq.~(\ref{frequency}). For the classical system, the commutator is
replaced by the Poisson brackets. By diagonalizing the frequency
matrix, we obtain the normal coordinates out of the slow variables for
the individual slow modes. The nonzero eigenvalue means the appearance
of the propagating mode for the associated slow variables, while the
zero eigenvalue corresponds to the diffusive mode.

The $L^0$ in the dissipative term gives the Onsager (transport)
coefficient. Within the present framework, $L^0$ is given by the
correlation among only the short wavelength fluctuations, and thus
does not show any anomalous behavior in itself. In general, however,
the Onsager coefficient can be divergent at a critical point because
of the effect of the long wavelength fluctuation. In the mode coupling
theory, actually, the renormalization coming from the nonlinear term
can give rise to the divergence in the Onsager coefficient. In order
to emphasize that $L^0$ is a bare or unrenormalized quantity, the
suffix $0$ is put on $L$.

The nonlinear term, which represents the interactions among the slow
modes, is called the mode coupling term. It is known that the
couplings among the fluctuations of the slow variables become very
strong near a critical point \cite{fixman}. Thus the mode coupling
term is crucial for the critical dynamics. One of the effects of the
mode coupling term is the possible divergence of the Onsager
coefficient mentioned above. If the mode coupling term is dropped, the
theory reduces to the van Hove theory \cite{vanhove}.

We note that the coefficients in the kinetic equation are all given by
the statistical average in the equilibrium ensemble and determined by
the static properties of the system. The dynamic nature is contained
in the kinetic equation itself. Thus in the mode coupling theory, we
input the static information into the kinetic equation in order to
obtain the dynamic information. In particular, the criticality is
furnished by the singularity of the static susceptibilities. The
anomalous behavior at a critical point such as the critical slowing
down and the softening is induced by the divergence of the static
susceptibilities. One consequence is that the dynamic critical
exponents are expressed in term of the static critical exponents in
this description, as we will see later.

\subsection{Kinetic equation for the Heisenberg antiferromagnet}
\label{antiferromagnet}

For illustration of the general formalism in the previous section, we
here give the kinetic equation for the Heisenberg antiferromagnet in
the disordered phase \cite{kawaprogress,kawaannals,kawareview},
which would be useful for the comparison with the chiral phase
transition later.

The Hamiltonian of the Heisenberg antiferromagnet is given by
\begin{equation}
H=-\sum_{i\neq j}J_{ij}\vec{s}_i\cdot\vec{s}_j \:,
\label{Heisenbergham}
\end{equation}
where $\vec{s}_i$ is the spin operator vector put on the $i$th lattice
site which satisfies
\begin{equation}
\left[s_i^\alpha,s_j^\beta\right]=\text i\epsilon^{\alpha\beta\gamma}
\delta_{ij}s_i^\gamma \hspace{1cm}
\left(\alpha,\beta,\gamma=x,y,z\right).
\end{equation}
The $J_{ij}$ represents the exchange interaction and has a negative
sign for the antiferromagnet.

What we should do in the first place is to determine the slow
variables. The slow variables are made up of the staggered
magnetization (order parameter), the magnetization and the
energy (conserved quantities).

The Fourier transforms of the staggered magnetization
$\sigma_{\vec{k}}^\alpha$, the magnetization $\mu_{\vec{k}}^\alpha$
and the energy density $e_{\vec{k}}$ are given by
\begin{align}
\sigma_{\vec{k}}^\alpha
&=
\sum_i \eta^i e^{-\text i\vec{k}\cdot\vec{r}_i}
\left(s_i^\alpha-\left\langle s_i^\alpha\right\rangle\right),\\
\mu_{\vec{k}}^\alpha
&=
\sum_i e^{-\text i\vec{k}\cdot\vec{r}_i}
\left(s_i^\alpha-\left\langle s_i^\alpha\right\rangle\right),\\
e_{\vec{k}}
&= \sum_i e^{-\text i\vec{k}\cdot\vec{r}_i}
(-)\sum_j J_{ij}
\left(
\vec{s}_i\cdot\vec{s}_j
-\left\langle \vec{s}_i\cdot\vec{s}_j\right\rangle
\right).
\end{align}
where $\eta^i$ has an opposite sign for the nearest neighbors.

The frequency matrix $\omega_{\vec{q}}$ can be calculated by the
commutation relation between the spin variables. We find that the
matrix elements are all zero. This indicates that all the slow modes
are diffusive. Especially note that the slow mode for the staggered
magnetization is diffusive as it should be.

The dissipation terms can be found from the macroscopic hydrodynamics.
Since the magnetization and the energy are conserved quantities, we
have
\begin{align}
k_{\text B}L_{\mu\vec{k},\mu\vec{k}}^0/\chi_{\mu\vec{k}}
&= k^2L_{\mu}^0/\chi_{\mu\vec{k}} \:,\\
k_{\text B}L_{e\vec{k},e\vec{k}}^0/\chi_{e\vec{k}}
&= k^2\lambda^0/C_{\vec{k}} \:,
\end{align}
which define the `bare' Onsager coefficients. In particular,
$\lambda^0$ is the `bare' thermal conductivity. The $C_{\vec{k}}$ is
the $\vec{k}$-dependent heat capacity per spin. For the staggered
magnetization which is not a conserved quantity, we have
\begin{equation}
k_{\text B}L_{\sigma\vec{k},\sigma\vec{k}}^0/\chi_{\sigma\vec{k}}
=L_{\sigma}^0/\chi_{\sigma\vec{k}} \:.
\end{equation}
The susceptibilities $\chi$'s are defined by
\begin{align}
\chi_{\sigma\vec{k}}
&= \frac{1}{N}\left(\sigma_{\vec{k}}^\alpha,\sigma_{-\vec{k}}^\alpha\right),\\
\chi_{\mu\vec{k}}
&= \frac{1}{N}\left(\mu_{\vec{k}}^\alpha,\mu_{-\vec{k}}^\alpha\right),\\
\chi_{e\vec{k}}
&= \frac{1}{N}\left(e_{\vec{k}},e_{-\vec{k}}\right),
\end{align}
where $N$ is the total number of the spin site in the system.
The coefficients of the nonlinear term, $\Omega_{j;lm}$, can be
computed with the spin commutation relation.

Finally we can obtain the kinetic equation,
\begin{align}
\frac{\text d}{\text d t}\vec{\sigma}_{\vec{q}}
&=
-\frac{L_{\sigma}^0}{\chi_{\sigma\vec{q}}}\vec{\sigma}_{\vec{q}}
+\frac{k_{\text B}T}{N}\sum_{\vec{k}}
\left(
\frac{1}{\chi_{\sigma\vec{k}}}-\frac{1}{\chi_{\mu\vec{q}-\vec{k}}}
\right)
\vec{\sigma}_{\vec{k}}\times\vec{\mu}_{\vec{q}-\vec{k}}
+f_{\vec{q}}^\sigma \:,
\label{anti1}\\
\frac{\text d}{\text d t}\vec{\mu}_{\vec{q}}
&=
-q^2\frac{L_{\mu}^0}{\chi_{\mu\vec{q}}}\vec{\mu}_{\vec{q}}
+\frac{k_{\text B}T}{2N}\sum_{\vec{k}}
\left(
\frac{1}{\chi_{\mu\vec{k}}}-\frac{1}{\chi_{\mu\vec{q}-\vec{k}}}
\right)
\vec{\mu}_{\vec{k}}\times\vec{\mu}_{\vec{q}-\vec{k}}\notag\\
&\qquad\qquad\qquad\qquad
+\frac{k_{\text B}T}{2N}\sum_{\vec{k}}
\left(
\frac{1}{\chi_{\sigma\vec{k}}}-\frac{1}{\chi_{\sigma\vec{q}-\vec{k}}}
\right)
\vec{\sigma}_{\vec{k}}\times\vec{\sigma}_{\vec{q}-\vec{k}}
+f_{\vec{q}}^\mu \:,
\label{anti2}\\
\frac{\text d}{\text d t}e_{\vec{q}}
&=
-q^2\frac{\lambda^0}{C_{\vec{q}}}e_{\vec{q}}
+f_{\vec{q}}^e \:.
\label{anti3}
\end{align}
Note that the energy mode decouples from the magnetization and
staggered magnetization modes.

\section{Slow modes near the chiral phase transition in the
$\mathrm{O}(2)$ sigma model}
\label{O(2)}

In this section, we apply the mode coupling theory to the linear sigma
model for investigation of the chiral phase transition. We examine
what kind of slow modes, diffusive or propagating modes, appears near
the critical point.  For that purpose, it is sufficient to consider
the frequency matrix $\omega$ in the kinetic equation. If the matrix
element gives a finite value, it means that there appears a
propagating mode in the associated slow variable. The zero matrix
element indicates a diffusive mode. For simplicity, we consider the
$\mathrm{O}(2)$ linear sigma model. The Lagrangian is
\begin{equation}
\mathcal{L}=\frac{1}{2}
\left[
\left(\partial_\mu\phi_1\right)^2
+\left(\partial_\mu\phi_2\right)^2
\right]
-\frac{1}{2}\mu^2
\left({\phi_1}^2+{\phi_2}^2\right)
-\frac{\lambda}{4}\left({\phi_1}^2+{\phi_2}^2\right)^2,
\end{equation}
where $\phi_{a}$ with $a=1,2$ is the meson field, and $\mu$ and
$\lambda$ are the bare mass and coupling constant respectively. For
definiteness, we assume that the $\phi_1$ takes a finite vacuum
expectation value $\left\langle\phi_1\right\rangle=\phi$ in the
ordered phase. Thus the fluctuations of $\phi_1$ and $\phi_2$
correspond to the $\sigma$ and $\pi$ meson modes, respectively.

As we argued in Subsec.~\ref{difference}, the slow variables for the
chiral phase transition are composed of the following; the meson field
$\phi(\vec{x},t)$ (order parameter), the canonical momentum
$\pi(\vec{x},t)$, the chiral charge density $Q(\vec{x},t)$, the energy
density $E(\vec{x},t)$ and the momentum density $P^i(\vec{x},t)$. The
last three of these slow variables are the conserved quantities. Note
that there is only one chiral charge, which is contrasted with the six
charges in the $\mathrm{O}(4)$ case. The microscopic expressions for
$Q(\vec{x},t), E(\vec{x},t), P^i(\vec{x},t)$ are given by
\begin{align}
Q(\vec{x},t)
&=
\left(\pi_1\phi_2-\pi_2\phi_1\right)(\vec{x},t),\\
E(\vec{x},t)
&=
\left[
\frac{1}{2}\pi_a\mbox{}^2
+\frac{1}{2}\left(\vec{\nabla}\phi_a\right)^2
+\frac{1}{2}\mu^2\phi_a\mbox{}^2
+\frac{\lambda}{4}\left(\phi_a\mbox{}^2\right)^2
\right](\vec{x},t),\\
P^i(\vec{x},t)
&=
-\left(\pi_a\nabla^i\phi_a\right)(\vec{x},t).
\end{align}

Here, as usual, we expand $\phi_a(\vec{x},t)$ and $\pi_a(\vec{x},t)$
into the normal modes in terms of the creation and annihilation
operators.
\begin{align}
\phi_a(\vec{x},t)
&=
\frac{1}{\sqrt{V}}\sum_{\vec{k}}\frac{1}{\sqrt{2\omega_{a\vec{k}}}}
\left(a_{a\vec{k}}(t)+a_{a-\vec{k}}^{\dagger}(t)\right)
e^{\text i\vec{k}\cdot\vec{x}},\\
\pi_a(\vec{x},t)
&=
\frac{1}{\sqrt{V}}\sum_{\vec{k}}(-\text i)\sqrt{\frac{\omega_{a\vec{k}}}{2}}
\left(a_{a\vec{k}}(t)-a_{a-\vec{k}}^{\dagger}(t)\right)
e^{\text i\vec{k}\cdot\vec{x}},
\end{align}
where $V$ is the volume of the system. The dispersion relation
$\omega_{a\vec{k}}$ is generally complicated as a consequence of the
interaction. For the free system ($\lambda=0$), the Klein-Gordon
equation gives $\omega_{\vec{k}}= \sqrt{\vec{k}^2+\mu^2}$. We will
determine $\omega_{a\vec{k}}$ at the final stage of consideration by
imposing a consistency condition.

The usual quantization conditions
\begin{align}
&\left[
\phi_a(\vec{x},t)-\left\langle\phi_a\right\rangle, \pi_b(\vec{y},t)
\right]
=\text i\delta_{ab}\delta(\vec{x}-\vec{y}),
\label{commutation1}
\\
&\left[\phi_a(\vec{x},t)-\left\langle\phi_a\right\rangle,
\phi_b(\vec{y},t)-\left\langle\phi_b\right\rangle\right]
=\left[\pi_a(\vec{x},t),\pi_b(\vec{y},t)\right]=0,
\label{commutation1.5}
\\
\intertext{or}
&\left[a_{a\vec{k}}(t),a_{b\vec{k}'}^{\dagger}(t)\right]
=\delta_{ab}\delta_{\vec{k}\vec{k}'},
\label{commutation2}
\\
&\left[a_{a\vec{k}}(t),a_{b\vec{k}'}(t)\right]
=\left[a_{a\vec{k}}^{\dagger}(t),a_{b\vec{k}'}^{\dagger}(t)\right]
=0,
\end{align}
provide us with the algebra to calculate the frequency matrix.

The Fourier components of the fluctuation of the slow variables are
given by
\begin{align}
\phi_{a\vec{q}}(t)
&=
\frac{1}{\sqrt{V}}\int\text{d}^3x\, e^{-\text i\vec{q}\cdot\vec{x}}
\left(
\phi_a(\vec{x},t)-\left\langle\phi_a\right\rangle_{\rm eq}
\right),\\
\pi_{a\vec{q}}(t)
&=
\frac{1}{\sqrt{V}}\int\text{d}^3x\, e^{-\text i\vec{q}\cdot\vec{x}}
\pi_a(\vec{x},t),\\
Q_{\vec{q}}(t)
&=\epsilon_{ab}
\frac{1}{\sqrt{V}}\int\text{d}^3x\, e^{-\text i\vec{q}\cdot\vec{x}}
\pi_a\phi_b(\vec{x},t),\\
E_{\vec{q}}(t)
&=
\frac{1}{\sqrt{V}}\int\text{d}^3x\, e^{-\text i\vec{q}\cdot\vec{x}}
\biggl[\Bigl(
\frac{1}{2}\pi_a\mbox{}^2
+\frac{1}{2}\left(\vec{\nabla}\phi_a\right)^2
+\frac{1}{2}\mu^2\phi_a\mbox{}^2 \nonumber\\
&\qquad +\frac{\lambda}{4}\left(\phi_a\mbox{}^2\right)^2
\Bigr)(\vec{x},t)
-\left\langle\mbox{the same expression}\right\rangle_{\rm eq}
\biggr], \\
P_{\vec{q}}^i(t)
&=
\frac{1}{\sqrt{V}}\int\text{d}^3x\, e^{-\text i\vec{q}\cdot\vec{x}}
\left(-\pi_a\nabla^i\phi_a\right)(\vec{x},t),
\end{align}
where $\epsilon_{ab}$ is the total antisymmetric tensor with
$\epsilon_{12}=1$. Equivalently we have
\begin{align}
\phi_{a\vec{q}}(t)
=&
\frac{1}{\sqrt{2\omega_{a\vec{k}}}}
\left(
\left(a_{a\vec{q}}+a_{a-\vec{q}}^{\dagger}\right)
-\left\langle a_{a\vec{q}}+a_{a-\vec{q}}^{\dagger}\right\rangle_{\rm eq}
\right),\\
\pi_{a\vec{q}}(t)
=&
(-\text i)\sqrt{\frac{\omega_{a\vec{q}}}{2}}
\left(a_{a\vec{q}}(t)-a_{a-\vec{q}}^{\dagger}(t)\right),\\
Q_{\vec{q}}(t)
=&
\epsilon_{ab}
\frac{1}{\sqrt{V}}\sum_{\vec{k}}\frac{-\text i}{2}
\sqrt{\frac{\omega_{a\vec{k}}}{\omega_{b\vec{q}-\vec{k}}}}
\left(a_{a\vec{k}}-a_{a-\vec{k}}^{\dagger}\right)
\left(a_{b\vec{q}-\vec{k}}+a_{b\vec{k}-\vec{q}}^{\dagger}\right),\\
E_{\vec{q}}(t)
=&
\frac{1}{2}\frac{1}{\sqrt{V}}\sum_{\vec{k}}\frac{-1}{2}
\sqrt{\omega_{a\vec{k}}\omega_{a\vec{q}-\vec{k}}}
\left(a_{a\vec{k}}-a_{a-\vec{k}}^{\dagger}\right)
\left(a_{a\vec{q}-\vec{k}}-a_{a\vec{k}-\vec{q}}^{\dagger}\right)
\nonumber\\
&+ \!\frac{1}{2}\frac{1}{\sqrt{V}}\sum_{\vec{k}}
\left[-\vec{k}\!\cdot\!(\vec{q}\!-\!\vec{k})\!+\!\mu^2\right]
\frac{1}{2}
\frac{1}{\sqrt{\omega_{a\vec{k}}\omega_{a\vec{q}-\vec{k}}}}
\!\left(a_{a\vec{k}}\!+\!a_{a-\vec{k}}^{\dagger}\right)
\left(a_{a\vec{q}-\vec{k}}\!+\!a_{a\vec{k}-\vec{q}}^{\dagger}\right)
\nonumber\\
&+ \frac{\lambda}{4}\frac{1}{V^{\frac{3}{2}}}
\sum_{\vec{k_1}\cdots\vec{k_4}}\frac{1}{4}
\delta_{\vec{q},\vec{k_1}+\cdots+\vec{k_4}}
\frac{1}{\sqrt{
\omega_{a\vec{k_1}}\omega_{a\vec{k_2}}
\omega_{b\vec{k_3}}\omega_{b\vec{k_4}}}}
\nonumber\\
&\qquad\qquad\times
\left(a_{a\vec{k_1}}+a_{a-\vec{k_1}}^{\dagger}\right)
\left(a_{a\vec{k_2}}+a_{a-\vec{k_2}}^{\dagger}\right)
\left(a_{b\vec{k_3}}+a_{b-\vec{k_3}}^{\dagger}\right)
\left(a_{b\vec{k_4}}+a_{b-\vec{k_4}}^{\dagger}\right)
\nonumber\\
&- \left\langle\mbox{the same expression}\right\rangle_{\rm eq}
\delta_{\vec{q},0} \:,
\\
P_{\vec{q}}^i(t)
=&
-\frac{1}{\sqrt{V}}\sum_{\vec{k}}\frac{1}{2}
(q-k)^i
\sqrt{\frac{\omega_{a\vec{k}}}{\omega_{b\vec{q}-\vec{k}}}}
\left(a_{a\vec{k}}-a_{a-\vec{k}}^{\dagger}\right)
\left(a_{a\vec{q}-\vec{k}}+a_{a\vec{k}-\vec{q}}^{\dagger}\right).
\end{align}

The frequency matrix element is given by the commutator of each slow
variable in the mode coupling theory. By noting that the commutator
between $\phi_a$ and $\pi_a$ is not zero, we can see a propagating
mode appearing from these two slow variables. This propagating mode
should and can be identified with the meson mode, which will be
discussed later. Note that if we do not include the canonical momentum
into the group of the slow variables just following the usual
prescription for determination of the slow variable, this propagating
mode would not appear and the order parameter fluctuation would
undesirably show diffusive behavior. The canonical momentum plays an
essential role in the chiral phase transition.

The inclusion of $\pi_a$ is found to be reasonable if we consider the
microscopic equation of motion, that is, the Heisenberg equation. The
Heisenberg equation for the meson field consists of the two equations
for the variables of $\phi$ and $\pi$ which are conjugate to each
other;
\begin{align}
\text i\frac{\text d}{\text d t}\phi(\vec{x},t)
&=
\left[\phi(\vec{x},t),H\right],\\
\text i\frac{\text d}{\text d t}\pi(\vec{x},t)
&=
\left[\pi(\vec{x},t),H\right],
\end{align}
where $H$ is the Hamiltonian of the system. From these two equations,
we obtain the Klein-Gordon equation for the meson dynamics.
\footnote{We note that the Klein-Gordon equation is of second order
with respect to the time derivative while the Heisenberg equation is
of first order. In order to reproduce the Klein-Gordon equation from
the Heisenberg equation, it is inevitable to set up the simultaneous
equations for two variables.}
Since our kinetic equation in the mode coupling theory is based on the
Mori equation which is derived from the Heisenberg equation, it is
natural that we need $\pi_a$ as the degree of freedom for a
description of the meson dynamics.

Turning to the case of the antiferromagnet, we realize that the
Heisenberg equation for the spin variables is
\begin{align}
&\text i\frac{\text d s_i^{\alpha}}{\text d t}
=\left[s_i^{\alpha}, H\right]
\hspace{2cm}(\alpha=x,y,z)
\end{align}
with $H$ being given by Eq.\ (\ref{Heisenbergham}).
Thus the staggered magnetization $\vec{\sigma}=(\sigma^x, \sigma^y,
\sigma^z)$ corresponding to the spin variable
$\vec{s}=(s^x, s^y,s^z)$ should be the degree of freedom for the
dynamics of the antiferromagnet. In fact, from the variable of the
staggered magnetization arises a diffusive mode as it should be, as we
have already seen in Sec.~\ref{review}.

The canonical momentum as a slow variable is itself nothing new
in the chiral phase transition. We know at least two examples of
critical points in which the canonical momentum enters into the
member of the slow variable.

One is the $\lambda$ transition in the superfluid. The order parameter
of the transition is given by the wavefunction of the Bose condensate
$\Phi$. Because the wavefunction is a complex variable, the order
parameter consists of the two components, \ie, $\Phi$ and
$\Phi^\dagger$. The wavefunction $\Phi$ obeys the non-relativistic
Sch\"{o}dinger equation, and as is well known, the canonical momentum
conjugate to $\Phi$ is given by $\Phi^\dagger$.
From the variables of $\Phi$ and $\Phi^\dagger$ (and the entropy
density), there appears the second sound mode which is a propagating
mode (and one diffusive mode) in the superfluid transition.

The other example is associated with the spin wave mode in the
ferromagnet. In the ordered phase, supposing that the spontaneous
magnetization points in the $z$ direction, the transverse
fluctuations of the magnetization, $M_x$ and $M_y$, make up the spin
wave. We can regard the two variables of $M_{\pm}=M_x\pm\text i M_y$
as conjugate to each other.

Note that in the above two examples, the conjugate variable can be
included into the member of the slow variable just by collecting the
order parameter and the conserved quantity. In the chiral phase
transition, however, the canonical momentum cannot be the slow
variable by the prescription. This is nothing but the new feature of the
chiral phase transition that we have not encountered in the
systems so far considered in Ref.~\cite{Hohenberg:1977ym}.

Alternatively to $\phi_a$ and $\pi_a$, we can take
\begin{equation}
\frac{1}{\sqrt{2\omega_{a\vec{q}}}}
\left(a_{a\vec{q}}-\left\langle a_{a\vec{q}}\right\rangle\right)
\hspace{1cm}\mbox{and}\hspace{1cm}
\frac{1}{\sqrt{2\omega_{a\vec{q}}}}
\left(a_{a-\vec{q}}^{\dagger}-
\left\langle a_{a-\vec{q}}^{\dagger}\right\rangle
\right)
\end{equation}
for the two variables. These are obtained by dividing the meson
field $\phi_{a\vec{q}}$, and are just a liner transformation of
$\phi_a$ and $\pi_a$. In the following discussions, we will employ
$a_a$ and $a_a^\dagger$ instead of $\phi_a$ and $\pi_a$ for the two
degrees of freedom. This is because $a_a$ and $a_a^\dagger$ give rise
to the diagonal frequency matrix as we will see later and are more
convenient.

Thus for the two degrees of freedom to describe the meson mode, we
will adopt
\begin{align}
\phi_{a\vec{q}}(t)
&\equiv
\frac{1}{\sqrt{2\omega_{a\vec{q}}}}
\left(a_{a\vec{q}}-\left\langle a_{a\vec{q}}\right\rangle\right),\\
\phi_{a-\vec{q}}^{\dagger}(t)
&\equiv
\frac{1}{\sqrt{2\omega_{a\vec{q}}}}
\left(a_{a-\vec{q}}^{\dagger}-
\left\langle a_{a-\vec{q}}^{\dagger}\right\rangle
\right).
\end{align}

Now we calculate the frequency matrix $\omega_{jl}$ given by
\begin{equation}
\omega_{jl}=-k_{\text B}T\left(\chi_l\right)^{-1}
\left\langle\left[A_j, A_l^\dagger\right]\right\rangle
\end{equation}
for the slow variables of
\begin{equation}
A_j=\left\{\phi_{1\vec{q}},\phi_{1-\vec{q}}^{\dagger},
\phi_{2\vec{q}},\phi_{2-\vec{q}}^{\dagger},Q_{\vec{q}},
E_{\vec{q}},P_{\vec{q}}^i\right\},
\label{order} 
\end{equation}
where $j$ on $A_j$ specifies each slow variable.
The static susceptibilities for $\phi_{a\vec{q}}$ and
$\phi_{a-\vec{q}}^{\dagger}$ are given by
\begin{align}
\chi_{a\vec{q}}
&\equiv
\left(\phi_{a\vec{q}},\left(\phi_{a\vec{q}}\right)^\dagger\right),\\
\chi_{a^\dagger\vec{q}}
&\equiv
\left(\phi_{a-\vec{q}}^{\dagger},
\left(\phi_{a-\vec{q}}^{\dagger}\right)^\dagger\right)
=\left(\phi_{a-\vec{q}},\left(\phi_{a-\vec{q}}\right)^\dagger\right)
=\chi_{a\vec{q}},
\end{align}
where the last equality holds because the susceptibility is a function
of $|\vec{q}|^2$. The other susceptibilities are defined similarly by
\begin{align}
\chi_{Q\vec{q}}
&\equiv
\left(Q_{\vec{q}},Q_{\vec{q}}^\dagger\right),\\
\chi_{e\vec{q}}
&\equiv
\left(E_{\vec{q}},E_{\vec{q}}^\dagger\right),\\
\chi_{p\vec{q}}
&\equiv
\left(P_{\vec{q}}^i,P_{\vec{q}}^{i\dagger}\right),
\end{align}
where $\chi_{p\vec{q}}$ does not depend on $i$.

Firstly, let us consider the disordered phase, in which $\phi_1$
and $\phi_2$ are degenerated. The frequency matrix is calculated to be
\begin{equation}
\omega_{jl}=
\left(
\begin{array}{ccccccc}
-\frac{k_{\text B}T}{2\omega_{1\vec{q}}\chi_{1\vec{q}}}
& 0 & 0 & 0 & 0 & 0 & 0 \\
0 & \frac{k_{\text B}T}{2\omega_{1\vec{q}}\chi_{1\vec{q}}}
& 0 & 0 & 0 & 0 & 0 \\
0 & 0 &
-\frac{k_{\text B}T}{2\omega_{2\vec{q}}\chi_{2\vec{q}}}
& 0 & 0 & 0 & 0 \\
0 & 0 & 0 &
\frac{k_{\text B}T}{2\omega_{2\vec{q}}\chi_{2\vec{q}}}
& 0 & 0 & 0 \\
0 & 0 & 0 & 0 & 0 & 0 & 0 \\
0 & 0 & 0 & 0 & 0 & 0
& -\frac{D}{\chi_{p\vec{q}}}q^i \\
0 & 0 & 0 & 0 & 0
& -\frac{D}{\chi_{e\vec{q}}}q^i & 0
\end{array}
\right),
\end{equation}
where
\begin{equation}
D=k_{\text B}T\frac{1}{V}\sum_{\vec{k}}
2\left(\omega_{a\vec{k}}\mbox{}^2+\frac{1}{3}\vec{k}^2\right)
\chi_{a\vec{k}} \:,
\end{equation}
and the matrix elements are placed in the order of Eq.\ (\ref{order}).
We note that there appear propagating modes for $\phi_1$ and $\phi_2$,
which should be identified with the $\sigma$ and $\pi$ meson modes.
The kinetic equation without the dissipative and nonlinear terms is
\begin{equation}
\frac{\text d}{\text d t}\phi_{a\vec{q}}
=
-\text i\frac{k_{\text B}T}{2\omega_{a\vec{q}}\chi_{a\vec{q}}}
\phi_{a\vec{q}} \:,\qquad
\frac{\text d}{\text d t}\phi_{a-\vec{q}}^{\dagger}
=
\text i\frac{k_{\text B}T}{2\omega_{a\vec{q}}\chi_{a\vec{q}}}
\phi_{a-\vec{q}}^{\dagger} \:.
\label{eq:free_eq}
\end{equation}
Here we impose a consistency condition that the frequency
coincide with the dispersion $\omega_{a\vec{q}}$, that is,
\begin{equation}
\frac{k_{\text B}T}{2\omega_{a\vec{q}}\chi_{a\vec{q}}}
= \omega_{a\vec{q}}.
\label{renormcondition1}
\end{equation}
This is justified because in the mode coupling theory, the kinetic
equation without the dissipation and mode coupling term should
reproduce the Heisenberg equation \cite{kawaannals}. In fact,
the equations given by Eq.~(\ref{eq:free_eq}) with
Eq.~(\ref{renormcondition1}) substituted are combined into
\begin{equation}
 \begin{split}
 \frac{\mathrm{d}}{\mathrm{d}t}\bigl(\phi_{a\vec{q}}+
  \phi_{a-\vec{q}}^\dagger\bigr)
 &= -\mathrm{i}\omega_{a\vec{q}}\bigl(\phi_{a\vec{q}}-
  \phi_{a-\vec{q}}^\dagger\bigr), \\
 \frac{\mathrm{d}}{\mathrm{d}t}(-\mathrm{i})\omega_{a\vec{q}}
  \bigl(\phi_{a\vec{q}}-\phi_{a-\vec{q}}^\dagger\bigr)
 &= -\omega_{a\vec{q}}\mbox{}^2 \bigl(\phi_{a\vec{q}}+
  \phi_{a-\vec{q}}^\dagger\bigr),
 \end{split}
\end{equation}
which give the Heisenberg equation or the Klein-Gordon equation
with the dispersion relation $\omega_{a\vec{q}}$. The condition
(\ref{renormcondition1}) determines the dispersion relation as
\begin{equation}
\omega_{a\vec{q}}=\sqrt{\frac{k_{\text B}T}{2\chi_{a\vec{q}}}}.
\end{equation}
In order to find $\chi_{a\vec{q}}$, consider the susceptibility for
the field $\phi(\vec{x})$ itself. In the Ornstein-Zernike
approximation, the susceptibility is written as
\begin{equation}
\left(\phi(\vec{x}),\phi(\vec{y})^\dagger\right)
=\frac{1}{V}\sum_{\vec{k}}
\frac{k_{\text B}T Z}{\vec{k}^2+m^2}
e^{\text i\vec{k}\cdot(\vec{x}-\vec{y})},
\end{equation}
where $m$ is the static screening mass. On the other hand,
\begin{align}
\left(\phi(\vec{x}),\phi(\vec{y})^\dagger\right)
&=
\left(
\frac{1}{\sqrt{V}}\sum_{\vec{k}}
\left(\phi_{\vec{k}}+\phi_{-\vec{k}}^\dagger\right)
e^{\text i\vec{k}\cdot\vec{x}},
\frac{1}{\sqrt{V}}\sum_{\vec{k}'}
\left(\phi_{\vec{k}'}^\dagger+\phi_{-\vec{k}'}\right)
e^{-\text i\vec{k}'\cdot\vec{y}}
\right)\nonumber\\
&=
\frac{1}{V}\sum_{\vec{k}}2\chi_{\vec{k}}
e^{\text i\vec{k}\cdot(\vec{x}-\vec{y})}.
\end{align}
By comparing the two expressions, we find
\begin{equation}
2\chi_{\vec{k}}=\frac{k_{\text B}T Z}{\vec{k}^2+m^2}.
\end{equation}
Thus we obtain the dispersion relation
\begin{equation}
\omega_{a\vec{q}}=Z^{-\frac{1}{2}}\sqrt{\vec{q}^2+m^2}.
\end{equation}
We should note that the static screening mass $m$ plays the role
of the pole mass in this framework.
As we approach the critical point, $m$ goes to zero, which describes
the softening of the meson modes. The way of the softening, or the way
how $m$ gets small, can be determined by the static scaling law, \ie,
$m\sim\xi^{-1}$. The renormalization constant $Z$ yields the anomalous
dimension as $Z\sim\xi^{\eta}$, which leads to the dynamic critical
exponent $1-\eta/2$ as is already advertised. Although the present
argument does not take account of the nonlinear or mode coupling terms
which are important near the critical point, the result does not alter
even when they are properly taken care of. The full analysis will be
given in Sec.~\ref{exponent}.

We can see that the slow mode for the chiral charge is diffusive.
Moreover, the energy and the longitudinal component of the momentum
couple to give rise to a propagating mode, while the transverse
components give diffusive modes. The kinetic equations for the
energy and the momentum are written as
\begin{align}
\frac{\text d}{\text d t} E_{\vec{q}}
&=
-\text i\frac{D}{\chi_{p\vec{q}}}q^iP_{\vec{q}}^{\text Li},\\
\frac{\text d}{\text d t} P_{\vec{q}}^{\text Li}
&=
-\text i\frac{D}{\chi_{e\vec{q}}}q^iE_{\vec{q}},\\
\frac{\text d}{\text d t} P_{\vec{q}}^{\text Ti}
&= 0,
\end{align}
where we defined $P_{\vec{q}}^{\text Li}$ and $P_{\vec{q}}^{\text Ti}$
by
\begin{equation}
P_{\vec{q}}^i=
\left(\delta^{ij}-\frac{q^iq^j}{\vec{q}^2}\right)P_{\vec{q}}^j
+\frac{q^iq^j}{\vec{q}^2}P_{\vec{q}}^j
\equiv P_{\vec{q}}^{\text Ti}+P_{\vec{q}}^{\text Li}.
\end{equation}
We note that here is the difference from the Heisenberg
antiferromagnet in which the momentum is absent. In the Heisenberg
antiferromagnet, the slow mode for the energy density is diffusive,
while the chiral system gives the propagating mode for the energy
because of the presence of the momentum density. What is this
propagating mode? It would be essentially a sound wave, but not
exactly the same as the usual first sound of the normal fluid. For
comparison, consider the slow modes in the normal fluid
\cite{kawareview,kadanoff,forster}, in which the slow variables are
the particle number density $N_{\vec{q}}$, the energy density
$E_{\vec{q}}$, and the velocity $u_{\vec{q}}^i$. The $N_{\vec{q}}$ and
$E_{\vec{q}}$ may be transformed to the entropy density $S_{\vec{q}}$
and the pressure $p_{\vec{q}}$ through
\begin{align}
S_{\vec{q}}
&=
\frac{1}{T}\left(E_{\vec{q}}-mhN_{\vec{q}}\right),\\
p_{\vec{q}}
&=
\frac{m}{\rho\chi_s(\vec{q})}N_{\vec{q}}-\frac{1}{\alpha_s(\vec{q})}
S_{\vec{q}},
\end{align}
where $\rho$ and $h$ are the equilibrium values of the density and the
enthalpy per unit mass, $m$ is the mass of a molecule, and
$\chi_s(\vec{q})$ and $\alpha_s(\vec{q})$ are the $\vec{q}$-dependent
adiabatic compressibility and adiabatic thermal expansion
coefficients. The kinetic equation shows that the modes for
$S_{\vec{q}}$ and $u_{\vec{q}}^{\text T}$ are diffusive while
$p_{\vec{q}}$ and $u_{\vec{q}}^{\text L}$ combine to give the first
sound mode. See Table~\ref{aboveTc}.
\begin{table}[b]
\begin{center}
 \begin{tabular}{|ccl|ccccl|} \hline
  \multicolumn{3}{|c|}{Chiral system} &
  \multicolumn{5}{|c|}{Normal fluid} \\ \hline
  $ 
  \begin{array}{c}
   Q \\
    \left.
    \begin{array}{c@{\,}} E \\ P^{\text L} \end{array}
   \right\} \\
   P^{\text T}
  \end{array}
  $
  &
  $
  \hspace{-4mm}
  \begin{array}{c}
  \cdots \\
   \begin{array}{c} {} \\ {} \end{array} \cdots
   \begin{array}{c} {} \\ {} \end{array} \\
   \cdots
  \end{array}
  $
  &
  $
  \hspace{-4mm}
  \begin{array}{c}
   \mbox{Diffusive mode} \\
   \begin{array}{c} {} \\ {} \end{array} \mbox{Propagating mode}\\
   \mbox{Diffusive mode}
  \end{array}
  $
  &
  $
  \begin{array}{c}
	   \left.  \begin{array}{c@{\,}} N \\ E \end{array} \right\}
  \\ {} \\ {}
  \end{array}
  $
  &
  $
  \hspace{-6mm}
  \begin{array}{c}
   \begin{array}{c}
	\rightarrow \left\{ \begin{array}{c}{} \\ {} \end{array} \right.
   \end{array}
   \\ {} \\ {}
  \end{array}
  $
  &
  $
  \hspace{-9mm}
  \begin{array}{c}
   S \\
   \left. \begin{array}{c} p \\ u^{\text L} \end{array}\right\}\\
   u^{\text T}
  \end{array}
  $
  &
  $
  \hspace{-7mm}
  \begin{array}{c}
   \cdots \\
   \begin{array}{c}{} \\ {} \end{array}  \cdots
   \begin{array}{c}{} \\ {} \end{array}  \\
   \cdots
    \end{array}
  $
  &
  $
  \hspace{-5mm}
  \begin{array}{l}
   \mbox{Diffusive mode} \\
   \begin{array}{l} {} \\ {} \end{array} \mbox{First sound} \\
   \mbox{Diffusive mode}
  \end{array}
  $
  \\ \hline
 \end{tabular}
\end{center}
\caption{Comparison of the chiral system with the normal fluid.}
\label{aboveTc}
\end{table}
Thus our propagating mode is to be compared with the first sound in
the normal fluid. If the first sound is called the pressure wave, our
propagating mode may be called the energy wave mode. If it is allowed
to identify the chiral charge in the chiral system with the number
density in the normal fluid, we can say that the basis that
diagonalizes the equation in each system is different. The basis of
the one system can be obtained by the linear transformation from that
of the other system.

Now we move to the ordered phase. The frequency matrix in the ordered
phase is found to be
\begin{equation}
\omega_{jl}=
\left(
\begin{array}{ccccccc}
-\frac{k_{\text B}T}{2\omega_{1\vec{q}}\chi_{1\vec{q}}}
& 0 & 0 & 0 & 0
& \frac{-B}{\chi_{e\vec{q}}} & 0 \\
0 & \frac{k_{\text B}T}{2\omega_{1\vec{q}}\chi_{1\vec{q}}}
& 0 & 0 & 0
& \frac{B^*}{\chi_{e\vec{q}}} & 0 \\
0 & 0 &
-\frac{k_{\text B}T}{2\omega_{2\vec{q}}\chi_{2\vec{q}}}
& 0 &
\frac{A}{\chi_{Q\vec{q}}} & 0 & 0 \\
0 & 0 & 0 &
\frac{k_{\text B}T}{2\omega_{2\vec{q}}\chi_{2\vec{q}}}
& \frac{-A^*}{\chi_{Q\vec{q}}} & 0 & 0 \\
0 & 0 &
\frac{A^*}{\chi_{2\vec{q}}} & \frac{-A}{\chi_{2\vec{q}}} & 0 & 0 & 0 \\
\frac{-B^*}{\chi_{1\vec{q}}} & \frac{B}{\chi_{1\vec{q}}} & 0 & 0 & 0 & 0
& -\frac{D}{\chi_{p\vec{q}}}q^i \\
0 & 0 & 0 & 0 & 0
& -\frac{D}{\chi_{e\vec{q}}}q^i & 0
\end{array}
\right),
\end{equation}
where
\begin{align}
A
=&
k_{\text B}T\frac{\text i}{2\sqrt{V}}\frac{1}{\sqrt{2\omega_{2\vec{0}}}}
\left\langle a_{1\vec{0}}+a_{1\vec{0}}^{\dagger}\right\rangle,\\
B
=&
k_{\text B}T\frac{1}{2\sqrt{V}}\frac{1}{\sqrt{2\omega_{1\vec{q}}}}
\mu^2
\frac{1}{\sqrt{
\omega_{1\vec{q}}\omega_{1\vec{0}}}}
\left\langle a_{1\vec{0}}+a_{1\vec{0}}^{\dagger}\right\rangle\nonumber\\
&+
k_{\text B}T\frac{\lambda}{4}\frac{1}{V^{\frac{3}{2}}}
\frac{1}{\sqrt{2\omega_{1\vec{q}}}}
\sum_{\vec{k}_1\cdots\vec{k}_4}
\delta_{\vec{q},-\vec{k}_1}
\delta_{-\vec{q},\vec{k}_1\cdots\vec{k}_4}
\frac{1}{\sqrt{
\omega_{1\vec{k}_1}\omega_{1\vec{k}_2}\omega_{b\vec{k}_3}
\omega_{b\vec{k}_4}}}
\nonumber\\
&\hspace{3cm}\times
\left\langle
\left(a_{1\vec{k}_2}+a_{1-\vec{k}_2}^{\dagger}\right)
\left(a_{b\vec{k}_3}+a_{b-\vec{k}_3}^{\dagger}\right)
\left(a_{b\vec{k}_4}+a_{b-\vec{k}_4}^{\dagger}\right)
\right\rangle .
\end{align}
The $\phi_1$ mode, \ie, the sigma meson mode can be treated in the
same way as in the disordered phase. The $\phi_2$ mode now couples to
the chiral charge $Q$. We concentrate on the $3\times3$ matrix of
$\left\{\phi_{2\vec{q}},\phi_{2-\vec{q}}^{\dagger},Q_{\vec{q}}\right\}$.
It can be diagonalized readily to give the eigenvalues
\begin{equation}
\lambda = 0,\
\pm\sqrt{
\left(\frac{k_{\text B}T}{2\omega_{2\vec{q}}\chi_{2\vec{q}}}\right)^2
+\frac{2|A|^2}{\chi_{2\vec{q}}\chi_{Q\vec{q}}}}.
\end{equation}
The zero eigenvalue gives a diffusive mode. The resulting propagating
mode should be regarded as the $\phi_2$ mode modified by the
interaction or the coupling with the chiral charge. Thus we impose the
consistency condition that the eigenvalue be identical to
$\omega_{2\vec{q}}$. Moreover since the $\phi_2$ mode must be the
Nambu-Goldstone mode, we should have $\omega_{2\vec{q}}=uq$, where $u$
is the pion velocity \cite{Son:2001ff}.
Thus we have
\begin{equation}
\sqrt{
\left(\frac{k_{\text B}T}{2\omega_{2\vec{q}}\chi_{2\vec{q}}}\right)^2
+\frac{2|A|^2}{\chi_{2\vec{q}}\chi_{Q\vec{q}}}}
=\omega_{2\vec{q}}=uq.
\end{equation}
Furthermore if we assume
\begin{equation}
\frac{2\chi_{2\vec{q}}}{k_{\text B}T}
=\frac{Z_{\pi}}{\vec{q}^2},
\end{equation}
then in the approximation of $|\vec{q}|\ll 1$, we find
\begin{equation}
\frac{1}{u^2Z_{\pi}^2}+
\frac{4|A|_{\vec{q}=0}^2}{k_{\text B}T\chi_{Q\vec{q}=0}}
\frac{1}{Z_{\pi}}=u^2,
\end{equation}
from which the renormalization constant $Z_{\pi}$ can be determined
if $u$ is given. This treatment for the pion mode is essentially the
same as that for the second sound mode in the superfluid transition
\cite{kawaannals}.

We can see that in the ordered phase, the modes of the energy
and momentum couple to the sigma meson mode through the
nondiagonal elements.

For summary, let us compare the ordered phase with the superfluid
\cite{kawaannals,forster,hohenmartin}. See Table~\ref{belowTc}.
\begin{table}[b]
\begin{center}
\begin{tabular}{|ccl|ccl|} \hline
\multicolumn{3}{|c|}{Chiral system in the ordered phase} &
\multicolumn{3}{|c|}{Superfluid} \\ \hline
 $\phi_1$ & $\cdots$ & $\sigma\ \mbox{mode}$ &  &  &  \\
 $ \left.
  \begin{array}{l@{\,}}
   \phi_2 \\ Q
  \end{array}
  \right\}
 $
 & $\cdots$ 
 &
 \hspace{-3mm}
 $
 \left(
 \begin{array}{l}
  \pi\ \mbox{mode (NG mode)}\\
  +\ \mbox{Diffusive mode}
 \end{array}\right.
 $
 &
 $
 \left.
 \begin{array}{c@{\,}}
  \Phi, \Phi^\dagger \\ S
 \end{array}
 \right\}
 $
 &  $\cdots$
 &
 \hspace{-3mm}
 $
 \left(
 \begin{array}{l}
  \mbox{Second sound}\ \mbox{(NG mode)}\\
  +\ \mbox{Diffusive mode}
 \end{array}\right.
 $
 \\
 $
 \left.
 \begin{array}{l@{\,}}
  E \\ p^{\text L}
 \end{array}
 \right\}
 $
 & $\cdots$ & Energy wave
 &
 $
 \left.
 \begin{array}{c}
  p \\ u^{\text L}
 \end{array}
 \right\}
 $
 & $\cdots$ & First sound
 \\
 $p^{\text T}$ & $\cdots$ & Diffusive mode
 & $u^{\text T}$ & $\cdots$ & Diffusive mode
 \\ \hline
\end{tabular}
\end{center}
\caption{Comparison of the ordered phase of the chiral system
with the superfluid.}
\label{belowTc}
\end{table}
In the superfluid, the order parameter $(\Phi, \Phi^\dagger)$ and the
entropy $S$ constitute the second sound wave mode and the diffusive
mode. If we call the second sound the entropy wave, the $\pi$ mode may
be viewed as the chiral charge wave in the analogy with the
superfluid.

Finally, we consider the hydrodynamic modes of the chiral system away
from the critical point. In general, the conserved quantitiy and the
transverse component of the order parameter in the ordered phase are
slow variables even away from the critical point \cite{forster}, the
latter of which is slow because of a dynamical reason rather than a
kinematical one.  The fluctuations of those variables make up the slow
(or hydrodynamic) modes and appear in the spectral functions as narrow
peaks though their widths are finite for the finite wavelength
fluctuations. In the chiral system, the conserved quantities, $Q$, $E$
and $P^i$, are the slow variables, so that the slow modes found in the
above consideration will be responsible for the hydrodynamics of the
system. In the disordered phase, since the meson fields $\phi_1$ and
$\phi_2$ are not conserved quantities, the fluctuations of $\phi_1$
and $\phi_2$ do not make hydrodynamic mode and decay (or relax) in
the microscopic time scale. Thus the $\sigma$ and $\pi$ meson modes
will disappear away from the critical point \cite{Hatsuda:1985eb}. In
the ordered phase, $\phi_2$ being the transverse fluctuation of the
order parameter is the dynamically slow variable. Thus the $\pi$ meson
mode appears as a hydrodynamic mode. The $\sigma$ meson mode, on the
other hand, is not a hydrodynamic mode and decays in the microscopic
time scale. This is, of course, consistent with the fact that we can
not observe the sigma meson as a narrow peak in the vacuum.

\section{Kinetic equation for the chiral phase transition
above $T_{\text c}$ in the $\mathrm{O}(4)$ linear sigma model}
\label{O(4)}

In this section, we give the kinetic equation for the $\mathrm{O}(4)$
linear sigma model. For simplicity, we consider the disordered phase.
The emphasis will be put on the effect of the propagating mode of the
order parameter, which is replaced by the diffusive mode the itinerant
antiferromagnet.

The Lagrangian of the
$\mathrm{O}(4)$ linear sigma model is given by
\begin{equation}
\mathcal{L}=\frac{1}{2}
\left(\partial_\mu\phi_\alpha\right)^2
-\frac{1}{2}\mu^2
\phi_\alpha\mbox{}^2
-\frac{\lambda}{4}\left(\phi_\alpha\mbox{}^2\right)^2,
\end{equation}
where $\alpha$ runs from 0 to 3 and $\phi_\alpha$ denotes the meson
fields, $\phi_\alpha=\left(\phi_0,\phi_1,\phi_2,\phi_3\right)$. For
the pion field, we may use
$\phi_a=\left(\phi_1,\phi_2,\phi_3\right)$ with $a$ running from 1 to
3. The conserved currents for the chiral
$\mathrm{SU}_{\text{L}}(2)\otimes\mathrm{SU}_{\text{R}}(2)$
transformation are the vector and axial vector currents given by
\begin{align}
V_a^{\mu}
&=
\epsilon_{abc}\phi_b\partial^\mu\phi_c,\\
A_a^{\mu}
&=
\phi_a\partial^\mu\phi_0-\phi_0\partial^\mu\phi_a,
\end{align}
where $\epsilon_{abc}$ is the total antisymmetric tensor with
$\epsilon_{123}=1$. The chiral charges are then
\begin{align}
Q^V_a(t)
&=
\int\text{d}^3x\, V_a^0(\vec{x},t)
=\int\text{d}^3x\, \epsilon_{abc}\phi_b\partial_t\phi_c,\\
Q^A_a(t)
&=
\int\text{d}^3x\, A_a^0(\vec{x},t)
=\int\text{d}^3x
\left(\phi_a\partial_t\phi_0-\phi_0\partial_t\phi_a\right),
\end{align}
which may be accommodated in the antisymmetric tensor
$Q_{\alpha\beta}(t)$, \ie,
\begin{equation}
Q_{\alpha\beta}(t)=\int\text{d}^3x\,
\epsilon_{\alpha\beta\gamma\delta}
\phi_\gamma\partial_t\phi_\delta,
\end{equation}
where $\epsilon_{\alpha\beta\gamma\delta}$ is the total antisymmetric
tensor with $\epsilon_{0123}=1$. The $Q^V_a$ and $Q^A_a$  are related
to $Q_{\alpha\beta}$ as
\begin{equation}
Q_{0a}=Q^V_a \:, \qquad
Q_{ab}=-\epsilon_{abc}Q^A_c.
\label{eq:qvqa}
\end{equation}

The meson field $\phi_\alpha$ and its conjugate momentum $\pi_\alpha$
are expanded in terms of the creation and annihilation operators,
\begin{align}
\phi_\alpha(\vec{x},t)
&=
\frac{1}{\sqrt{V}}\sum_{\vec{k}}\frac{1}{\sqrt{2\omega_{\vec{k}}}}
\left(a_{\alpha\vec{k}}(t)+a_{\alpha-\vec{k}}^{\dagger}(t)\right)
e^{\text i\vec{k}\cdot\vec{x}},\\
\pi_\alpha(\vec{x},t)
&=
\frac{1}{\sqrt{V}}\sum_{\vec{k}}(-\text i)\sqrt{\frac{\omega_{\vec{k}}}{2}}
\left(a_{\alpha\vec{k}}(t)-a_{\alpha-\vec{k}}^{\dagger}(t)\right)
e^{\text i\vec{k}\cdot\vec{x}}.
\end{align}
We note that the dispersion $\omega_{\vec{k}}$ does not depend on
$\alpha$ because all the $\sigma$ and $\pi$ mesons are degenerated in
the disordered phase. We also note that the six charges, $Q^V_a$ and
$Q^A_a$, are degenerated in this phase.

The quantization conditions
\begin{align}
&\left[\phi_\alpha(\vec{x},t), \pi_\beta(\vec{y},t)\right]
=\text i\delta_{\alpha\beta}\delta(\vec{x}-\vec{y}),\\
&\left[\phi_\alpha(\vec{x},t), \phi_\beta(\vec{y},t)\right]
=\left[\pi_\alpha(\vec{x},t), \pi_\beta(\vec{y},t)\right]=0,
\\
\intertext{or}
&\left[a_{\alpha\vec{k}}(t),a_{\beta\vec{k}'}^{\dagger}(t)\right]
=\delta_{\alpha\beta}\delta_{\vec{k}\vec{k}'} \:,
\\
&\left[a_{\alpha\vec{k}}(t),a_{\beta\vec{k}'}(t)\right]
=\left[a_{\alpha\vec{k}}^{\dagger}(t),a_{\beta\vec{k}'}^{\dagger}(t)\right]
=0
\end{align}
provide us with the rule to calculate the coefficients in the
kinetic equation.

The slow variables that consist of the Fourier components of the
order parameter and its conjugate, the chiral charges, the
energy density and the momentum density are defined by
\begin{align}
\phi_{\alpha\vec{q}}(t)
&=
\frac{1}{\sqrt{2\omega_{\vec{q}}}}
a_{\alpha\vec{q}}(t),\\
\phi_{\alpha-\vec{q}}^{\dagger}(t)
&=
\frac{1}{\sqrt{2\omega_{\vec{q}}}}
a_{\alpha-\vec{q}}^{\dagger}(t),\\
Q_{\alpha\beta\vec{q}}(t)
&=
\epsilon_{\alpha\beta\gamma\delta}
\frac{1}{\sqrt{V}}\int\text{d}^3x\, e^{-\text i\vec{q}\cdot\vec{x}}
\phi_\gamma\pi_\delta(\vec{x},t),\\
E_{\vec{q}}(t)
&=
\frac{1}{\sqrt{V}}\int\text{d}^3x\, e^{-\text i\vec{q}\cdot\vec{x}}
\biggl[
\biggl\{
\frac{1}{2}\pi_\alpha\mbox{}^2
+\frac{1}{2}\left(\vec{\nabla}\phi_\alpha\right)^2
+\frac{1}{2}\mu^2\phi_\alpha\mbox{}^2 \notag\\
&\qquad\qquad
+\frac{\lambda}{4}\left(\phi_\alpha\mbox{}^2\right)^2
\biggr\}(\vec{x},t)
-\left\langle\mbox{the same expression}\right\rangle_{\text{eq}}
\biggr],&\\
P_{\vec{q}}^i(t)
&=
\frac{1}{\sqrt{V}}\int\text{d}^3x\, e^{-\text i\vec{q}\cdot\vec{x}}
\left(-\pi_\alpha\nabla^i\phi_\alpha\right)(\vec{x},t).
\end{align}
In the same way as Eq.~(\ref{eq:qvqa}), we can obtain $Q_{a\vec{q}}^V$
and $Q_{a\vec{q}}^A$ from $Q_{\alpha\beta\vec{q}}$ by
\begin{equation}
Q_{0a\vec{q}}=Q_{a\vec{q}}^V \:, \qquad
Q_{ab\vec{q}}=-\epsilon_{abc}Q_{c\vec{q}}^A.
\end{equation}
The $Q_{\alpha\beta\vec{q}}$, $E_{\vec{q}}$ and $P_{\vec{q}}^i$ may be
written in terms of the creation and annihilation operators as
\begin{align}
Q_{\alpha\beta\vec{q}}(t)
=&
\epsilon_{\alpha\beta\gamma\delta}
\frac{1}{\sqrt{V}}\sum_{\vec{k}}\frac{-\text i}{2}
\sqrt{\frac{\omega_{\vec{q}-\vec{k}}}{\omega_{\vec{k}}}}
\left(a_{\gamma\vec{k}}+a_{\gamma-\vec{k}}^{\dagger}\right)
\left(a_{\delta\vec{q}-\vec{k}}-a_{\delta\vec{k}-\vec{q}}^{\dagger}\right),\\
E_{\vec{q}}(t)
=&
\frac{1}{2}\frac{1}{\sqrt{V}}\sum_{\vec{k}}\frac{-1}{2}
\sqrt{\omega_{\vec{k}}\omega_{\vec{q}-\vec{k}}}
\left(a_{\alpha\vec{k}}-a_{\alpha-\vec{k}}^{\dagger}\right)
\left(a_{\alpha\vec{q}-\vec{k}}-a_{\alpha\vec{k}-\vec{q}}^{\dagger}\right)
\nonumber\\
&+ \frac{1}{2}\frac{1}{\sqrt{V}}\sum_{\vec{k}}
\left[-\vec{k}\!\cdot\!(\vec{q}\!-\!\vec{k})\!+\!\mu^2\right]\frac{1}{2}
\frac{1}{\sqrt{\omega_{\vec{k}}\omega_{\vec{q}-\vec{k}}}}
\left(a_{\alpha\vec{k}}\!+\!a_{\alpha-\vec{k}}^{\dagger}\right)
\left(a_{\alpha\vec{q}-\vec{k}}\!+\!a_{\alpha\vec{k}-\vec{q}}^{\dagger}\right)
\nonumber\\
&+ \frac{\lambda}{4}\frac{1}{V^{\frac{3}{2}}}
\sum_{\vec{k_1}\cdots\vec{k_4}}\frac{1}{4}
\delta_{\vec{q},\vec{k_1}+\cdots+\vec{k_4}}
\frac{1}{\sqrt{
\omega_{\vec{k_1}}\omega_{\vec{k_2}}
\omega_{\vec{k_3}}\omega_{\vec{k_4}}}}
\nonumber\\
&\qquad\times
\left(a_{\alpha\vec{k_1}}+a_{\alpha-\vec{k_1}}^{\dagger}\right)
\left(a_{\alpha\vec{k_2}}+a_{\alpha-\vec{k_2}}^{\dagger}\right)
\left(a_{\beta\vec{k_3}}+a_{\beta-\vec{k_3}}^{\dagger}\right)
\left(a_{\beta\vec{k_4}}+a_{\beta-\vec{k_4}}^{\dagger}\right)
\nonumber\\
&\qquad -\left\langle\mbox{the same expression}\right\rangle_{\text{eq}}
\delta_{\vec{q},0} \:,
\\
P_{\vec{q}}^i(t)
=&
-\frac{1}{\sqrt{V}}\sum_{\vec{k}}\frac{1}{2}
\sqrt{\frac{\omega_{\vec{q}-\vec{k}}}{\omega_{\vec{k}}}}
k^i
\left(a_{\alpha\vec{q}-\vec{k}}-a_{\alpha\vec{k}-\vec{q}}^{\dagger}\right)
\left(a_{\alpha\vec{k}}+a_{\alpha-\vec{k}}^{\dagger}\right).
\end{align}

We shall now derive the kinetic equation for each variable, the general
form of which is given in Eqs.\ (\ref{finalkinetic}),
(\ref{frequency}) and (\ref{coupling}).

We note that the second term in the expression of $\Omega_{j;lm}$
involves the frequency $\omega_{\vec{q}}^{jp}$ and the presence of the
propagating mode leads to appearance of the second term. Moreover the
first term in $\Omega_{j;lm}$ is also influenced by the presence of
the propagating mode. Thus we expect a substantial difference in the
kinetic equation between the chiral phase transition and the itinerant
antiferromagnet because the propagating mode associated with the order
parameter is absent in the latter.

The susceptibilities are defined by
\begin{align}
\chi_{\alpha\vec{q}}
&\equiv
\left(\phi_{\alpha\vec{q}},
\left(\phi_{\alpha\vec{q}}\right)^\dagger\right)
\equiv\chi_{\phi\vec{q}} \:,
\nonumber\\
\chi_{\alpha^\dagger\vec{q}}
&\equiv
\left(\phi_{\alpha-\vec{q}}^{\dagger},
\left(\phi_{\alpha-\vec{q}}^{\dagger}\right)^\dagger\right)
=\chi_{\phi\vec{q}} \:,
\nonumber\\
\chi_{Q_{\alpha\beta}Q_{\alpha\beta}\vec{q}}
&\equiv
\left(Q_{\alpha\beta\vec{q}},
\left(Q_{\alpha\beta\vec{q}}\right)^\dagger\right)
\equiv\chi_{Q\vec{q}} \:,\\
\chi_{e\vec{q}}
&\equiv
\left(E_{\vec{q}},\left(E_{\vec{q}}\right)^\dagger\right)
\equiv k_{\text B}T^2C_{\vec{q}} \:,
\nonumber\\
\chi_{P^iP^i\vec{q}}
&\equiv
\left(P_{\vec{q}}^i,\left(P_{\vec{q}}^i\right)^\dagger\right)
\equiv\chi_{p\vec{q}} \:,\nonumber
\end{align}
where $C_{\vec{q}}$ is the $\vec{q}$-dependent heat capacity. Note that
the susceptibilities are not dependent on the indices $\alpha$ or
$i=x,y,z$.

To obtain $\omega_{\vec{q}}^{jl}$ and $\Omega_{j;lm}$, we need the
commutation relations for the variables, which will be listed in
Appendix~\ref{app:commutation}.

Now we write down the kinetic equation for each slow variable.

$\bullet$ Order parameter --- $\phi_{\alpha\vec{q}}$
\begin{align}
&\frac{\text d}{\text d t}\phi_{\alpha\vec{q}}
=
\left(-\text i\omega_{\vec{q}}-\frac{L_{\phi}^0}{2\chi_{\phi\vec{q}}}\right)
\phi_{\alpha\vec{q}} \nonumber\\
&\quad + \text i\sum_{\vec{k}}
\biggl[
\frac{1}{2}\sum_{\beta}\sum_{\alpha'\beta'}
\mathcal{V}\left(_{\vec{q},\ \vec{k},\ \vec{q}-\vec{k}}
^{\alpha,\ \beta,\ Q_{\alpha'\beta'}}\right)
\phi_{\beta\vec{k}} Q_{\alpha'\beta'\vec{q}-\vec{k}}
+
\mathcal{V}\left(_{\vec{q},\ \vec{k},\ \vec{q}-\vec{k}}
^{\alpha,\ \alpha,\ e}\right)
\phi_{\alpha\vec{k}} E_{\vec{q}-\vec{k}}
\nonumber\\
&\quad + \sum_i
\mathcal{V}\left(_{\vec{q},\ \vec{k},\ \vec{q}-\vec{k}}
^{\alpha,\ \alpha,\ P^i}\right)
\phi_{\alpha\vec{k}} P_{\vec{q}-\vec{k}}^i
+
\frac{1}{2}\sum_{\beta}\sum_{\alpha'\beta'}
\mathcal{V}\left(_{\vec{q},\ \vec{k},\ \vec{q}-\vec{k}}
^{\alpha,\ \beta^\dagger,\ Q_{\alpha'\beta'}}\right)
\phi_{\beta-\vec{k}}^{\dagger} Q_{\alpha'\beta'\vec{q}-\vec{k}}
\nonumber\\
&\quad +
\mathcal{V}\left(_{\vec{q},\ \vec{k},\ \vec{q}-\vec{k}}
^{\alpha,\ \alpha^\dagger,\ e}\right)
\phi_{\alpha-\vec{k}}^{\dagger} E_{\vec{q}-\vec{k}}
+\sum_i
\mathcal{V}\left(_{\vec{q},\ \vec{k},\ \vec{q}-\vec{k}}
^{\alpha,\ \alpha^\dagger,\ P^i}\right)
\phi_{\alpha-\vec{k}}^{\dagger} P_{\vec{q}-\vec{k}}^i
\biggr] +f_{\vec{q}}^\phi \:,
\end{align}
where the dispersion $\omega_{\vec{q}}$ is given by
$\omega_{\vec{q}}=Z^{-\frac{1}{2}}\sqrt{\vec{q}^2+m^2}$ in the
Ornstein-Zernike approximation. We have defined the `bare' Onsager
coefficient $L_{\phi}^0$ for $\phi_{\alpha\vec{q}}$ and
$\phi_{\alpha-\vec{q}}^{\dagger}$ by
\begin{equation}
k_{\text B}L_{\alpha\vec{q},\alpha\vec{q}}^0/\chi_{\alpha\vec{q}}
=
k_{\text B}L_{\alpha^\dagger\vec{q},\alpha^\dagger\vec{q}}^0
/\chi_{\alpha^\dagger\vec{q}}
=
L_\phi^0/2\chi_{\phi\vec{q}}.
\end{equation}
The coefficient of the nonlinear terms,
$\mathcal{V}_{j;lm}\equiv\Omega_{j;lm}/\left(\chi_l\chi_m\right)^{\frac{1}{2}}$,
are given by
\begin{equation}
\begin{split}
&\mathcal{V}\left(_{\vec{q},\ \vec{k},\ \vec{q}-\vec{k}}
^{\alpha,\ \beta,\ Q_{\alpha'\beta'}}\right) \\
&= -k_{\text B}T
\left\{
\frac{-\text i}{2\sqrt{V}}\epsilon_{\alpha\beta\alpha'\beta'}
\sqrt{\frac{\omega_{\vec{k}}}{\omega_{\vec{q}}}}
\left(
\sqrt{\frac{\omega_{\vec{q}}}{\omega_{\vec{k}}}}
+\sqrt{\frac{\omega_{\vec{k}}}{\omega_{\vec{q}}}}
\right)
\frac{1}{\chi_{Q\vec{q}-\vec{k}}}
-
\frac{
\left(\phi_{\alpha\vec{q}},\phi_{\beta\vec{k}}^{\dagger}
Q_{\alpha'\beta'\vec{k}-\vec{q}}\right)
}
{
2\omega_{\vec{q}}\chi_{\phi\vec{q}}\chi_{\phi\vec{k}}
\chi_{Q\vec{q}-\vec{k}}
}
\right\},
\end{split}
\end{equation}
\
\begin{equation}
\mathcal{V}\left(_{\vec{q},\ \vec{k},\ \vec{q}-\vec{k}}
^{\alpha,\ \beta,\ e}\right)
=
-k_{\text B}T
\left\{
\delta_{\alpha\beta}\frac{k_{\text B}T}{\sqrt{V}}
\frac{1}{2\omega_{\vec{k}}\chi_{\phi\vec{k}}\chi_{e\vec{q}-\vec{k}}}
-
\frac{
\left(\phi_{\alpha\vec{q}},\phi_{\beta\vec{k}}^{\dagger}
E_{\vec{k}-\vec{q}}\right)
}
{
2\omega_{\vec{q}}\chi_{\phi\vec{q}}\chi_{\phi\vec{k}}
\chi_{e\vec{q}-\vec{k}}
}
\right\},
\label{coefficient}
\end{equation}
\
\begin{equation}
\begin{split}
&\mathcal{V}\left(_{\vec{q},\ \vec{k},\ \vec{q}-\vec{k}}
^{\alpha,\ \beta,\ P^i}\right) \\
&= -k_{\text B}T
\left\{
\delta_{\alpha\beta}\frac{1}{2\sqrt{V}}
\sqrt{\frac{\omega_{\vec{k}}}{\omega_{\vec{q}}}}
\left(
\sqrt{\frac{\omega_{\vec{q}}}{\omega_{\vec{k}}}}k^i
+\sqrt{\frac{\omega_{\vec{k}}}{\omega_{\vec{q}}}}q^i
\right)
\frac{1}{\chi_{p\vec{q}-\vec{k}}}
-
\frac{
\left(\phi_{\alpha\vec{q}},\phi_{\beta\vec{k}}^{\dagger}
P_{\vec{k}-\vec{q}}^i\right)
}
{
2\omega_{\vec{q}}\chi_{\phi\vec{q}}\chi_{\phi\vec{k}}
\chi_{p\vec{q}-\vec{k}}
}
\right\},
\end{split}
\end{equation}
\
\begin{equation}
\begin{split}
&\mathcal{V}\left(_{\vec{q},\ \vec{k},\ \vec{q}-\vec{k}}
^{\alpha,\ \beta^\dagger,\ Q_{\alpha'\beta'}}\right) \\
&= -k_{\text B}T
\left\{
\frac{-\text i}{2\sqrt{V}}\epsilon_{\alpha\beta\alpha'\beta'}
\sqrt{\frac{\omega_{\vec{k}}}{\omega_{\vec{q}}}}
\left(
\sqrt{\frac{\omega_{\vec{q}}}{\omega_{\vec{k}}}}
-\sqrt{\frac{\omega_{\vec{k}}}{\omega_{\vec{q}}}}
\right)
\frac{1}{\chi_{Q\vec{q}-\vec{k}}}
-
\frac{
\left(\phi_{\alpha\vec{q}},\phi_{\beta-\vec{k}}
Q_{\alpha'\beta'\vec{k}-\vec{q}}\right)
}
{
2\omega_{\vec{q}}\chi_{\phi\vec{q}}\chi_{\phi\vec{k}}
\chi_{Q\vec{q}-\vec{k}}
}
\right\},
\end{split}
\end{equation}
\
\begin{equation}
\mathcal{V}\left(_{\vec{q},\ \vec{k},\ \vec{q}-\vec{k}}
^{\alpha,\ \beta^\dagger,\ e}\right)
=
-k_{\text B}T(-)
\frac{
\left(\phi_{\alpha\vec{q}},\phi_{\beta-\vec{k}}
E_{\vec{k}-\vec{q}}\right)
}
{
2\omega_{\vec{q}}\chi_{\phi\vec{q}}\chi_{\phi\vec{k}}
\chi_{e\vec{q}-\vec{k}}
}, \qquad\qquad\qquad\qquad\qquad
\end{equation}
\
\begin{equation}
\begin{split}
&\mathcal{V}\left(_{\vec{q},\ \vec{k},\ \vec{q}-\vec{k}}
^{\alpha,\ \beta^\dagger,\ P^i}\right) \\
&= -k_{\text B}T
\left\{
\delta_{\alpha\beta}\frac{1}{2\sqrt{V}}
\sqrt{\frac{\omega_{\vec{k}}}{\omega_{\vec{q}}}}
\left(
\sqrt{\frac{\omega_{\vec{q}}}{\omega_{\vec{k}}}}k^i
-\sqrt{\frac{\omega_{\vec{k}}}{\omega_{\vec{q}}}}q^i
\right)
\frac{1}{\chi_{p\vec{q}-\vec{k}}}
-
\frac{
\left(\phi_{\alpha\vec{q}},\phi_{\beta-\vec{k}}
P_{\vec{k}-\vec{q}}^i\right)
}
{
2\omega_{\vec{q}}\chi_{\phi\vec{q}}\chi_{\phi\vec{k}}
\chi_{p\vec{q}-\vec{k}}
}
\right\}.
\end{split}
\end{equation}
For illustration, we shall give an explicit derivation of
$\mathcal{V}\left(_{\vec{q},\ \vec{k},\ \vec{q}-\vec{k}}^{\alpha,\ \beta,\ e}\right)$
in Eq.~(\ref{coefficient}).
\begin{align}
&\mathcal{V}\left(_{\vec{q},\ \vec{k},\ \vec{q}-\vec{k}}
^{\alpha,\ \beta,\ e}\right)
\nonumber\\
&=
\frac{-k_{\text B}T}
{\left(\chi_{\phi\vec{k}}\chi_{e\vec{q}-\vec{k}}\right)}
\left\{
\left\langle\left[\phi_{\alpha\vec{q}},
\phi_{\beta\vec{k}}^{\dagger}
E_{\vec{q}-\vec{k}}^\dagger\right]\right\rangle
-\left\langle\left[\phi_{\alpha\vec{q}},
\left(\phi_{\alpha\vec{q}}\right)^\dagger\right]\right\rangle
\left(\chi_{\phi\vec{q}}\right)^{-1}
\left(\phi_{\alpha\vec{q}},
\phi_{\beta\vec{k}}^{\dagger}
E_{\vec{q}-\vec{k}}^\dagger\right)
\right\}
\nonumber\\
&=
\frac{-k_{\text B}T}
{\left(\chi_{\phi\vec{k}}\chi_{e\vec{q}-\vec{k}}\right)}
\left\{
\left\langle
\left[\phi_{\alpha\vec{q}}, \phi_{\beta\vec{k}}^{\dagger}
\right]E_{\vec{q}-\vec{k}}^\dagger
\right\rangle
+
\left\langle
\phi_{\beta\vec{k}}^{\dagger}
\left[\phi_{\alpha\vec{q}}, E_{\vec{q}-\vec{k}}^\dagger\right]
\right\rangle
-
\frac{
\left(\phi_{\alpha\vec{q}},
\phi_{\beta\vec{k}}^{\dagger}E_{\vec{k}-\vec{q}}\right)
}
{2\omega_{\vec{q}}\chi_{\phi\vec{q}}}
\right\}.
\label{Vabe}
\end{align}
The first term in the curly brackets vanishes. Note that the third
term appears because of the presence of the propagating mode (the
meson mode). For the second term, we use the approximation that
\begin{equation}
\left[A_{\vec{k}},B_{\vec{k}'}\right]\cong
\left[A_{\vec{k}+\vec{l}},B_{\vec{k}'-\vec{l}}\right],
\label{approx}
\end{equation}
where $|\vec{k}|,|\vec{k}'|,|\vec{l}|\ll 1$, which has a ground in the
locality of the microscopic interaction \cite{kawareview}. With this
approximation, we can manipulate the second term in Eq.~(\ref{Vabe});
\begin{align}
&\mbox{(second term)}
=
\left\langle
\phi_{\beta\vec{k}}^{\dagger}
\left[\phi_{\alpha\vec{q}}, E_{\vec{k}-\vec{q}}\right]
\right\rangle
\nonumber\\
&\qquad \cong
\left\langle
\phi_{\beta\vec{k}}^{\dagger}
\left[\phi_{\alpha\vec{k}}, E_{\vec{0}}\right]
\right\rangle
=
\left\langle
\phi_{\beta\vec{k}}^{\dagger}
\left[\phi_{\alpha\vec{k}}, \frac{1}{\sqrt{V}}\mathcal{H}\right]
\right\rangle
\cong
\frac{1}{\sqrt{V}}
\left(
\phi_{\beta\vec{k}}^{\dagger},
\left[\phi_{\alpha\vec{k}}, \mathcal{H}\right]
\right)
\nonumber\\
&\qquad =
\frac{1}{\sqrt{V}}
\left(
\text{i}\dot{\phi}_{\alpha\vec{k}}, \phi_{\beta\vec{k}}^{\dagger}
\right)
=
\frac{1}{\sqrt{V}}
k_{\text B}T
\left\langle
\left[\phi_{\alpha\vec{k}}, \phi_{\beta\vec{k}}^{\dagger}\right]
\right\rangle
=
\frac{1}{\sqrt{V}}
k_{\text B}T\delta_{\alpha\beta}
\frac{1}{2\omega_{\vec{k}}},
\end{align}
where $\mathcal{H}$ is the Hamiltonian and we have used the Heisenberg
equation $\text i\dot{\mathcal{O}}=\left[\mathcal{O},\mathcal{H}\right]$
and the relation
$\text i(\dot{A},B)=k_{\text B}T \left\langle\left[A,B\right]\right\rangle$.
Thus we obtain Eq.~(\ref{coefficient}). We note that the second term
of Eq.~(\ref{Vabe}) is proportional to the frequency of the meson mode
so that it vanishes if there is no propagating mode. Thus we see that
the coefficient of the coupling of the order parameter and the energy
density remains even in the disordered phase owing to the presence of
the propagating mode of the order parameter. This is contrasted with
the antiferromagnet, in which the mode of the order parameter is
diffusive. In fact, in the antiferromagnet, the energy mode decouples
from the other modes in the disordered phase as we have seen in
Subsec.~\ref{antiferromagnet} \cite{kawaannals,kawareview}. We also
note that the second term in Eq.\ (\ref{coefficient}) involving the
three point correlation has arisen due to the propagating mode.

$\bullet$ Order parameter conjugate --- $\phi_{\alpha-\vec{q}}^{\dagger}$

Similarly, we obtain the kinetic equation for
$\phi_{\alpha-\vec{q}}^{\dagger}$ \:.
\begin{align}
&\frac{\text d}{\text d t}\phi_{\alpha-\vec{q}}^{\dagger}
=
\left(\text i\omega_{\vec{q}}-\frac{L_{\phi}^0}{2\chi_{\phi\vec{q}}}\right)
\phi_{\alpha-\vec{q}}^{\dagger} \nonumber\\
&\quad + \text i\sum_{\vec{k}}
\biggl[
\frac{1}{2}\sum_{\beta}\sum_{\alpha'\beta'}
\mathcal{V}\left(_{\vec{q},\ \vec{k},\ \vec{q}-\vec{k}}
^{\alpha^\dagger,\ \beta,\ Q_{\alpha'\beta'}}\right)
\phi_{\beta\vec{k}} Q_{\alpha'\beta'\vec{q}-\vec{k}}
+
\mathcal{V}\left(_{\vec{q},\ \vec{k},\ \vec{q}-\vec{k}}
^{\alpha^\dagger,\ \alpha,\ e}\right)
\phi_{\alpha\vec{k}} E_{\vec{q}-\vec{k}}
\nonumber\\
&\quad + \sum_i
\mathcal{V}\left(_{\vec{q},\ \vec{k},\ \vec{q}-\vec{k}}
^{\alpha^\dagger,\ \alpha,\ P^i}\right)
\phi_{\alpha\vec{k}} P_{\vec{q}-\vec{k}}^i
+
\frac{1}{2}\sum_{\beta}\sum_{\alpha'\beta'}
\mathcal{V}\left(_{\vec{q},\ \vec{k},\ \vec{q}-\vec{k}}
^{\alpha^\dagger,\ \beta^\dagger,\ Q_{\alpha'\beta'}}\right)
\phi_{\beta-\vec{k}}^{\dagger} Q_{\alpha'\beta'\vec{q}-\vec{k}}
\nonumber\\
&\quad +
\mathcal{V}\left(_{\vec{q},\ \vec{k},\ \vec{q}-\vec{k}}
^{\alpha^\dagger,\ \alpha^\dagger,\ e}\right)
\phi_{\alpha-\vec{k}}^{\dagger} E_{\vec{q}-\vec{k}}
+\sum_i
\mathcal{V}\left(_{\vec{q},\ \vec{k},\ \vec{q}-\vec{k}}
^{\alpha^\dagger,\ \alpha^\dagger,\ P^i}\right)
\phi_{\alpha-\vec{k}}^{\dagger} P_{\vec{q}-\vec{k}}^i
\biggr] +f_{\vec{q}}^\phi
\end{align}
with the coefficients
\begin{equation}
\begin{split}
&\mathcal{V}\left(_{\vec{q},\ \vec{k},\ \vec{q}-\vec{k}}
^{\alpha^\dagger,\ \beta,\ Q_{\alpha'\beta'}}\right) \\
&= -k_{\text B}T
\left\{
\frac{-\text i}{2\sqrt{V}}\epsilon_{\alpha\beta\alpha'\beta'}
\sqrt{\frac{\omega_{\vec{k}}}{\omega_{\vec{q}}}}
\left(
\sqrt{\frac{\omega_{\vec{q}}}{\omega_{\vec{k}}}}
-\sqrt{\frac{\omega_{\vec{k}}}{\omega_{\vec{q}}}}
\right)
\frac{1}{\chi_{Q\vec{q}-\vec{k}}}
+
\frac{
\left(\phi_{\alpha-\vec{q}}^{\dagger},\phi_{\beta\vec{k}}^{\dagger}
Q_{\alpha'\beta'\vec{k}-\vec{q}}\right)
}
{
2\omega_{\vec{q}}\chi_{\phi\vec{q}}\chi_{\phi\vec{k}}
\chi_{Q\vec{q}-\vec{k}}
}
\right\},
\end{split}
\end{equation}
\
\begin{equation}
\mathcal{V}\left(_{\vec{q},\ \vec{k},\ \vec{q}-\vec{k}}
^{\alpha^\dagger,\ \beta,\ e}\right)
=
-k_{\text B}T
\frac{
\left(\phi_{\alpha-\vec{q}}^{\dagger},\phi_{\beta\vec{k}}^{\dagger}
E_{\vec{k}-\vec{q}}\right)
}
{
2\omega_{\vec{q}}\chi_{\phi\vec{q}}\chi_{\phi\vec{k}}
\chi_{e\vec{q}-\vec{k}}
}, \qquad\qquad\qquad\qquad\qquad\qquad\quad
\end{equation}
\
\begin{equation}
\begin{split}
&\mathcal{V}\left(_{\vec{q},\ \vec{k},\ \vec{q}-\vec{k}}
^{\alpha^\dagger,\ \beta,\ P^i}\right) \\
&= -k_{\text B}T
\left\{
\delta_{\alpha\beta}\frac{1}{2\sqrt{V}}
\sqrt{\frac{\omega_{\vec{k}}}{\omega_{\vec{q}}}}
\left(
\sqrt{\frac{\omega_{\vec{q}}}{\omega_{\vec{k}}}}k^i
-\sqrt{\frac{\omega_{\vec{k}}}{\omega_{\vec{q}}}}q^i
\right)
\frac{1}{\chi_{p\vec{q}-\vec{k}}}
+
\frac{
\left(\phi_{\alpha-\vec{q}}^{\dagger},\phi_{\beta\vec{k}}^{\dagger}
P_{\vec{k}-\vec{q}}^i\right)
}
{
2\omega_{\vec{q}}\chi_{\phi\vec{q}}\chi_{\phi\vec{k}}
\chi_{p\vec{q}-\vec{k}}
}
\right\},
\end{split}
\end{equation}
\
\begin{equation}
\begin{split}
&\mathcal{V}\left(_{\vec{q},\ \vec{k},\ \vec{q}-\vec{k}}
^{\alpha^\dagger,\ \beta^\dagger,\ Q_{\alpha'\beta'}}\right) \\
&= -k_{\text B}T
\left\{
\frac{-\text i}{2\sqrt{V}}\epsilon_{\alpha\beta\alpha'\beta'}
\sqrt{\frac{\omega_{\vec{k}}}{\omega_{\vec{q}}}}
\left(
\sqrt{\frac{\omega_{\vec{q}}}{\omega_{\vec{k}}}}
+\sqrt{\frac{\omega_{\vec{k}}}{\omega_{\vec{q}}}}
\right)
\frac{1}{\chi_{Q\vec{q}-\vec{k}}}
+
\frac{
\left(\phi_{\alpha-\vec{q}}^{\dagger},\phi_{\beta-\vec{k}}
Q_{\alpha'\beta'\vec{k}-\vec{q}}\right)
}
{
2\omega_{\vec{q}}\chi_{\phi\vec{q}}\chi_{\phi\vec{k}}
\chi_{Q\vec{q}-\vec{k}}
}
\right\},
\end{split}
\end{equation}
\
\begin{equation}
\mathcal{V}\left(_{\vec{q},\ \vec{k},\ \vec{q}-\vec{k}}
^{\alpha^\dagger,\ \beta^\dagger,\ e}\right)
=
-k_{\text B}T
\left\{
-\delta_{\alpha\beta}\frac{k_{\text B}T}{\sqrt{V}}
\frac{1}{2\omega_{\vec{k}}\chi_{\phi\vec{k}}\chi_{e\vec{q}-\vec{k}}}
+
\frac{
\left(\phi_{\alpha-\vec{q}}^{\dagger},\phi_{\beta-\vec{k}}
E_{\vec{k}-\vec{q}}\right)
}
{
2\omega_{\vec{q}}\chi_{\phi\vec{q}}\chi_{\phi\vec{k}}
\chi_{e\vec{q}-\vec{k}}
}
\right\}, \qquad\quad
\end{equation}
\
\begin{equation}
\begin{split}
&\mathcal{V}\left(_{\vec{q},\ \vec{k},\ \vec{q}-\vec{k}}
^{\alpha^\dagger,\ \beta^\dagger,\ P^i}\right) \\
&= -k_{\text B}T
\left\{
\delta_{\alpha\beta}\frac{1}{2\sqrt{V}}
\sqrt{\frac{\omega_{\vec{k}}}{\omega_{\vec{q}}}}
\left(
\sqrt{\frac{\omega_{\vec{q}}}{\omega_{\vec{k}}}}k^i
+\sqrt{\frac{\omega_{\vec{k}}}{\omega_{\vec{q}}}}q^i
\right)
\frac{1}{\chi_{p\vec{q}-\vec{k}}}
+
\frac{
\left(\phi_{\alpha-\vec{q}}^{\dagger},\phi_{\beta-\vec{k}}
P_{\vec{k}-\vec{q}}^i\right)
}
{
2\omega_{\vec{q}}\chi_{\phi\vec{q}}\chi_{\phi\vec{k}}
\chi_{p\vec{q}-\vec{k}}
}
\right\}.
\end{split}
\end{equation}

$\bullet$ Chiral charges --- $Q_{\alpha\beta\vec{q}}$

Since all the six charges should behave in the same manner,
we take specifically $Q_{1\vec{q}}^V=Q_{01\vec{q}}$. Its
kinetic equation is given by
\begin{align}
\frac{\text d}{\text d t}Q_{1\vec{q}}^{V}
=&
-q^2\frac{L_Q^0}{\chi_{Q\vec{q}}}
Q_{1\vec{q}}^{V} \nonumber\\
&+ \text i\sum_{\vec{k}}
\biggl[
\mathcal{V}\left(_{\vec{q},\ \vec{k},\ \vec{q}-\vec{k}}
^{Q_{01},\ 2,\ 3}\right)
\phi_{2\vec{k}} \phi_{3\vec{q}-\vec{k}}
+
\mathcal{V}\left(_{\vec{q},\ \vec{k},\ \vec{q}-\vec{k}}
^{Q_{01},\ 2,\ 3^{\dagger}}\right)
\phi_{2\vec{k}} \phi_{3\vec{k}-\vec{q}}^{\dagger}
\nonumber\\
&+
\mathcal{V}\left(_{\vec{q},\ \vec{k},\ \vec{q}-\vec{k}}
^{Q_{01},\ 2^{\dagger},\ 3}\right)
\phi_{2-\vec{k}}^{\dagger} \phi_{3\vec{q}-\vec{k}}
+
\mathcal{V}\left(_{\vec{q},\ \vec{k},\ \vec{q}-\vec{k}}
^{Q_{01},\ 2^\dagger,\ 3^\dagger}\right)
\phi_{2-\vec{k}}^{\dagger} \phi_{3\vec{k}-\vec{q}}^{\dagger}
\nonumber\\
&+
\mathcal{V}\left(_{\vec{q},\ \vec{k},\ \vec{q}-\vec{k}}
^{Q_{01},\ Q^{V}_2,\ Q^{V}_3}\right)
Q_{2\vec{k}}^V Q_{3\vec{q}-\vec{k}}^V
+
\mathcal{V}\left(_{\vec{q},\ \vec{k},\ \vec{q}-\vec{k}}
^{Q_{01},\ Q^A_2,\ Q^A_3}\right)
Q_{2\vec{k}}^A Q_{3\vec{q}-\vec{k}}^A
\nonumber\\
&+
\sum_i
\mathcal{V}\left(_{\vec{q},\ \vec{k},\ \vec{q}-\vec{k}}
^{Q_{01},\ Q^V_1,\ P^i}\right)
Q_{1\vec{k}}^V P_{\vec{q}-\vec{k}}^i
\biggr]
+f_{\vec{q}}^Q
\end{align}
with the coefficients
\begin{align}
&\mathcal{V}\left(_{\vec{q},\ \vec{k},\ \vec{q}-\vec{k}}
^{Q_{01},\ 2,\ 3}\right)
=
\mathcal{V}\left(_{\vec{q},\ \vec{k},\ \vec{q}-\vec{k}}
^{Q_{01},\ 2^\dagger,\ 3^\dagger}\right)
\nonumber\\
&\qquad =
-k_{\text B}T\frac{\text i}{2\sqrt{V}}
\left(
\sqrt{\frac{\omega_{\vec{k}}}{\omega_{\vec{q}-\vec{k}}}}
-\sqrt{\frac{\omega_{\vec{q}-\vec{k}}}{\omega_{\vec{k}}}}
\right)
\left(
\sqrt{\frac{\omega_{\vec{q}-\vec{k}}}{\omega_{\vec{k}}}}
\frac{1}{\chi_{\phi\vec{k}}}
+\sqrt{\frac{\omega_{\vec{k}}}{\omega_{\vec{q}-\vec{k}}}}
\frac{1}{\chi_{\phi\vec{q}-\vec{k}}}
\right),
\\
&\mathcal{V}\left(_{\vec{q},\ \vec{k},\ \vec{q}-\vec{k}}
^{Q_{01},\ 2,\ 3^{\dagger}}\right)
=
\mathcal{V}\left(_{\vec{q},\ \vec{k},\ \vec{q}-\vec{k}}
^{Q_{01},\ 2^{\dagger},\ 3}\right)
\nonumber\\
&\qquad =
-k_{\text B}T\frac{\text i}{2\sqrt{V}}
\left(
\sqrt{\frac{\omega_{\vec{k}}}{\omega_{\vec{q}-\vec{k}}}}
+\sqrt{\frac{\omega_{\vec{q}-\vec{k}}}{\omega_{\vec{k}}}}
\right)
\left(
\sqrt{\frac{\omega_{\vec{q}-\vec{k}}}{\omega_{\vec{k}}}}
\frac{1}{\chi_{\phi\vec{k}}}
-\sqrt{\frac{\omega_{\vec{k}}}{\omega_{\vec{q}-\vec{k}}}}
\frac{1}{\chi_{\phi\vec{q}-\vec{k}}}
\right),
\\
&\mathcal{V}\left(_{\vec{q},\ \vec{k},\ \vec{q}-\vec{k}}
^{Q_{01},\ Q^{V}_2,\ Q^{V}_3}\right)
=
\mathcal{V}\left(_{\vec{q},\ \vec{k},\ \vec{q}-\vec{k}}
^{Q_{01},\ Q^A_2,\ Q^A_3}\right)
=
-k_{\text B}T\frac{\text i}{\sqrt{V}}
\left(
\frac{1}{\chi_{Q\vec{k}}}
-\frac{1}{\chi_{Q\vec{q}-\vec{k}}}
\right),
\\
&\mathcal{V}\left(_{\vec{q},\ \vec{k},\ \vec{q}-\vec{k}}
^{Q_{01},\ Q^V_1,\ P^i}\right)
=
-k_{\text B}T\frac{1}{\sqrt{V}}q^i
\frac{1}{\chi_{p\vec{q}-\vec{k}}}.
\end{align}
We have defined the `bare' Onsager coefficient $L_Q^0$ for the chiral
charge by
\begin{equation}
k_{\text B}L_{Q_{\alpha\beta}\vec{q},Q_{\alpha\beta}\vec{q}}^0
/\chi_{Q_{\alpha\beta}\vec{q}}
=q^2L_Q^0/\chi_{Q\vec{q}},
\end{equation}
which involves $q^2$ because the chiral charge is a conserved
quantity.

$\bullet$ Energy --- $E_{\vec{q}}$
\begin{align}
\frac{\text d}{\text d t}E_{\vec{q}}
=&
\text i\omega_{ep^i\vec{q}}P_{\vec{q}}^i
-\lambda^0
\left(
\frac{q^2}{C_{\vec{q}}}E_{\vec{q}}
-\frac{T}{\rho +p}\text iq^i\frac{\partial P_{\vec{q}}^i}{\partial t}
\right)
\nonumber\\
&+ \frac{\text i}{2}\sum_{\vec{k}\alpha}
\mathcal{V}\left(_{\vec{q},\ \vec{k},\ \vec{q}-\vec{k}}
^{e,\ \alpha,\ \alpha}\right)
\phi_{\alpha\vec{k}} \phi_{\alpha\vec{q}-\vec{k}}
\nonumber\\
&+ \text i\sum_{\vec{k}\alpha}
\mathcal{V}\left(_{\vec{q},\ \vec{k},\ \vec{q}-\vec{k}}
^{e,\ \alpha,\ \alpha^\dagger}\right)
\left(
\phi_{\alpha\vec{k}} \phi_{\alpha\vec{k}-\vec{q}}^{\dagger}
-\left\langle
\phi_{\alpha\vec{k}} \phi_{\alpha\vec{k}-\vec{q}}^{\dagger}
\right\rangle
\right)
\nonumber\\
&+ \frac{\text i}{2}\sum_{\vec{k}\alpha}
\mathcal{V}\left(_{\vec{q},\ \vec{k},\ \vec{q}-\vec{k}}
^{e,\ \alpha^\dagger,\ \alpha^\dagger}\right)
\phi_{\alpha-\vec{k}}^{\dagger} \phi_{\alpha\vec{k}-\vec{q}}^{\dagger}
\nonumber\\
&+ \text i\sum_{\vec{k}i}
\mathcal{V}\left(_{\vec{q},\ \vec{k},\ \vec{q}-\vec{k}}
^{e,\ e,\ P^i}\right)
E_{\vec{k}} P_{\vec{q}-\vec{k}}^i
+f_{\vec{q}}^e \:,
\label{kineticeqE}
\end{align}
The $\rho$ and $p$ in the dissipation term are the proper energy
density and the pressure in the equilibrium state respectively. The
$\lambda^0$ is the `bare' heat conductivity and $C_{\vec{q}}$ is the
$\vec{q}$-dependent specific heat. The dissipation terms for the
energy and the momentum are derived from the relativistic
hydrodynamics. The derivation will be given in
Appendix~\ref{app:hydro}.

The frequency is given by
\begin{equation}
\omega_{ep^i\vec{q}}
=
-\Theta q^i/\chi_{p\vec{q}},
\end{equation}
and the coefficients are computed as
\begin{align}
\mathcal{V}\left(_{\vec{q},\ \vec{k},\ \vec{q}-\vec{k}}
^{e,\ \alpha,\ \alpha}\right)
&=
k_{\text B}T
\Theta q^i
\frac{
\left(P_{\vec{q}}^i,\phi_{\alpha\vec{k}}^{\dagger}
\phi_{\alpha\vec{q}-\vec{k}}^{\dagger}\right)
}
{
\chi_{p\vec{q}}\chi_{\phi\vec{k}}\chi_{\phi\vec{q}-\vec{k}}
},
\\
\mathcal{V}\left(_{\vec{q},\ \vec{k},\ \vec{q}-\vec{k}}
^{e,\ \alpha,\ \alpha^\dagger}\right)
&=
-k_{\text B}T
\left\{
-\frac{k_{\text B}T}{\sqrt{V}}
\frac{1}{2\omega_{\vec{k}}}
\frac{1}{\chi_{\phi\vec{k}}\chi_{\phi\vec{q}-\vec{k}}}
-\Theta q^i
\frac{
\left(P_{\vec{q}}^i,\phi_{\alpha\vec{k}}^{\dagger}
\phi_{\alpha\vec{k}-\vec{q}}\right)
}
{
\chi_{p\vec{q}}\chi_{\phi\vec{k}}\chi_{\phi\vec{q}-\vec{k}}
}
\right\},
\\
\mathcal{V}\left(_{\vec{q},\ \vec{k},\ \vec{q}-\vec{k}}
^{e,\ \alpha^\dagger,\ \alpha^\dagger}\right)
&=
k_{\text B}T
\Theta q^i
\frac{
\left(P_{\vec{q}}^i,\phi_{\alpha-\vec{k}}
\phi_{\alpha\vec{k}-\vec{q}}\right)
}
{
\chi_{p\vec{q}}\chi_{\phi\vec{k}}\chi_{\phi\vec{q}-\vec{k}}
},
\\
\mathcal{V}\left(_{\vec{q},\ \vec{k},\ \vec{q}-\vec{k}}
^{e,\ e,\ P^i}\right)
&\!=\!\!
-k_{\text B}T
\left\{\!
\frac{1}{\sqrt{V}}(q\!+\!k)^i
\frac{1}{\chi_{e\vec{k}}}
\!-\!\frac{k_{\text B}T}{\sqrt{V}}
\Theta k^i
\frac{1}{\chi_{e\vec{k}}\chi_{p\vec{q}-\vec{k}}}
\!+\!\Theta q^j
\frac{
\left(P_{\vec{q}}^j,E_{-\vec{k}} P_{\vec{k}-\vec{q}}^i\right)
}
{
\chi_{p\vec{q}}\chi_{e\vec{k}}\chi_{p\vec{q}-\vec{k}}
}
\right\},
\end{align}
where
\begin{equation}
\Theta\equiv
\frac{8}{V}\sum_{\vec{k}}
\left({\omega_{\vec{k}}}^2+\frac{1}{3}\vec{k}^2\right)
\chi_{\phi\vec{k}} \:.
\end{equation}

$\bullet$ Momentum --- $P_{\vec{q}}^i$
\begin{align}
&\frac{\text d}{\text d t}P_{\vec{q}}^i
=
\text i\omega_{p^ie\vec{q}}E_{\vec{q}}
-\lambda^0
\left(
-\frac{\text iq^i}{C_{\vec{q}}}\frac{\partial E_{\vec{q}}}{\partial t}
-\frac{T}{\rho +p}\frac{\partial^2 P_{\vec{q}}^i}{\partial t^2}
\right)
\nonumber\\
&\quad - \frac{\eta^0}{\rho +p}
\left(
q^2P_{\vec{q}}^i+\frac{1}{3}q^i(\vec{q}\cdot\vec{P}_{\vec{q}})
\right)
-\frac{\zeta^0}{\rho +p}q^i(\vec{q}\cdot\vec{P}_{\vec{q}})
\nonumber\\
&\quad + \frac{\text i}{2}\sum_{\vec{k}\alpha}
\mathcal{V}\left(_{\vec{q},\ \vec{k},\ \vec{q}-\vec{k}}
^{P^i,\ \alpha,\ \alpha}\right)
\phi_{\alpha\vec{k}} \phi_{\alpha\vec{q}-\vec{k}}
+ \text i\sum_{\vec{k}\alpha}
\mathcal{V}\left(_{\vec{q},\ \vec{k},\ \vec{q}-\vec{k}}
^{P^i,\ \alpha,\ \alpha^\dagger}\right)
\left(
\phi_{\alpha\vec{k}} \phi_{\alpha\vec{k}-\vec{q}}^{\dagger}
-\left\langle
\phi_{\alpha\vec{k}} \phi_{\alpha\vec{k}-\vec{q}}^{\dagger}
\right\rangle
\right)
\nonumber\\
&\quad + \frac{\text i}{2}\sum_{\vec{k}\alpha}
\mathcal{V}\left(_{\vec{q},\ \vec{k},\ \vec{q}-\vec{k}}
^{P^i,\ \alpha^\dagger,\ \alpha^\dagger}\right)
\phi_{\alpha-\vec{k}}^{\dagger} \phi_{\alpha\vec{k}-\vec{q}}^{\dagger}
+ \frac{\text i}{2}\sum_{\vec{k}Q}
\mathcal{V}\left(_{\vec{q},\ \vec{k},\ \vec{q}-\vec{k}}
^{P^i,\ Q,\ Q}\right)
\left(
Q_{\vec{k}} Q_{\vec{q}-\vec{k}}
-\left\langle
Q_{\vec{k}} Q_{\vec{q}-\vec{k}}
\right\rangle
\right)
\nonumber\\
&\quad + \frac{\text i}{2}\sum_{\vec{k}}
\mathcal{V}\left(_{\vec{q},\ \vec{k},\ \vec{q}-\vec{k}}
^{P^i,\ e,\ e}\right)
\left(
E_{\vec{k}} E_{\vec{q}-\vec{k}}
-\left\langle
E_{\vec{k}} E_{\vec{q}-\vec{k}}
\right\rangle
\right)
\nonumber\\
&\quad + \frac{\text i}{2}\sum_{\vec{k}jl}
\mathcal{V}\left(_{\vec{q},\ \vec{k},\ \vec{q}-\vec{k}}
^{P^i,\ P^j,\ P^l}\right)
\left(
P_{\vec{k}}^j P_{\vec{q}-\vec{k}}^l
-\left\langle
P_{\vec{k}}^j P_{\vec{q}-\vec{k}}^l
\right\rangle
\right)
+f_{\vec{q}}^p
\label{kineticeqP}
\end{align}
The $\eta^0$ and $\zeta^0$ in the dissipation terms are the `bare'
shear and bulk viscosities. The dissipation terms are derived in
Appendix~\ref{app:hydro}.

The frequency is given by
\begin{equation}
\omega_{p^ie\vec{q}}
=
-\Theta q^i/\chi_{e\vec{q}},
\end{equation}
and the coefficients are computed likewise as
\begin{equation}
\begin{split}
&\mathcal{V}\left(_{\vec{q},\ \vec{k},\ \vec{q}-\vec{k}}
^{P^i,\ \alpha,\ \alpha}\right)
=
-k_{\text B}T
\Biggl[
-\frac{1}{2\sqrt{V}}
\sqrt{\frac{\omega_{\vec{q}-\vec{k}}}{\omega_{\vec{k}}}}
\left(
\sqrt{\frac{\omega_{\vec{k}}}{\omega_{\vec{q}-\vec{k}}}}(q-k)^i
+\sqrt{\frac{\omega_{\vec{q}-\vec{k}}}{\omega_{\vec{k}}}}k^i
\right)
\frac{1}{\chi_{\phi\vec{k}}}
\\
&\quad - \frac{1}{2\sqrt{V}}
\sqrt{\frac{\omega_{\vec{k}}}{\omega_{\vec{q}-\vec{k}}}}
\left(
\sqrt{\frac{\omega_{\vec{q}-\vec{k}}}{\omega_{\vec{k}}}}k^i
+\sqrt{\frac{\omega_{\vec{k}}}{\omega_{\vec{q}-\vec{k}}}}(q-k)^i
\right)
\frac{1}{\chi_{\phi\vec{q}-\vec{k}}}
-\Theta q^i
\frac{
\left(E_{\vec{q}},\phi_{\alpha\vec{k}}^{\dagger}
\phi_{\alpha\vec{q}-\vec{k}}^{\dagger}\right)
}
{
\chi_{e\vec{q}}\chi_{\phi\vec{k}}\chi_{\phi\vec{q}-\vec{k}}
}
\Biggl],
\end{split}
\end{equation}
\
\begin{equation}
\begin{split}
&\mathcal{V}\left(_{\vec{q},\ \vec{k},\ \vec{q}-\vec{k}}
^{P^i,\ \alpha,\ \alpha^\dagger}\right)
=
-k_{\text B}T
\Biggl[
-\frac{1}{2\sqrt{V}}
\sqrt{\frac{\omega_{\vec{q}-\vec{k}}}{\omega_{\vec{k}}}}
\left(
\sqrt{\frac{\omega_{\vec{k}}}{\omega_{\vec{q}-\vec{k}}}}(q-k)^i
-\sqrt{\frac{\omega_{\vec{q}-\vec{k}}}{\omega_{\vec{k}}}}k^i
\right)
\frac{1}{\chi_{\phi\vec{k}}}
\\
&\quad - \frac{1}{2\sqrt{V}}
\sqrt{\frac{\omega_{\vec{k}}}{\omega_{\vec{q}-\vec{k}}}}
\left(
\sqrt{\frac{\omega_{\vec{q}-\vec{k}}}{\omega_{\vec{k}}}}k^i
-\sqrt{\frac{\omega_{\vec{k}}}{\omega_{\vec{q}-\vec{k}}}}(q-k)^i
\right)
\frac{1}{\chi_{\phi\vec{q}-\vec{k}}}
-\Theta q^i
\frac{
\left(E_{\vec{q}},\phi_{\alpha\vec{k}}^{\dagger}
\phi_{\alpha\vec{k}-\vec{q}}\right)
}
{
\chi_{e\vec{q}}\chi_{\phi\vec{k}}\chi_{\phi\vec{q}-\vec{k}}
}
\Biggr],
\end{split}
\end{equation}
\
\begin{equation}
\begin{split}
&\mathcal{V}\left(_{\vec{q},\ \vec{k},\ \vec{q}-\vec{k}}
^{P^i,\ \alpha^\dagger,\ \alpha^\dagger}\right)
=
-k_{\text B}T
\Biggl[
-\frac{1}{2\sqrt{V}}
\sqrt{\frac{\omega_{\vec{q}-\vec{k}}}{\omega_{\vec{k}}}}
\left(
\sqrt{\frac{\omega_{\vec{k}}}{\omega_{\vec{q}-\vec{k}}}}(q-k)^i
+\sqrt{\frac{\omega_{\vec{q}-\vec{k}}}{\omega_{\vec{k}}}}k^i
\right)
\frac{1}{\chi_{\phi\vec{k}}}
\\
&\quad - \frac{1}{2\sqrt{V}}
\sqrt{\frac{\omega_{\vec{k}}}{\omega_{\vec{q}-\vec{k}}}}
\left(
\sqrt{\frac{\omega_{\vec{q}-\vec{k}}}{\omega_{\vec{k}}}}k^i
+\sqrt{\frac{\omega_{\vec{k}}}{\omega_{\vec{q}-\vec{k}}}}(q-k)^i
\right)
\frac{1}{\chi_{\phi\vec{q}-\vec{k}}}
-\Theta q^i
\frac{
\left(E_{\vec{q}},\phi_{\alpha-\vec{k}}
\phi_{\alpha\vec{k}-\vec{q}}\right)
}
{
\chi_{e\vec{q}}\chi_{\phi\vec{k}}\chi_{\phi\vec{q}-\vec{k}}
}
\Biggr],
\end{split}
\end{equation}
\
\begin{equation}
\mathcal{V}\left(_{\vec{q},\ \vec{k},\ \vec{q}-\vec{k}}
^{P^i,\ Q,\ Q}\right)
=
-k_{\text B}T
\left[
\frac{1}{\sqrt{V}}
\left(
k^i\frac{1}{\chi_{Q\vec{k}}}+(q-k)^i\frac{1}{\chi_{Q\vec{q}-\vec{k}}}
\right)
-\Theta q^i
\frac{
\left(E_{\vec{q}},Q_{-\vec{k}}Q_{\vec{k}-\vec{q}}\right)
}
{
\chi_{e\vec{q}}\chi_{Q\vec{k}}\chi_{Q\vec{q}-\vec{k}}
}
\right],
\end{equation}
\
\begin{equation}
\mathcal{V}\left(_{\vec{q},\ \vec{k},\ \vec{q}-\vec{k}}
^{P^i,\ e,\ e}\right)
=
-k_{\text B}T
\Theta q^i
\left[
\frac{k_{\text B}T}{\sqrt{V}}
\frac{1}{\chi_{e\vec{k}}\chi_{e\vec{q}-\vec{k}}}
-
\frac{
\left(E_{\vec{q}},E_{-\vec{k}}E_{\vec{k}-\vec{q}}\right)
}
{
\chi_{e\vec{q}}\chi_{e\vec{k}}\chi_{E\vec{q}-\vec{k}}
}
\right], \qquad\qquad\qquad
\end{equation}
\
\begin{equation}
\begin{split}
&\mathcal{V}\left(_{\vec{q},\ \vec{k},\ \vec{q}-\vec{k}}
^{P^i,\ P^j,\ P^l}\right)
=
-k_{\text B}T
\Biggl[
\frac{1}{\sqrt{V}}
\left\{
(q^j\delta^{il}+k^i\delta^{jl})
\frac{1}{\chi_{p\vec{k}}}
+(q^l\delta^{ij}+(q-k)^i\delta^{jl})
\frac{1}{\chi_{p\vec{q}-\vec{k}}}
\right\}
\\
&\quad -\Theta q^i
\frac{
\left(E_{\vec{q}},P_{-\vec{k}}^j P_{\vec{k}-\vec{q}}^l\right)
}
{
\chi_{e\vec{q}}\chi_{p\vec{k}}\chi_{p\vec{q}-\vec{k}}
}
\Biggr].
\end{split}
\end{equation}

We note that there are time derivatives in the dissipation terms
for $E_{\vec{q}}$ and $P_{\vec{q}}^i$. These time derivatives
may be eliminated by means of the kinetic equations themselves.
We can perform it simply by replacing $\dot{E}_{\vec{q}}$ and
$\dot{P}_{\vec{q}}^i$ with the associated frequency terms. This
is because the time derivative of $E_{\vec{q}}$ and
$P_{\vec{q}}^i$ and the frequency term are of second order with
respect to the small quantities, that is, the wavenumber and
the fluctuation of the slow variables, whereas the dissipation
term and the nonlinear term are of higher order. Thus we have
\begin{align}
\frac{\text d}{\text d t}E_{\vec{q}}
=&
-\text iq^j\frac{\Theta}{\chi_{p\vec{q}}}P_{\vec{q}}^j
-\frac{q^2\lambda^0}{C_{\vec{q}}}
\left(
1-\frac{T}{\rho +p}\frac{\Theta}{k_{\text B}T^2}
\right)E_{\vec{q}},
\\
\frac{\text d}{\text d t}P_{\vec{q}}^i
=&
-\text iq^i\frac{\Theta}{\chi_{e\vec{q}}}E_{\vec{q}}
+\lambda^0\frac{q^i}{C_{\vec{q}}}
\frac{\Theta}{\chi_{p\vec{q}}}
\left(
1-\frac{T}{\rho +p}\frac{\Theta}{k_{\text B}T^2}
\right)
(\vec{q}\cdot\vec{P}_{\vec{q}})
\nonumber\\
&- \frac{\eta^0}{\rho +p}
\left(
q^2P_{\vec{q}}^i+\frac{1}{3}q^i(\vec{q}\cdot\vec{P}_{\vec{q}})
\right)
-\frac{\zeta^0}{\rho +p}q^i(\vec{q}\cdot\vec{P}_{\vec{q}}),
\end{align}
where the dissipation term and the nonlinear term are omitted.
\newpage

\section{Dynamic critical exponents}
\label{exponent}

We have derived the kinetic equation for the chiral phase transition
in the previous section. For the full analysis, the numerical
calculation would be involved. As an application without the numerical
work, we shall estimate the dynamic critical exponents from the
kinetic equation in this section. The dynamic critical exponents
specify the way how the typical inverse time scales of each slow mode,
\ie, the widths and the frequencies are scaled. The
fluctuation-dissipation theorem may allow us to imagine that the
characteristic scaling law of the dynamic fluctuation of the slow
variable originates from the characteristic behavior of the
fluctuation in equilibrium states. Then it is reasonable to expect
that there are some relations between the dynamic and static critical
exponents. Actually, the dynamic exponents are given in terms of the
static ones within the mode coupling theory as we will see below. We
also discuss the anomaly of the Onsager coefficients, which are caused
by renormalization from the nonlinear terms.

The analysis goes exactly in the same way as in
Ref.~\cite{kawareview}. In Sec.~\ref{preliminary}, we prepare the
general formalism for obtaining the dynamic critical exponents. In
Sec.~\ref{chiral exponent}, we apply the formalism to the chiral phase
transition.

\subsection{General formalism}
\label{preliminary}

In order to obtain the dynamic critical exponents, we have to know the
characteristic scaling behavior of the dynamic fluctuations of the
slow variables. In the mode coupling theory, the scaling properties of
the dynamic fluctuations are given by those of the static ones as a
starting point. The amplitude of the characteristic static fluctuation
of the slow variable $A_j$ is measured by its static susceptibility
$\chi_j=\left\langle|A_j|^2\right\rangle$. We define the ``reduced''
slow variable $\tilde{A}_j$ by
\begin{equation}
\tilde{A}_j=A_j/\!\sqrt{\chi_j}.
\end{equation}
We also rescale the kinetic equation (\ref{finalkinetic}) to obtain
the reduced kinetic equation for $\tilde{A}_j$,
\begin{equation}
\frac{\text d}{\text d t}\tilde{A}_j(t)
=\sum_l\left(\text i\tilde{\omega}_{jl}-
\frac{k_{\text B}L_{jl}^0}{\sqrt{\chi_j\chi_l}}
\right)\tilde{A}_l(t)
+ \frac{\text i}{2}\sum_{lm}\tilde{\Omega}_{j;lm}
\left(
\tilde{A}_l\tilde{A}_m
-\left\langle \tilde{A}_l\tilde{A}_m\right\rangle
\right)
+\tilde{f}_j,
\label{reduced kinetic eq}
\end{equation}
where
\begin{align}
\tilde{\omega}_{jl}
&\equiv
-k_{\text B}T
\left\langle\left[
\tilde{A}_j,\tilde{A}_l
\right]\right\rangle
=\sqrt{\frac{\chi_l}{\chi_j}}\omega_{jl},
\label{reduced_frequency}
\\
\tilde{\Omega}_{j;lm}
&\equiv
-k_{\text B}T
\left\{
\left\langle\left[
\tilde{A}_j, \tilde{A}_l\tilde{A}_m
\right]\right\rangle
-\sum_p\left\langle\left[
\tilde{A}_j, \tilde{A}_p
\right]\right\rangle
\left(\tilde{A}_p, \tilde{A}_l\tilde{A}_m
\right)
\right\}
=\Omega_{j;lm}/\!\sqrt{\chi_j}
\label{reduced_coupling}
\\
\tilde{f}_j
&\equiv
f_j/\!\sqrt{\chi_j}.
\end{align}
The reduced variables have a finite amplitude of order unity, so that
the information of the characteristic scaling laws of the fluctuations
of the slow variables is now pushed away into the coefficients of the
reduced kinetic equation, which are given in
Eqs.~(\ref{reduced_frequency}) and (\ref{reduced_coupling}).

We consider the time correlation function for the reduced variable
defined by
\begin{equation}
g_{jl}(t)=\left\langle \tilde{A}_j(t)\tilde{A}_l(0)^\dagger
\right\rangle,
\hspace{1cm}(t\geq 0)
\end{equation}
which we will call simply the propagator. The physical information is
contained in those propagators. The dynamics of the reduced variables
is subject to the kinetic equation (\ref{reduced kinetic eq}). By
solving it, we can calculate the propagator.

In order to analyze the propagator, it is convenient to introduce
the second quantized formalism, in which we can make the field
theoretical techniques and considerations.
In the second quantized formalism, the propagator is written as
\begin{equation}
g_{jl}(t)
=\langle 0|
\alpha_j e^{-\text it\mathcal{H}}\tilde{\alpha}_l
|0\rangle.
\hspace{1cm}(t\geq 0)
\label{propagator in second quantized}
\end{equation}

$\alpha_j$ and $\tilde{\alpha}_j$ are the operators that create and
annihilate the mode $j$. We have defined the associated ``vacuum''
state $|0\rangle$, in which no $j$ modes are present. The
``Hamiltonian'' $\mathcal{H}$ governs the dynamics of the modes and is
given by
\begin{align}
\mathcal{H}&=\mathcal{H}_0+\mathcal{H}',
\\
\mathcal{H}_0
&= -\sum_{jl}
\left[
\tilde{\omega}_{jl}
+\text i\left(\chi_j\chi_l\right)^{-\frac{1}{2}}L_{jl}^0
\right]
\tilde{\alpha}_j^0 \alpha_l^0,
\label{Hfree}
\\
\mathcal{H}'
&=-\frac{1}{2}\sum_{jlm}
\left(
\tilde{\Omega}_{j;lm}^{*}
\tilde{\alpha}_{l}\tilde{\alpha}_{m}\alpha_{j}
+
\tilde{\Omega}_{j;lm}
\tilde{\alpha}_{j}\alpha_{l}\alpha_{m}
\right),
\label{Hint}
\end{align}
where $\mathcal{H}_0$ and $\mathcal{H}'$ are the free and interaction
``Hamiltonians'' and correspond to the linear and nonlinear terms in
the kinetic equation respectively. We note that the free
``Hamiltonian'' is not Hermitian because of the dissipation
term. In the following, we will use $\mathcal{H}_{0jl}^N$ and
$\mathcal{H}_{0jl}^D$ to denote the two kinds of terms in
Eq.~(\ref{Hfree}), which describe non-dissipative and dissipative
processes respectively. In the second quantized formalism, the
expression of the propagator Eq.~(\ref{propagator in second
quantized}) can have the interpretation that the mode $l$ is created
at $t=0$, which turns into the mode $j$ through the mixing and
coupling with the other modes to be annihilated at a later time $t$.

We note that in this formalism, the reduced coefficients
$\tilde{\omega}$'s and $\tilde{\Omega}$'s, which contain the
information of the characteristic scaling properties of the
fluctuations, appear in the ``Hamiltonian'' $\mathcal{H}$. Thus the
``Hamiltonian'' determines the scaling law of the slow modes.

Our present goal is to obtain the dynamic critical exponents or the
characteristic time scales of each mode. For that purpose, it becomes
necessary to consider each mode separately and derive the effective
``Hamiltonians'' for each mode. We can obtain the effective
``Hamiltonian'' by introducing the projection operator that separates
the subspace of the interesting variables from that of the other
variables. We denote the set of the slow variables for which
we want the effective ``Hamiltonian'' by $\{a\}$ and the eliminated
set by $\{b\}$. The ``vacuum'' state is then a product of the
``vacuum'' states for $\{a\}$ and $\{b\}$ denoted by $|0_a\rangle$ and
$|0_b\rangle$,
\begin{equation}
|0\rangle=|0_a\rangle|0_b\rangle.
\end{equation}
The projection operator $\mathcal{P}_b$ is defined by
$\mathcal{P}_b |\cdots\rangle=|0_b\rangle\langle 0_b|\cdots\rangle$,
which projects arbitrary states of $\{b\}$ into the ground state
$|0_b\rangle$.

Consider the one-sided Fourier transform of the propagator for the
slow variables $j$ and $l$ that belong to the set $\{a\}$;
\begin{equation}
\hat{g}_{jl}(\omega)\equiv
\int_0^\infty\text{d}t\,  e^{\text i\omega t}g_{jl}(t)
=\text i
\langle 0|
\alpha_j\frac{1}{\omega-\mathcal{H}}\tilde{\alpha}_l
|0\rangle
\end{equation}
We can manipulate it into
\begin{equation}
\hat{g}_{jl}(\omega)
=\text i
\langle 0_a|
\alpha_j
\frac{1}{\omega-\mathcal{H}_a(\omega)}
\tilde{\alpha}_l
|0_a\rangle
\label{reducedpropagator}
\end{equation}
in order to have the frequency dependent effective ``Hamiltonian''
$\mathcal{H}_a(\omega)$, which acts on the variables $\{a\}$ with the
interactions with the variables $\{b\}$ included and is given by
\begin{align}
\mathcal{H}_a(\omega)
&=\mathcal{H}_{0a}
+
\langle 0_b|
\mathcal{H}'
|0_b\rangle
+
\langle 0_b|
\mathcal{H}'
\left[
\omega-\left(1-\mathcal{P}_b\right)\mathcal{H}
\right]^{-1}
\left(1-\mathcal{P}_b\right)
\mathcal{H}'
|0_b\rangle.
\label{effectivehamiltonian}
\end{align}

Now what we have to do for investigation of the slow modes is to know
the scaling properties of the coefficients $\tilde{\omega}$'s and
$\tilde{\Omega}$'s. We see that it is necessary and sufficient to have
the scaling behaviors of the commutators of the reduced variables,
\ie, $[\tilde{A},\tilde{A}]$, which we will examine for the chiral
phase transition in the next subsection.

\subsection{Investigation of the chiral phase transition}
\label{chiral exponent}

Firstly, we consider the scaling behavior of the amplitude of the slow
variable $A_j$ itself. For the chiral phase transition, $A_j$ consists
of $\{\phi_{\alpha\vec{k}},\phi_{\alpha-\vec{k}}^{\dagger},
Q_{\vec{k}},E_{\vec{k}},P_{\vec{k}}^i\}$. These variables are the
Fourier transforms of the corresponding density, which we will denote
as $A_j(\vec{x})$. The $A_j(\vec{x})$ exhibits characteristic anomaly
at the critical point and is scaled in terms of the correlation length
$\xi$. To make it explicit, we write $A_j(\vec{x},\xi,V)$ for
$A_j(\vec{x})$ where we have also included the possible dependence on
the volume $V$. Applying the block spin transformation with the block
size $L$ on  $A_j(\vec{x},\xi,V)$, we have the scaled amplitude of the
fluctuation as
\begin{equation}
A_j(\vec{x},\xi,V)=L^{-x_j}A_j(\vec{x}/L,\xi/L,V/L^d),
\label{eq:scale}
\end{equation}
where $x_j$ is the exponent to be determined. For the Fourier
component which is defined by
\begin{equation}
A(\vec{k})=\frac{1}{\sqrt{V}}\int
\text{d}^d x\, e^{-\text i\vec{k}\cdot\vec{x}}A(\vec{x}),
\end{equation}
we then have
\begin{equation}
A_j(\vec{k},\xi,V)=L^{\frac{d}{2}-x_j}A_j(L\vec{k},\xi/L,V/L^d).
\label{eq:scale2}
\end{equation}
The exponent $x_j$ can be determined by the (static) susceptibility
$\chi_j$:
\begin{equation}
L^{d-2x_j}
\left\langle
|A_j(\vec{0},\xi/L,V/L^d)|^2
\right\rangle
=
\left\langle
|A_j(\vec{0},\xi,V)|^2
\right\rangle
=\chi_{j,\vec{k}=\vec{0}}\sim\xi^{\frac{\gamma_j}{\nu}},
\end{equation}
where $\gamma_j$ and $\nu$ are the static critical exponents.
When choosing $L=\xi$, we find immediately
\begin{equation}
x_j=\frac{1}{2}\left(d-\frac{\gamma_j}{\nu}\right).
\end{equation}
For the chiral phase transition, we have
\begin{equation}
\gamma_\phi=\gamma,\qquad
\gamma_Q=0,\qquad
\gamma_e=\alpha,\qquad
\gamma_p=0
\end{equation}
leading to
\begin{align}
x_\phi&=\frac{1}{2}\left(d-\frac{\gamma}{\nu}\right)
=\frac{\beta}{\nu},\\
x_Q&=\frac{d}{2},\\
x_e&=\frac{1}{2}\left(d-\frac{\alpha}{\nu}\right)
=d-\frac{1}{\nu},\\
x_p&=\frac{d}{2},
\end{align}
where we have used the scaling relations
\begin{equation}
2-\alpha=d\nu,\qquad \alpha+2\beta+\gamma=2.
\end{equation}
We note that the static critical exponents $\alpha,\beta,\gamma,\nu$
are the same as those of the ferro- and antiferromagnet because the
chiral phase transition is in the same static universality class as
these systems.

Thus the amplitudes of fluctuation $A_j$ have the characteristic
scaling properties as follows;
\begin{align}
\phi_\alpha(\vec{k},\xi,V)
&=\xi^{\frac{\gamma}{2\nu}}\phi_\alpha(\xi\vec{k},1,V/\xi^d),
\nonumber\\
Q_{\alpha\beta}(\vec{k},\xi,V)
&=\xi^0Q_{\alpha\beta}(\xi\vec{k},1,V/\xi^d),
\nonumber\\
E(\vec{k},\xi,V)
&=\xi^{\frac{\alpha}{2\nu}}E(\xi\vec{k},1,V/\xi^d),
\label{scaling}
\\
P^i(\vec{k},\xi,V)
&=\xi^0P^i(\xi\vec{k},1,V/\xi^d).\nonumber
\end{align}

We now discuss the scaling property of the commutators of
$\tilde{A}_j$. If the exponent of $[A_j,A_l]$ is denoted as $x_{jl}$,
namely $[A_j,A_l]$ scales as $\xi^{d/2-x_{jl}}$ in the same sense as
Eqs.~(\ref{scaling}), then we find that $[\tilde{A}_j,\tilde{A}_l]$
scales as $\xi^{x_j+x_l-d/2-x_{jl}}$, that is,
\begin{equation}
[\tilde{A}_j,\tilde{A}_l]\sim
\xi^{x_j+x_l-\frac{d}{2}-x_{jl}}.
\end{equation}
The computation of $x_{jl}$ for
$\{\phi_{\alpha\vec{k}},\phi_{\alpha-\vec{k}}^{\dagger},
Q_{\vec{k}},E_{\vec{k}},P_{\vec{k}}^i\}$ in the chiral phase
transition is performed in Appendix~\ref{app:xjl}. Using those
results, we find
\begin{equation}
\begin{split}
& [\tilde{\phi},\tilde{\phi^\dagger}]
 \sim \xi^{-\frac{\gamma}{2\nu}},
 \qquad
[\tilde{\phi},\tilde{\phi}]
 = [\tilde{\phi^\dagger},\tilde{\phi^\dagger}]=0, \\
& [\tilde{\phi},\tilde{Q}]
 \sim \xi^{-\frac{d}{2}},
 \qquad
[\tilde{\phi},\tilde{E}]
 \sim \xi^{-\frac{d}{2}-\frac{1}{\nu}},
 \qquad
[\tilde{\phi},\tilde{P}]
 \sim \xi^{-\frac{d}{2}-1}, \\
& [\tilde{Q},\tilde{Q}]
 \sim \xi^{-\frac{d}{2}},
 \qquad
[\tilde{Q},\tilde{E}]
 \sim \xi^{-\frac{d}{2}-\frac{1}{\nu}},
 \qquad
[\tilde{Q},\tilde{P}]
 \sim \xi^{-\frac{d}{2}-1}, \\
& [\tilde{E},\tilde{E}]
 \sim \xi^{\frac{d}{2}-1-\frac{2}{\nu}}
 =\xi^{-\frac{d}{2}-1-\frac{\alpha}{\nu}},
 \qquad
[\tilde{E},\tilde{P}]
 \sim \xi^{-\frac{d}{2}-1-\frac{1}{\nu}}
 =\xi^{-d-1-\frac{\alpha}{2\nu}}, \\
&[\tilde{P},\tilde{P}]
 \sim \xi^{-\frac{d}{2}-1}.
\end{split}
\end{equation}

We are now ready to derive the scaling behavior of the coefficients in
the kinetic equation or in the ``Hamiltonian.'' The nonvanishing
frequencies of the propagating modes are those of the meson mode and
the energy wave mode. The scaling properties for the reduced
frequencies are
\begin{align}
\tilde{\omega}_{\phi\vec{q}}
&=
-k_{\text B}T
\left\langle\left[
\tilde{\phi}_{\vec{q}},\tilde{\phi}_{-\vec{q}}^\dagger
\right]\right\rangle
\sim\xi^{-\frac{\gamma}{2\nu}},
\\
\tilde{\omega}_{ep^{\text L}\vec{q}}
&=
-k_{\text B}T
\left\langle\left[
\tilde{E}_{\vec{q}},\tilde{P}_{\vec{q}}^{\text L\dagger}
\right]\right\rangle
\sim\xi^{-\frac{d}{2}-1-\frac{1}{\nu}}.
\end{align}
For the scaling properties of the Onsager coefficients, we have
\begin{equation}
L_{\phi}^0\sim\xi^0,\hspace{1cm}
L_{Q}^0\sim\xi^{-2},\hspace{1cm}
\lambda^0\sim\eta^0\sim\zeta^0\sim\xi^{-2}
\end{equation}
because $\phi$ is not conserved while $Q$, $E$, and $P^i$ are
conserved. We also note that
\begin{equation}
\chi_{\phi}\sim\xi^{\frac{\gamma}{\nu}},\hspace{1cm}
\chi_Q\sim\chi_p\sim\xi^0,\hspace{1cm}
\chi_e\sim\xi^{\frac{\alpha}{\nu}}.
\end{equation}
Thus we can find the scaling properties of the free
``Hamiltonian'' in the second quantized formalism.
The free ``Hamiltonian'' $\mathcal{H}_0$ is separated into
several parts:
\begin{equation}
\mathcal{H}_0
=\mathcal{H}_{0\phi}^N
+\mathcal{H}_{0\phi}^D
+\mathcal{H}_{0Q}^D
+\mathcal{H}_{0ep^{\text L}}^N
+\mathcal{H}_{0e}^D
+\mathcal{H}_{0p^{\text T}}^D
+\mathcal{H}_{0p^{\text L}}^D,
\end{equation}
the scaling properties of which are found to be
\begin{equation}
\begin{split}
& \mathcal{H}_{0\phi}^N
 \sim\xi^{-\frac{\gamma}{2\nu}},\qquad
\mathcal{H}_{0\phi}^D
 \sim\xi^{-\frac{\gamma}{\nu}},\qquad
\mathcal{H}_{0Q}^D
 \sim\xi^{-2}, \\
&\mathcal{H}_{0ep^{\text L}}^N
 \sim\xi^{-\frac{d}{2}-1-\frac{1}{\nu}},\qquad
\mathcal{H}_{0e}^D
 \sim\xi^{-2-\frac{\alpha}{\nu}}, \\
&\mathcal{H}_{0p^{\text T}}^D
 \sim\xi^{-2},\qquad
\mathcal{H}_{0p^{\text L}}^D
 \sim\xi^{-2-\frac{\alpha}{\nu}}+\xi^{-2}.
\end{split}
\end{equation}

We next turn to $\tilde{\Omega}_{j;lm}$. This is contained in the
interaction ``Hamiltonian'' $\mathcal{H}'$ and determines its
scaling property. We note that $\tilde{\Omega}_{j;lm}$ 
involves the commutators of $\tilde{A}_j$. Hence it would be
useful to divide the interaction ``Hamiltonian'' as
\begin{align}
\mathcal{H}'
&=
\mathcal{H}'_{\phi}+\mathcal{H}'_{Q}
+\mathcal{H}'_{e}+\mathcal{H}'_{p}
\\
\intertext{with}
\mathcal{H}'_{\phi}
&=
\mathcal{H}'_{[\phi,\phi]}
+\mathcal{H}'_{[\phi,Q]}
+\mathcal{H}'_{[\phi,E]}
+\mathcal{H}'_{[\phi,P]},
\nonumber\\
\mathcal{H}'_{Q}
&=
\mathcal{H}'_{[Q,\phi]}
+\mathcal{H}'_{[Q,Q]}
+\mathcal{H}'_{[Q,E]}
+\mathcal{H}'_{[Q,P]},
\nonumber\\
\mathcal{H}'_{e}
&=
\mathcal{H}'_{[E,\phi]}
+\mathcal{H}'_{[E,Q]}
+\mathcal{H}'_{[E,E]}
+\mathcal{H}'_{[E,P]},
\\
\mathcal{H}'_{p}
&=
\mathcal{H}'_{[P,\phi]}
+\mathcal{H}'_{[P,Q]}
+\mathcal{H}'_{[P,E]}
+\mathcal{H}'_{[P,P]},\nonumber
\end{align}
where the scaling properties of each term are
\begin{equation}
\begin{split}
& \mathcal{H}'_{[\phi,\phi]} \sim\xi^{-\frac{\gamma}{2\nu}},\qquad
 \mathcal{H}'_{[\phi,Q]} \sim\xi^{-\frac{d}{2}},\qquad
 \mathcal{H}'_{[\phi,E]} \sim\xi^{-\frac{d}{2}-\frac{1}{\nu}},\qquad
 \mathcal{H}'_{[\phi,P]} \sim\xi^{-\frac{d}{2}-1}, \\
&\mathcal{H}'_{[Q,Q]} \sim\xi^{-\frac{d}{2}},\qquad
 \mathcal{H}'_{[Q,E]} \sim\xi^{-\frac{d}{2}-\frac{1}{\nu}},\qquad
 \mathcal{H}'_{[Q,P]} \sim\xi^{-\frac{d}{2}-1}, \\
&\mathcal{H}'_{[E,E]} \sim\xi^{\frac{d}{2}-1-\frac{2}{\nu}},\qquad
 \mathcal{H}'_{[E,P]} \sim\xi^{-\frac{d}{2}-1-\frac{1}{\nu}},\qquad
 \mathcal{H}'_{[P,P]} \sim\xi^{-\frac{d}{2}-1}.
\end{split}
\end{equation}

We are now in a position to discuss the dynamic behavior of each slow
mode. We start with the dynamic behavior of the meson mode or the
order parameter fluctuation. The information of the dynamic behavior
is contained in the reduced propagator
\begin{equation}
g_{\phi\vec{q}\alpha}(t)
=
\langle 0|
\alpha_{\phi\vec{q}\alpha} e^{-\text it\mathcal{H}}
\tilde{\alpha}_{\phi\vec{q}\alpha}
|0\rangle,
\end{equation}
where $\alpha_{\phi\vec{q}\alpha}$ and
$\tilde{\alpha}_{\phi\vec{q}\alpha}$ are the annihilation and creation
operators for the mode $\phi_{\alpha\vec{q}}$. From
Eqs.~(\ref{reducedpropagator}) and (\ref{effectivehamiltonian}), we
find that the Fourier transform of $g_{\phi\vec{q}\alpha}(t)$ is
written as
\begin{equation}
\hat{g}_{\phi\vec{q}\alpha}(\omega)
=\text i
\langle 0_\phi|
\alpha_{\phi\vec{q}\alpha}
\frac{1}{\omega-\mathcal{H}_\phi(\omega)}
\tilde{\alpha}_{\phi\vec{q}\alpha}
|0_\phi\rangle,
\end{equation}
where
\begin{align}
\mathcal{H}_\phi(\omega)
=&\mathcal{H}_{0\phi}^N
+\mathcal{H}_{0\phi}^D
+\langle 0_{Qep}|
\mathcal{H}'
|0_{Qep}\rangle
\nonumber\\
&+
\langle 0_{Qep}|
\mathcal{H}'
\frac{1}
{
\omega-\left(1-\mathcal{P}_{Qep}\right)\mathcal{H}
}
\left(1-\mathcal{P}_{Qep}\right)
\mathcal{H}'
|0_{Qep}\rangle.
\label{Heffphi}
\end{align}
The vacuum states $|0_\phi\rangle$ and $|0_{Qep}\rangle$ are defined
with respect to the modes $\{\phi\}$ and $\{Q,e,p\}$ respectively and
$\mathcal{P}_{Qep}$ is the projection operator onto
$|0_{Qep}\rangle$.

Here we need to use the general property of the interaction
``Hamiltonian'' \cite{kawareview}. The general argument tells us that
the interaction ``Hamiltonian'' involving the commutator of a mode $j$
has at least one creation or annihilation operator of $j$. This means
that
\begin{equation}
\langle 0_j|
\mathcal{H}'_{[j,l]}
|0_j\rangle=0.
\end{equation}
Moreover it is proved that the interaction ``Hamiltonian'' with
$\Omega_{j;lm}$, \ie, $\mathcal{H}'_j$ has at least one creation
operator as well as one annihilation operator of $j$. Thus both
$\mathcal{H}'_j|0_j\rangle$ and $\langle 0_j|\mathcal{H}'_j$ vanish.

With these in mind, we find
\begin{equation}
\langle 0_{Qep}|
\mathcal{H}'
|0_{Qep}\rangle
=
\langle 0_{Qep}|
\mathcal{H}'_{[\phi,\phi]}
|0_{Qep}\rangle
\sim\xi^{-\frac{\gamma}{2\nu}}.
\end{equation}
As for the last term in Eq.\ (\ref{Heffphi}), we note that at least
one mode among $\{Q,e,p\}$ must be excited in intermediate states
because of the presence of $1-\mathcal{P}_{Qep}$. In order to make the
last term as large as possible, it is desirable that the denominator
is small. Because the ``Hamiltonian'' for the $\phi$ mode,
$\mathcal{H}_{0\phi}^N\sim\mathcal{H}'_{[\phi,\phi]}\sim\xi^{-\gamma/2\nu}$,
is larger than that of the other modes, the denominator would become
larger in the absence of the $\phi$ mode in the intermediate
state. However this is not allowed because
\begin{equation}
\langle 0_{Qep}|
\mathcal{H}'
|0_\phi\rangle
=
\langle 0_\phi|
\mathcal{H}'
|0_{Qep}\rangle
=0.
\end{equation}
In the end, it turns out that the largest contribution is found in the
presence of the $\phi$ mode in the intermediate state to be
$(\mathcal{H}'_{[\phi,\phi]})^2/(\mathcal{H}_{0\phi}^N+\mathcal{H}'_{[\phi,\phi]})\sim\xi^{-\gamma/2\nu}$.
Thus the two interaction ``Hamiltonian''s in the effective
``Hamiltonian'' for the $\phi$ mode are determined by the $\phi$ mode
itself. This indicates that the $\phi$ mode almost decouples from the
other modes.

We have clarified the scaling property of the effective
``Hamiltonian'' for the $\phi$ mode,
\begin{align}
\mathcal{H}_\phi
&=\mathcal{H}_{0\phi}^N
+\mathcal{H}_{0\phi}^D
+\langle 0_{Qep}|
\mathcal{H}'
|0_{Qep}\rangle
\nonumber\\
&\qquad +
\langle 0_{Qep}|
\mathcal{H}'
\frac{1}
{
\omega-\left(1-\mathcal{P}_{Qep}\right)\mathcal{H}
}
\left(1-\mathcal{P}_{Qep}\right)
\mathcal{H}'
|0_{Qep}\rangle
\nonumber\\
&\sim
\xi^{-\frac{\gamma}{2\nu}}
+\xi^{-\frac{\gamma}{\nu}}
+\xi^{-\frac{\gamma}{2\nu}}
+\xi^{-\frac{\gamma}{2\nu}}.
\end{align}
The dynamics of the $\phi$ mode is dominated by
$\mathcal{H}_{0\phi}^N$ and the two interaction terms, all of which
scale as $\xi^{-\gamma/2\nu}$. This means that
$g_{\phi\vec{q}\alpha}(t)$ has the following form,
\begin{equation}
g_{\phi\vec{q}\alpha}(t)
=F\left(t\Gamma_{\phi\vec{q}\alpha}(\xi),q\xi\right)
\end{equation}
with the characteristic frequency of the form,
\begin{equation}
\Gamma_{\phi\vec{q}\alpha}(\xi)
=q^{\frac{\gamma}{2\nu}}f(q\xi).
\end{equation}
Thus we obtain the dynamic critical exponent $z$ for the $\phi$ mode
\begin{equation}
z_{\phi}=\frac{\gamma}{2\nu}=1-\frac{\eta}{2}.
\end{equation}
When we use the values of the static critical exponents, $\eta=0.03$
\cite{bnm}, we find $z_\phi\cong0.98$, which is contrasted with the
value $z_N=d/2=3/2$ of the staggered magnetization mode in the
antiferromagnet.

Recently Boyanovsky and de Vega have calculated the dynamic critical
exponent $z_\phi$ in a different approach \cite{Boyanovsky:2001pa}.
They apply the renormalization group method directly to the
microscopic theory, \ie, the field theory of the O($N$) linear sigma
model and obtain
\begin{equation}
z=1+\epsilon\frac{N+2}{(N+8)^2}+\mathcal{O}(\epsilon^2).
\end{equation}
For $N=4$ and $\epsilon=1$, the above expression gives
$z_\phi\cong1.04$, which is compatible with our result rather than the
value for the antiferromagnet.

We note that the fluctuation effect
reduces the value to less than unity in our framework, while the
opposite occurs in the calculation in Ref.~\cite{Boyanovsky:2001pa}.
The further close investigations of the kinetic equation, \ie, the
application of the dynamic renormalization group method to the kinetic
equation, may reveal the critical exponent larger than unity. Or
it might be possible that the inclusion of higher order terms in the
$\epsilon$ expansion and the proper resummation of those terms in the
approach in Ref.~\cite{Boyanovsky:2001pa} turn its tendency to the
opposite. In any case, it needs more involved investigation to settle
whether the dynamic critical exponent is larger or less than unity.

We note that if we employ the values in the large $N$ limit for
$\eta$, that is, $\eta=0$, we have $z_\phi=1$. This is reasonable
because the fluctuation is suppressed in the large $N$ limit.

We find that the dissipation term  $\mathcal{H}_{0\phi}^D$ is smaller
than the interaction terms. Thus the dissipation term is overwhelmed
or renormalized by the interaction terms, which results in the same
scaling pattern as the frequency term, that is,
$\sim\xi^{-\gamma/2\nu}$. As a consequence, the Onsager coefficient
diverges at the critical point as
\begin{equation}
L_\phi\sim\xi^{\frac{\gamma}{2\nu}}.
\end{equation}
What this means is that the width of the meson mode becomes narrower
just in the same manner as that of the softening when we approach the
critical point. This is not necessarily trivial. In fact, one of the
motivations of the calculation in Ref.~\cite{Boyanovsky:2001pa} is
based on the suspicion that the meson mode may not be a good
quasiparticle mode with a narrow width near the chiral phase
transition. The calculation shows that the meson mode indeed becomes a
good quasiparticle near the critical point, which agrees with our
result.

As a final comment on the $\phi$ mode, we give the comparison
with the antiferromagnet. In the antiferromagnet, the scaling
property of the commutator of the staggered magnetization
$\sigma$ is given by $[\tilde{\sigma},\tilde{\sigma}]\sim\xi^{-d/2}$
which leads to $\mathcal{H}'\sim\xi^{-d/2}$. The dissipation term
$\mathcal{H}_{0\sigma}^D$ scales as
$\mathcal{H}_{0\sigma}^D\sim\xi^{-\gamma/\nu}$ so that it is
renormalized by $\mathcal{H}'$ to scale as $\sim\xi^{-d/2}$. This
gives the critical exponent $z_\sigma=d/2$, which is also predicted by
the renormalization group method \cite{hhma}.

Next we consider the chiral charge $Q$. Using
Eqs.~(\ref{reducedpropagator}) and (\ref{effectivehamiltonian}),
we have the propagator of the $Q$ mode as
\begin{equation}
\hat{g}_{Q\vec{q}\alpha\beta}(\omega)
=\text i
\langle 0_Q|
\alpha_{Q\vec{q}\alpha\beta}
\frac{1}{\omega-\mathcal{H}_Q(\omega)}
\tilde{\alpha}_{Q\vec{q}\alpha\beta}
|0_Q\rangle,
\end{equation}
where
\begin{align}
\mathcal{H}_Q(\omega)
=&\mathcal{H}_{0Q}^D
+\langle 0_{\phi ep}|
\mathcal{H}'
|0_{\phi ep}\rangle
\nonumber\\
&+
\langle 0_{\phi ep}|
\mathcal{H}'
\frac{1}
{
\omega-\left(1-\mathcal{P}_{\phi ep}\right)\mathcal{H}
}
\left(1-\mathcal{P}_{\phi ep}\right)
\mathcal{H}'
|0_{\phi ep}\rangle.
\label{HeffQ}
\end{align}
The second term reduces to
\begin{equation}
\langle 0_{\phi ep}|
\mathcal{H}'_{[Q,Q]}
|0_{\phi ep}\rangle
\sim\xi^{-\frac{d}{2}}.
\end{equation}
The similar argument to the $\phi$ mode dynamics applies to the third
term of Eq.~(\ref{HeffQ}). There arises the most important
contribution from the situation where the intermediate state has the
$Q$ mode as well as the $e$ or $p$ mode excitation but does not have
the $\phi$ mode. Because the $Q$ mode ``Hamiltonian,''
$\mathcal{H}'_{[Q,Q]}\sim\xi^{-d/2}$, is greater than those for the
$e$ and $p$ modes, the $Q$ mode dominates in the intermediate state.
Therefore the third term scales as
$(\mathcal{H}'_{[Q,Q]})^2/\mathcal{H}'_{[Q,Q]}\sim\xi^{-d/2}$. As in
the $\phi$ mode, the two interaction ``Hamiltonian''s in the effective
``Hamiltonian'' $\mathcal{H}_Q$ are determined only by the $Q$ mode,
which means that the $Q$ mode decouples from the other modes. Thus we
find
\begin{align}
\mathcal{H}_Q
&=\mathcal{H}_{0Q}^D
+\langle 0_{\phi ep}|
\mathcal{H}'
|0_{\phi ep}\rangle
+
\langle 0_{\phi ep}|
\mathcal{H}'
\frac{1}
{
\omega-\left(1-\mathcal{P}_{\phi ep}\right)\mathcal{H}
}
\left(1-\mathcal{P}_{\phi ep}\right)
\mathcal{H}'
|0_{\phi ep}\rangle
\nonumber\\
&\sim
\xi^{-2}+\xi^{-\frac{d}{2}}+\xi^{-\frac{d}{2}},
\end{align}
which gives us the dynamic critical exponent for the $Q$ mode,
\begin{equation}
z_Q=\frac{d}{2}.
\label{Qmode}
\end{equation}
We see that the dissipation term receives the renormalization from the
interaction terms and the Onsager coefficient diverges as
\begin{equation}
L_Q\sim\xi^{2-\frac{d}{2}}.
\end{equation}
The behavior of the $Q$ mode is quite the same as that of the
magnetization $\mu$ in the antiferromagnet which is the counterpart of
the $Q$ mode in the chiral phase transition. The magnetization is
renormalized by the interaction terms to have the critical exponent
$z_\mu=d/2$, which is the same as that of the staggered magnetization.
In this sense, the $Q$ mode may be regarded as a vestige of the
antiferromagnet.

The third mode to be considered is the $p^{\text T}$ mode. The
discussion goes in the similar way to before.
The propagator for $p^{\text T}$ is written as
\begin{equation}
\hat{g}_{p^{\text T}\vec{q}i}(\omega)
=\text i
\langle 0_{p^{\text T}}|
\alpha_{p^{\text T}\vec{q}i}
\frac{1}{\omega-\mathcal{H}_{p^{\text T}}(\omega)}
\tilde{\alpha}_{p^{\text T}\vec{q}i}
|0_{p^{\text T}}\rangle,
\end{equation}
where
\begin{align}
\mathcal{H}_{p^{\text T}}(\omega)
=&\mathcal{H}_{0p^{\text T}}^D
+\langle 0_{\phi Qep^{\text L}}|
\mathcal{H}'
|0_{\phi Qep^{\text L}}\rangle
\nonumber\\
&+
\langle 0_{\phi Qep^{\text L}}|
\mathcal{H}'
\frac{1}
{
\omega-\left(1-\mathcal{P}_{\phi Qep^{\text L}}\right)\mathcal{H}
}
\left(1-\mathcal{P}_{\phi Qep^{\text L}}\right)
\mathcal{H}'
|0_{\phi Qep^{\text L}}\rangle.
\label{Heffpt}
\end{align}
The scaling property of the second term in
$\mathcal{H}_{p^{\text T}}(\omega)$ is given by
\begin{equation}
\langle 0_{\phi Qep^{\text L}}|
\mathcal{H}'
|0_{\phi Qep^{\text L}}\rangle
=
\langle 0_{\phi Qep^{\text L}}|
\mathcal{H}'_{[p^{\text T},p^{\text T}]}
|0_{\phi Qep^{\text L}}\rangle
\sim\xi^{-\frac{d}{2}-1}.
\end{equation}
For the third term, we have the most important contribution coming
from the process with the $p^{\text T}$ as well as the $e$ or
$p^{\text L}$ modes but without the $\phi$ and $Q$ modes in the
intermediate state, which scales as $\sim\xi^{-d+\alpha/\nu}$. Thus we
find
\begin{align}
\mathcal{H}_{p^{\text T}}
&=\mathcal{H}_{0p^{\text T}}^D
+\langle 0_{\phi Qep^{\text L}}|
\mathcal{H}'
|0_{\phi Qep^{\text L}}\rangle
\nonumber\\
&\qquad +
\langle 0_{\phi Qep^{\text L}}|
\mathcal{H}'
\frac{1}
{
\omega-\left(1-\mathcal{P}_{\phi Qep^{\text L}}\right)\mathcal{H}
}
\left(1-\mathcal{P}_{\phi Qep^{\text L}}\right)
\mathcal{H}'
|0_{\phi Qep^{\text L}}\rangle
\nonumber\\
& \sim
\xi^{-2}
+\xi^{-\frac{d}{2}-1}
+\xi^{-d+\frac{\alpha}{\nu}}.
\end{align}
In the intermediate state in the third term, the $e$ and $p^{\text L}$
modes couple to the $p^{\text T}$ mode. The most dominant term in the
effective ``Hamiltonian'' is, however, not the interaction terms
but the free term. Thus the $p^{\text T}$ mode again closes its
dynamics within itself. The $p^{\text T}$ mode scales as
$\sim\xi^{-2}$ without renormalization, leading to the critical
exponent
\begin{equation}
z_{p^{\text T}}=2.
\label{pTmode}
\end{equation}
Since the renormalization is not effective, the Onsager
coefficient of the transverse momentum, \ie, the shear viscosity, does
not diverge at the critical point. This may be compared with the
transverse velocity mode in the normal fluid in which the shear
viscosity shows the logarithmic divergence. The logarithmic divergence
is a consequence of renormalization from the third term with the
$\xi^{-2}$ scaling. In the case of the normal fluid, there is a very
slow mode, that is, the entropy mode which scales as $\xi^{-d}$.
Because the intermediate state is dominated by this entropy mode, the
denominator of the third term becomes very small, which results in the
larger third term than ours.

Finally we consider the energy wave mode composed of the $e$ and
$p^{\text L}$ variables. The propagator is found to be
\begin{equation}
\hat{g}_{ep^{\text L}\vec{q}i}(\omega)
=\text i
\langle 0_{ep^{\text L}}|
\alpha_{ep^{\text L}\vec{q}i}
\frac{1}{\omega-\mathcal{H}_{ep^{\text L}}(\omega)}
\tilde{\alpha}_{ep^{\text L}\vec{q}i}
|0_{ep^{\text L}}\rangle,
\end{equation}
where
\begin{align}
\mathcal{H}_{ep^{\text L}}(\omega)
=&\mathcal{H}_{0ep^{\text L}}^N
+\mathcal{H}_{0e}^D
+\mathcal{H}_{0p^{\text L}}^D
+\langle 0_{\phi Qp^{\text T}}|
\mathcal{H}'
|0_{\phi Qp^{\text T}}\rangle
\nonumber\\
&+
\langle 0_{\phi Qp^{\text T}}|
\mathcal{H}'
\frac{1}
{
\omega-\left(1-\mathcal{P}_{\phi Qp^{\text T}}\right)\mathcal{H}
}
\left(1-\mathcal{P}_{\phi Qp^{\text T}}\right)
\mathcal{H}'
|0_{\phi Qp^{\text T}}\rangle.
\label{Heffepl}
\end{align}
The first interaction term scales as
\begin{equation}
\langle 0_{\phi Qp^{\text T}}|
\mathcal{H}'
|0_{\phi Qp^{\text T}}\rangle
=
\langle 0_{\phi Qp^{\text T}}|
\mathcal{H}'_{[E,E]}
+\mathcal{H}'_{[E,P^{\text L}]}
+\mathcal{H}'_{[P^{\text L},P^{\text L}]}
|0_{\phi Qp^{\text T}}\rangle
\sim\xi^{\frac{d}{2}-1-\frac{2}{\nu}}.
\end{equation}
The most important contribution of the second interaction term arises
from the process in which the $e$ and $p^{\text L}$ modes as well as
the $p^{\text T}$ mode appear but the $\phi$ and $Q$ modes are absent
in the intermediate state, which gives the scaling behavior of the
last term $\sim\xi^{d-(4-\alpha)/\nu}$. Thus we find
\begin{align}
\mathcal{H}_{ep^{\text L}}
&=\mathcal{H}_{0ep^{\text L}}^N
+\mathcal{H}_{0e}^D
+\mathcal{H}_{0p^{\text L}}^D
+\langle 0_{\phi Qp^{\text T}}|
\mathcal{H}'
|0_{\phi Qp^{\text T}}\rangle
\nonumber\\
&\qquad+
\langle 0_{\phi Qp^{\text T}}|
\mathcal{H}'
\frac{1}
{
\omega-\left(1-\mathcal{P}_{\phi Qp^{\text T}}\right)\mathcal{H}
}
\left(1-\mathcal{P}_{\phi Qp^{\text T}}\right)
\mathcal{H}'
|0_{\phi Qp^{\text T}}\rangle
\nonumber\\
&\sim
\xi^{-\frac{d}{2}-1-\frac{1}{\nu}}
+\xi^{-2-\frac{\alpha}{\nu}}
+\left(\xi^{-2-\frac{\alpha}{\nu}}+\xi^{-2}\right)
+\xi^{\frac{d}{2}-1-\frac{2}{\nu}}
+\xi^{d-\frac{4-\alpha}{\nu}},
\end{align}
where the two terms in the parenthesis correspond to the terms
associated with the thermal conductivity $\lambda^0$ and the bulk
viscosity $\zeta^0$ in $\mathcal{H}_{0p^{\text L}}^D$. We see that the
non-dissipation term is renormalized by the interaction terms, but the
dissipation terms are the dominant terms and not renormalized, which
give the critical exponent
\begin{equation}
z_{ep^{\text L}}=2+\frac{\alpha}{\nu}\cong1.74
\label{epLmode}
\end{equation}
with $\alpha=-0.19$ and $\nu=0.73$ substituted \cite{bnm}. Thus the
timescale of the energy wave mode is determined by the $e$ and
$p^{\text L}$ modes themselves and is not affected by the other
modes. We note that the bulk viscosity does not diverge in contrast
with the normal fluid. In the normal fluid, the commutator
$[\tilde{p},\tilde{u}^{\text L}]$, that corresponds to
$[\tilde{E},\tilde{P}^{\text L}]$ in our system, is very large. It
enters into the non-dissipation term as well as into $\mathcal{H}'$,
the latter of which thus renormalizes the dissipation term. Because
the slow entropy mode dominates the denominator, the second
interaction term becomes even larger. Thus the bulk viscosity of the
normal fluid shows the very strong divergence. In the chiral phase
transition, $[\tilde{E},\tilde{P^{\text L}}]$ is somehow very small and
the intermediate state in the second interaction term is not occupied
by such a slow mode as the entropy mode.Thus the divergence of
$\lambda$ and $\zeta$ does not occur.

From the above discussions on the four modes, we notice that the
dynamic critical exponents, \ie, the time scales of each mode are all
different and all the modes decouple from each other. As mentioned in
Sec.~\ref{generalreview}, the original dynamic scaling hypothesis
proposed in Ref.~\cite{hhhypothesis} claims that all the exponents
are identical. In this sense, the ``original'' dynamic scaling
hypothesis is violated in the present system.

Although the four modes decouple completely in our system, this
should be regarded as an extreme case. In the antiferromagnet, the
modes of the staggered magnetization and the magnetization couple to
each other strongly. The critical exponents for the two modes are
identical, though the energy mode has a different critical exponent.
Thus the original dynamic scaling hypothesis holds partially in the
antiferromagnet.

We note that in the chiral system, the largest critical exponent or
the largest time scale is provided by the $p^{\text T}$ mode. This
means that the $p^{\text T}$ mode is the slowest among the four
modes. Thus it determines the relaxation time of the whole system. The
other modes in themselves are certainly of interest. In particular,
the softening and the narrowing of the meson mode have the many
physical implications. But when we consider the relaxation of the
system, it is dominated by the decay of the $p^{\text T}$ mode, and
the other modes are not relevant.

\section{Summary}
\label{summary}

We have discussed the dynamic aspect of the chiral phase transition.
The classification method of the dynamic universality class had been
established by Hohenberg and Halperin. The method is based on the
prescription of how to collect the slow variables in the system. The
prevailing prescription tells us to take the order parameter and
conserved quantities of the system for the slow variables. If we
follow the prescription, we are incorrectly led to identify the
dynamic universality class of the chiral phase transition with that of
the antiferromagnet and to find the dynamic critical exponent of the
order parameter fluctuation to be $z=d/2=1.5$, as Rajagopal and
Wilczek had argued.

We have then turned into the consideration of the slow mode rather
than the slow variable in order to find the crucial difference between
the chiral phase transition and the antiferromagnet. While the order
parameter fluctuation of the antiferromagnet in the disordered phase
is a diffusive mode, the meson field which is the order parameter of
the chiral phase transition gives apparently a propagating mode. We
have stated that in order to describe the meson mode appropriately, we
must include the canonical momentum conjugate to the meson field into
the member of the slow variables. Since the slow modes, and
accordingly the slow variables of the two systems, are different, it
is impossible to say that the dynamic universality class of the two
systems are identical. What this means is the breakdown of the
prescription for choosing the slow variables just by gathering the
order parameter and the conserved quantities, because the canonical
momentum for the meson field is neither the order parameter nor the
conserved quantity. In the chiral phase transition, the order
parameter and the conserved quantity does not furnish us with the full
slow variables required to describe the slow mode properly. We have
dangerous potentiality that the dynamic universality class given by
Hohenberg and Halperin's classification method is divided into finer
classes, if we consider the slow modes and their appropriate slow
variables in that dynamic universality class.

The necessity of the canonical momentum for the slow variable is
itself nothing strange. When we return to the microscopic eqution of
motion, \ie, the Heisenberg equation for the meson field, we can
immediately realize that the canonical momentum in addition to the
meson field is necessary to describe the meson dynamics.
Moreover we have mentioned that the canonical momentum as a slow
variable is not restricted to the chiral phase transition. In the
suprefluid and the ferromagnet, the canonical momentum plays the role
of the slow variable in order to describe the second sound and the
spin wave respectively. In the two systems, the canonical momentum is
just an order parameter or a conserved quantity, and can become the
slow variable just by the old prescription. In the chiral phase
transition, on the other, the canonical momentum is not the order
parameter or the conserved quantity. This is just the only new feature
of the chiral phase transition.

The above observation has inspired the necessity of reanalysis of the
chiral phase transition. We have employed the mode coupling theory for
that purpose.

Firstly, we have applied the theory to the O(2) linear sigma model and
found that the meson mode appears desirably from the meson field and
the canonical momentum.  We have also clarified the other slow modes
than the meson mode: The chiral charge gives a diffusive mode in the
disordered phase. In the ordered phase, it couples with the pion
field, which results in lifting the degeneracy of the pion mode with
the sigma mode. The energy and the longitudinal momentum are combined
to give the energy wave mode, which is a propagating mode and a
correspondent to a sound wave in a fluid. The transverse momenta
give diffusive modes.

We have then examined the O(4) linear sigma model in the mode coupling
theory in order to derive the kinetic equation and calculate the
dynamic critical exponents in the disordered phase. We have found the
exponent for the meson mode to be $z_{\phi}=1-\eta/2\cong 0.98$, which
is to be compared with the value of $d/2$ obtained assuming that the
chiral phase transition belongs to the dynamic universality class of
the antiferromagnet. The different dynamic critical exponents show
explicitly the different dynamic behavior. We have also calculated the
dynamic critical exponents of the modes other than the meson mode. The
results are given in Eqs.~(\ref{Qmode}), (\ref{pTmode}) and
(\ref{epLmode}). We have noted that the largest exponent is given by
the transverse momentum fluctuation, which determines the relaxation
time of the whole system in the chiral phase transition.

We have succeeded in describing the meson mode by including the
canonical momentum into the member of the slow variables. We were able
to find, even accidentally, the needed slow variable, that is, the
canonical momentum by returning to the microscopic Heisenberg
equation. However, we must admit that the argument does not go beyond
a heuristic one. We do not have any definite prescription to determine
the slow variables.
It will be our future problem to find out the way how to determine the
slow variables uniquely, if it exists. Once it is found, it
means that we have obtained a new classification method for the
dynamic universality class, which completes that of Hohenberg and
Halperin.

We note that a ``particle mode'' like the meson mode is inherent in
the relativistic system because it accompanies the anti-particle mode
as a partner. As we have seen, for the description of the ``particle
mode,'' we need the canonical momentum that is neither the order
parameter nor the conserved quantity. Thus the relativistic system
that involves the ``particle mode'' as a slow mode should be a
quite novel critical point that is not classified into any dynamic
universality classes of the non-relativistic systems considered by
Hohenberg and Halperin in Ref.~\cite{Hohenberg:1977ym}.
One of such a system would be the critical end point (CEP) in the QCD
phase diagram
\cite{Asakawa:1989bq,Berges:1998rc,Halasz:1998qr,Stephanov:1998dy,Fukushima:2002mp,Fujii:2004jt,Son:2004iv}.
Although the dynamic universality class of the CEP is discussed in
Ref.~\cite{Son:2004iv}, the argument is not sufficient in
respect that only the order parameter and the conserved quantities are
compared and the slow modes are not taken into account
explicitly. Other new and intriguing systems would be the tricritical
point, the critical point associated with the color-superconductivity,
the confinement transition, the electro-weak transition and so on. The
analysis of those systems including CEP should also be what we should
do in the future.

\section*{Acknowledgments}

Two of the authors (K.~O and K.~O) are grateful to all the members of
the nuclear theory group at Komaba in the university of Tokyo for
useful discussions. K.~Ohnishi thanks Prof.~T.~Hatsuda for precious
comments and discussions. He also thanks Dr.~M.~Ohtani for fruitfull
discussions. He had an opportunity to have stimulating discussions
with Prof.~T.~Kunihiro during stay at YITP as its visitor program. He
would like to express his gratitude to Pror.~Kunihiro. K.~F is
supported by Japan Society for the Promotion of Science for Young
Scientists.

\appendix
\section{Commutation relations of the slow variables}
\label{app:commutation}
In this appendix, we list up the commutation relations among the
slow variables.

\begin{align}
\left[\phi_{\alpha\vec{q}}, (\phi_{\alpha'\vec{q}'})^\dagger\right]
&=
\frac{1}{2\omega_{\vec{q}}}\delta_{\alpha\alpha'}\delta_{\vec{q}\vec{q}'}
\label{phiphi}
\\
\left[\phi_{\alpha\vec{q}},
(\phi_{\alpha'-\vec{q}'}^{\dagger})^\dagger\right]
&=0
\\
\left[\phi_{\alpha\vec{q}},
(Q_{\alpha'\beta'\vec{q}'})^\dagger\right]&
\nonumber\\
&\hspace{-2cm}=
\frac{-\text i}{2\sqrt{V}\sqrt{2\omega_{\vec{q}}}}
\epsilon_{\alpha\alpha'\beta'\gamma}
\left\{
\sqrt{\frac{\omega_{\vec{q}}}{\omega_{\vec{q}-\vec{q}'}}}
\left(a_{\gamma\vec{q}-\vec{q}'}
+a_{\gamma\vec{q}'-\vec{q}}^{\dagger}\right)
+\sqrt{\frac{\omega_{\vec{q}-\vec{q}'}}{\omega_{\vec{q}}}}
\left(a_{\gamma\vec{q}-\vec{q}'}
-a_{\gamma\vec{q}'-\vec{q}}^{\dagger}\right)
\right\}
\label{phiQ}
\\
\left[\phi_{\alpha\vec{q}},
(E_{\vec{q}'})^\dagger\right]
&=
\frac{1}{2\sqrt{V}\sqrt{2\omega_{\vec{q}}}}
\sqrt{\omega_{\vec{q}}\omega_{\vec{q}-\vec{q}'}}
\left(a_{\alpha\vec{q}-\vec{q}'}
-a_{\alpha\vec{q}'-\vec{q}}^{\dagger}\right)
\nonumber\\
&+
\frac{1}{2\sqrt{V}\sqrt{2\omega_{\vec{q}}}}
\frac{1}{\sqrt{
\omega_{\vec{q}}\omega_{\vec{q}-\vec{q}'}}}
\left[\vec{q}\cdot(\vec{q}-\vec{q}')+\mu^2\right]
\left(a_{\alpha\vec{q}-\vec{q}'}
+a_{\alpha\vec{q}'-\vec{q}}^{\dagger}\right)
\nonumber\\
&+
\frac{\lambda}{V^{\frac{3}{2}}\sqrt{2\omega_{\vec{q}}}}
\sum_{\vec{k}_1\cdots\vec{k}_4}
\frac{1}{4}
\frac{1}{\sqrt{
\omega_{\vec{k}_1}\cdots\omega_{\vec{k}_4}}}
\delta_{-\vec{q}', \vec{k}_1+\cdots+\vec{k}_4}
\delta_{-\vec{q}, \vec{k}_1}
\nonumber\\
&\hspace{3cm}\times
\left(a_{\alpha\vec{k}_2}
+a_{\alpha-\vec{k}_2}^{\dagger}\right)
\left(a_{\beta\vec{k}_3}
+a_{\beta-\vec{k}_3}^{\dagger}\right)
\left(a_{\beta\vec{k}_4}
+a_{\beta-\vec{k}_4}^{\dagger}\right)
\\
\left[\phi_{\alpha\vec{q}},
(P_{\vec{q}'}^i)^\dagger\right]&
\nonumber\\
&\hspace{-2cm}=
\frac{1}{2\sqrt{V}\sqrt{2\omega_{\vec{q}}}}
\left\{
\sqrt{\frac{\omega_{\vec{q}}}{\omega_{\vec{q}-\vec{q}'}}}(q-q')^i
\left(a_{\alpha\vec{q}-\vec{q}'}
+a_{\alpha\vec{q}'-\vec{q}}^{\dagger}\right)
+\sqrt{\frac{\omega_{\vec{q}-\vec{q}'}}{\omega_{\vec{q}}}}q^i
\left(a_{\alpha\vec{q}-\vec{q}'}
-a_{\alpha\vec{q}'-\vec{q}}^{\dagger}\right)
\right\}
\label{phiP}
\\
\left[\phi_{\alpha-\vec{q}}^{\dagger},
(\phi_{\alpha'-\vec{q}'}^{\dagger})^\dagger\right]
&=
-\frac{1}{2\omega_{\vec{q}}}\delta_{\alpha\alpha'}\delta_{\vec{q}\vec{q}'}
\\
\left[\phi_{\alpha-\vec{q}}^{\dagger},
(Q_{\alpha'\beta'\vec{q}'})^\dagger\right]
&=
-\left[\phi_{\alpha-\vec{q}},
Q_{\alpha'\beta'\vec{q}'}\right]^\dagger
\nonumber\\
&\hspace{-1.8cm}=
\frac{-\text i}{2\sqrt{V}\sqrt{2\omega_{\vec{q}}}}
\epsilon_{\alpha\alpha'\beta'\gamma}
\left\{
\sqrt{\frac{\omega_{\vec{q}}}{\omega_{\vec{q}-\vec{q}'}}}
\left(a_{\gamma\vec{q}-\vec{q}'}
+a_{\gamma\vec{q}'-\vec{q}}^{\dagger}\right)
-\sqrt{\frac{\omega_{\vec{q}-\vec{q}'}}{\omega_{\vec{q}}}}
\left(a_{\gamma\vec{q}-\vec{q}'}
-a_{\gamma\vec{q}'-\vec{q}}^{\dagger}\right)
\right\}
\end{align}
\begin{align}
\left[\phi_{\alpha-\vec{q}}^{\dagger},
(E_{\vec{q}'})^\dagger\right]
=&
-\left[\phi_{\alpha-\vec{q}},
E_{\vec{q}'}\right]^\dagger
\nonumber\\
=&
\frac{1}{2\sqrt{V}\sqrt{2\omega_{\vec{q}}}}
\sqrt{\omega_{\vec{q}}\omega_{\vec{q}-\vec{q}'}}
\left(a_{\alpha\vec{q}-\vec{q}'}
-a_{\alpha\vec{q}'-\vec{q}}^{\dagger}\right)
\nonumber\\
&-
\frac{1}{2\sqrt{V}\sqrt{2\omega_{\vec{q}}}}
\frac{1}{\sqrt{
\omega_{\vec{q}}\omega_{\vec{q}-\vec{q}'}}}
\left[\vec{q}\cdot(\vec{q}-\vec{q}')+\mu^2\right]
\left(a_{\alpha\vec{q}-\vec{q}'}
+a_{\alpha\vec{q}'-\vec{q}}^{\dagger}\right)
\nonumber\\
&-
\frac{\lambda}{V^{\frac{3}{2}}\sqrt{2\omega_{\vec{q}}}}
\sum_{\vec{k}_1\cdots\vec{k}_4}
\frac{1}{4}
\frac{1}{\sqrt{
\omega_{\vec{k}_1}\cdots\omega_{\vec{k}_4}}}
\delta_{-\vec{q}', \vec{k}_1+\cdots+\vec{k}_4}
\delta_{-\vec{q}, \vec{k}_1}
\nonumber\\
&\hspace{2cm}\times
\left(a_{\alpha\vec{k}_2}
+a_{\alpha-\vec{k}_2}^{\dagger}\right)
\left(a_{\beta\vec{k}_3}
+a_{\beta-\vec{k}_3}^{\dagger}\right)
\left(a_{\beta\vec{k}_4}
+a_{\beta-\vec{k}_4}^{\dagger}\right)
\\
\left[\phi_{\alpha-\vec{q}}^{\dagger},
(P_{\vec{q}'}^i)^\dagger\right]
=&
-\left[\phi_{\alpha-\vec{q}},
P_{\vec{q}'}^i\right]^\dagger
\nonumber\\
&\hspace{-2.5cm}=
\frac{1}{2\sqrt{V}\sqrt{2\omega_{\vec{q}}}}
\left\{
\sqrt{\frac{\omega_{\vec{q}}}{\omega_{\vec{q}-\vec{q}'}}}(q-q')^i
\left(a_{\alpha\vec{q}-\vec{q}'}
+a_{\alpha\vec{q}'-\vec{q}}^{\dagger}\right)
-\sqrt{\frac{\omega_{\vec{q}-\vec{q}'}}{\omega_{\vec{q}}}}q^i
\left(a_{\alpha\vec{q}-\vec{q}'}
-a_{\alpha\vec{q}'-\vec{q}}^{\dagger}\right)
\right\}
\\
\left[Q_{a\vec{q}}^V, (Q_{b\vec{q}'}^V)^\dagger\right]
=&
\frac{\text i}{\sqrt{V}}\epsilon_{abc}Q_{c\vec{q}-\vec{q}'}^V
\label{QQ}
\\
\left[Q_{a\vec{q}}^V, (Q_{b\vec{q}'}^A)^\dagger\right]
=&
\frac{\text i}{\sqrt{V}}\epsilon_{abc}Q_{c\vec{q}-\vec{q}'}^A
\\
\left[Q_{a\vec{q}}^A, (Q_{b\vec{q}'}^A)^\dagger\right]
=&
\frac{\text i}{\sqrt{V}}\epsilon_{abc}Q_{c\vec{q}-\vec{q}'}^V
\\
\left[Q_{\alpha\beta\vec{q}},
(E_{\vec{q}'})^\dagger \right]
=&
-\text i\epsilon_{\alpha\beta\gamma\delta}\frac{1}{V}\sum_{\vec{k}}
\frac{1}{2}
\frac{1}{\sqrt{
\omega_{\vec{k}}\omega_{\vec{q}-\vec{q}'-\vec{k}}}}
\nonumber\\
&\hspace{1cm}\times(\vec{q}-\vec{k})\cdot(\vec{q}-\vec{q}'-\vec{k})
\left(a_{\gamma\vec{k}}
+a_{\gamma-\vec{k}}^{\dagger}\right)
\left(a_{\delta\vec{q}-\vec{q}'\vec{k}}
+a_{\delta\vec{k}+\vec{q}'-\vec{q}}^{\dagger}\right)
\\
\left[Q_{\alpha\beta\vec{q}},
(P_{\vec{q}'}^i)^\dagger \right]
=&
\frac{1}{\sqrt{V}}q^iQ_{\alpha\beta\vec{q}-\vec{q}'}
\label{QP}
\\
\left[E_{\vec{q}}, (E_{\vec{q}'})^\dagger \right]
=&
\frac{1}{\sqrt{V}}(q+q')^i P_{\vec{q}-\vec{q}'}^i
\label{EE}
\\
\left[E_{\vec{q}}, (P_{\vec{q}'}^i)^\dagger \right]
=&
-\frac{1}{2}(q+q')^i
\frac{1}{V}\sum_{\vec{k}}\frac{1}{2}
\sqrt{\omega_{\vec{k}}\omega_{\vec{q}-\vec{q}'-\vec{k}}}
\left(a_{\alpha\vec{k}}-a_{\alpha-\vec{k}}^{\dagger}\right)
\left(a_{\alpha\vec{q}-\vec{q}'-\vec{k}}
-a_{\alpha\vec{k}+\vec{q}'-\vec{q}}^{\dagger}\right)
\nonumber\\
&+
\frac{1}{V}\sum_{\vec{k}}\frac{1}{2}
\frac{1}{\sqrt{
\omega_{\vec{k}}\omega_{\vec{q}-\vec{q}'-\vec{k}}}}
\left[-\vec{q}\cdot\vec{k}+\vec{k}^2+\mu^2\right](q-q'-k)^i
\nonumber\\
&\hspace{1cm}\times
\left(a_{\alpha\vec{k}}+a_{\alpha-\vec{k}}^{\dagger}\right)
\left(a_{\alpha\vec{q}-\vec{q}'-\vec{k}}
+a_{\alpha\vec{k}+\vec{q}'-\vec{q}}^{\dagger}\right)
\nonumber\\
&+
\frac{\lambda}{V^2}
\sum_{\vec{k}_1\cdots\vec{k}_4}
\frac{1}{4}
\frac{1}{\sqrt{
\omega_{\vec{k}_1}\cdots\omega_{\vec{k}_4}}}
k_4^i
\delta_{\vec{q}-\vec{q}', \vec{k}_1+\cdots+\vec{k}_4}
\nonumber\\
&\hspace{1cm}\times
\left(a_{\beta\vec{k}_1}+a_{\beta-\vec{k}_1}^{\dagger}\right)
\left(a_{\beta\vec{k}_2}+a_{\beta-\vec{k}_2}^{\dagger}\right)
\left(a_{\alpha\vec{k}_3}+a_{\alpha-\vec{k}_3}^{\dagger}\right)
\left(a_{\alpha\vec{k}_4}+a_{\alpha-\vec{k}_4}^{\dagger}\right)
\\
\left[P_{\vec{q}}^i,
(P_{\vec{q}'}^j)^\dagger \right]
=&
\frac{1}{V}
\left(
q^jP_{\vec{q}-\vec{q}'}^i+{q'}^{i}P_{\vec{q}-\vec{q}'}^j
\right)
\label{PP}
\end{align}

\section{Dissipation terms in the relativistic hydrodynamics}
\label{app:hydro}
The dissipation terms for $E_{\vec{q}}$ and $P_{\vec{q}}^i$ are
read from the relativistic hydrodynamic equation
\cite{weinberg}. We will perform it in this appendix.
The energy-momentum tensor for the imperfect fluid is given by
\begin{align}
T^{\alpha\beta}
&=
T_0^{\alpha\beta}+\Delta T^{\alpha\beta},
\intertext{where}
T_0^{\alpha\beta}&=p\eta^{\alpha\beta}+(p+\rho)U^\alpha U^\beta,
\\
\Delta T^{\alpha\beta}
&=
-\eta H^{\alpha\gamma}H^{\beta\delta}W_{\gamma\delta}
-\lambda\left(H^{\alpha\gamma}U^\beta
+H^{\beta\gamma}U^\alpha\right)Q_\gamma
-\zeta H^{\alpha\beta}
\frac{\partial U^\gamma}{\partial x^\gamma}.
\end{align}
The $T_0^{\alpha\beta}$ is for the perfect fluid and
$\Delta T^{\alpha\beta}$ takes care of the dissipation effects.
The $p$ and $\rho$ are the pressure and the proper energy
density. The coefficients $\lambda$, $\eta$, $\zeta$ are the heat
conductivity, the shear viscosity and the bulk viscosity, respectively.
The metric tensor $\eta^{\alpha\beta}$ are taken to be
$\eta^{\alpha\beta}=\text{diag}(-1,1,1,1)$. The four velocity
$U^\alpha =(1/\sqrt{1-v^2},\vec{v}/\sqrt{1-v^2})$ has the
property $U^\alpha U_\alpha=-1$ in this metric.
The $W_{\alpha\beta}$, $Q_\alpha$, $H_{\alpha\beta}$ are defined by
\begin{align}
W_{\alpha\beta}
&=
\frac{\partial U_\alpha}{\partial x^\beta}
+\frac{\partial U_\beta}{\partial x^\alpha}
-\frac{2}{3}\eta_{\alpha\beta}
\frac{\partial U^\gamma}{\partial x^\gamma},
\\
Q_\alpha
&=
\frac{\partial T}{\partial x^\alpha}
+T\frac{\partial U_\alpha}{\partial x^\beta}U^\beta,
\\
H_{\alpha\beta}
&=
\eta_{\alpha\beta}+U_\alpha U_\beta.
\end{align}
We define the energy density and the momentum density as
\begin{align}
E(\vec{x},t)&=T_0^{00}
=\frac{\rho+pv^2}{1-v^2},
\\
P^i(\vec{x},t)&=T_0^{i0}
=(p+\rho)\frac{v^i}{1-v^2}.
\label{momentum}
\end{align}
We can derive the relation between $U^\mu$ and $P^i$.
>From Eq.\ (\ref{momentum}), we see
\begin{equation}
P^i=(p+\rho)U^0 U^i.
\label{Purelation1}
\end{equation}
Moreover using $U^\alpha U_\alpha=-1$, we find
\begin{equation}
U^0
=\sqrt{1+\frac{1}{(p+\rho)^2}P^jP_j}
\cong 1+\frac{1}{2(p+\rho)^2}P^jP_j
\label{Purelation2}
\end{equation}
in the approximation of the small momentum fluctuation
$P^i \ll 1$.

Now consider the hydrodynamic equation. It is given by the
energy-momentum conservation,
\begin{equation}
0=
\frac{\partial T^{\alpha\beta}}{\partial x^\beta}
=
\frac{\partial T_0^{\alpha\beta}}{\partial x^\beta}
+\frac{\partial \Delta T^{\alpha\beta}}{\partial x^\beta},
\end{equation}
from which we find
\begin{align}
\frac{\partial}{\partial t}E
&=
-\frac{\partial P^j}{\partial x^j}
-\frac{\partial \Delta T^{0\beta}}{\partial x^\beta},
\\
\frac{\partial}{\partial t}P^i
&=
-\frac{\partial T_0^{ij}}{\partial x^j}
-\frac{\partial \Delta T^{i\beta}}{\partial x^\beta}.
\end{align}
The second terms in the right hand sides are the dissipation
terms, which we will compute in the following.
The $\Delta T^{\alpha\beta}$ are calculated to be
\begin{align}
\Delta T^{\alpha\beta}=
&- \lambda
\left[
2U^\alpha U^\beta U^\gamma
\frac{\partial T}{\partial x^\gamma}
+U^\alpha\frac{\partial T}{\partial x_\beta}
+U^\beta\frac{\partial T}{\partial x_\alpha}
+TU^\gamma
\frac{\partial \left(U^\alpha U^\beta\right)}{\partial x^\gamma}
\right]
\nonumber\\
&- \eta
\left[
\frac{\partial U^\alpha}{\partial x_\beta}
+\frac{\partial U^\beta}{\partial x_\alpha}
-\frac{2}{3}
\left(\eta^{\alpha\beta}+U^\alpha U^\beta\right)
\frac{\partial U^\gamma}{\partial x^\gamma}
+U^\gamma
\frac{\partial \left(U^\alpha U^\beta\right)}{\partial x^\gamma}
\right]
\nonumber\\
&- \zeta
\left(\eta^{\alpha\beta}+U^\alpha U^\beta\right)
\frac{\partial U^\gamma}{\partial x^\gamma}.
\end{align}
Thus the dissipation term for the energy density is
\begin{align}
-\frac{\partial \Delta T^{0\beta}}{\partial x^\beta}=
\frac{\partial}{\partial x^0}
\biggl[
&\lambda
\left(
2U^0 U^0 U^\gamma
\frac{\partial T}{\partial x^\gamma}
+2U^0\frac{\partial T}{\partial x_0}
+TU^\gamma
\frac{\partial \left(U^0 U^0\right)}{\partial x^\gamma}
\right)
\nonumber\\
+&\eta
\left(
2\frac{\partial U^0}{\partial x_0}
-\frac{2}{3}
\left(-1+U^0 U^0\right)
\frac{\partial U^\gamma}{\partial x^\gamma}
+U^\gamma
\frac{\partial \left(U^0 U^0\right)}{\partial x^\gamma}
\right)
\nonumber\\
+&\zeta
\left(-1+U^0 U^0\right)
\frac{\partial U^\gamma}{\partial x^\gamma}
\biggr]
\nonumber\\
+\frac{\partial}{\partial x^j}
\biggl[
&\lambda
\left(
2U^0 U^j U^\gamma
\frac{\partial T}{\partial x^\gamma}
+U^0\frac{\partial T}{\partial x_j}
+U^j\frac{\partial T}{\partial x_0}
+TU^\gamma
\frac{\partial \left(U^0 U^j\right)}{\partial x^\gamma}
\right)
\nonumber\\ 
+&\eta
\left(
\frac{\partial U^0}{\partial x_j}
+\frac{\partial U^j}{\partial x_0}
-\frac{2}{3}U^0 U^j
\frac{\partial U^\gamma}{\partial x^\gamma}
+U^\gamma
\frac{\partial \left(U^0 U^j\right)}{\partial x^\gamma}
\right)
\nonumber\\
+&\zeta
U^0 U^j\frac{\partial U^\gamma}{\partial x^\gamma}
\biggr]
\label{Edissipation}
\end{align}
Here the temperature $T$ is related to $E$ through the thermodynamic
relation
\begin{equation}
\delta T=\frac{1}{C}\delta E
\end{equation}
with $C$ being the specific heat. The $U^0$ and $U^j$ are 
translated to $P^j$ by Eqs.\ (\ref{Purelation1}) and
(\ref{Purelation2}). Moreover following the non-relativistic case,
we take only the terms up to the third order of the assumed
small quantities, that is,  the frequency and wavenumber
as well as the fluctuations $E$ and $P^i$. Since each term 
in Eq.\ (\ref{Edissipation}) involves two time and space
derivatives or the frequency and wavenumber, we have only to
take the first order of $E$ and $P^i$. Thus we find
\begin{align}
-\frac{\partial \Delta T^{\alpha\beta}}{\partial x^\beta}
&=
\lambda\frac{\partial}{\partial x^j}
\left[
\frac{1}{C}
\frac{\partial E}{\partial x_j}
+T\frac{\partial}{\partial t}
\left(\frac{1}{\rho+p}P^j\right)
\right]
\nonumber\\
&\cong
\lambda\frac{\partial}{\partial x^j}
\left(
\frac{1}{C}
\frac{\partial E}{\partial x_j}
+T_0\frac{1}{\rho_0+p_0}
\frac{\partial P^j}{\partial t}
\right),
\end{align}
where in the last step, we replaced $T$, $\rho$, $p$ with their
equilibrium values $T_0$, $\rho_0$, $p_0$.

Similarly, the dissipation term for the momentum density becomes
\begin{align}
-\frac{\partial \Delta T^{i\beta}}{\partial x^\beta}
=& \lambda
\frac{\partial}{\partial t}
\left[
\frac{1}{C}\frac{\partial E}{\partial x_i}
+T_0\frac{1}{\rho_0+p_0}\frac{\partial P^i}{\partial t}
\right]
\nonumber\\
&+\eta
\frac{1}{\rho_0+p_0}
\frac{\partial}{\partial x^j}
\left(
\frac{\partial P^i}{\partial x_j}
+\frac{\partial P^j}{\partial x_i}
-\frac{2}{3}\eta^{ij}
\frac{\partial P^l}{\partial x^l}
\right)
\nonumber\\
&+\zeta
\frac{1}{\rho_0+p_0}
\frac{\partial}{\partial x_i}
\frac{\partial P^l}{\partial x^l}.
\end{align}

The hydrodynamic equation with the dissipation term is thus
\begin{align}
\frac{\partial E}{\partial t}
=&
-\frac{\partial P^j}{\partial x^j}
+\lambda\frac{\partial}{\partial x^j}
\left(
\frac{1}{C}
\frac{\partial E}{\partial x_j}
+\frac{T_0}{\rho_0+p_0}
\frac{\partial P^j}{\partial t}
\right),
\\
\frac{\partial P^i}{\partial t}
=&
-\frac{\partial T_0^{ij}}{\partial x^j}
+\lambda\frac{\partial}{\partial t}
\left(
\frac{1}{C}\frac{\partial E}{\partial x_i}
+\frac{T_0}{\rho_0+p_0}\frac{\partial P^i}{\partial t}
\right)
\nonumber\\
&+\eta
\frac{1}{\rho_0+p_0}
\frac{\partial}{\partial x^j}
\left(
\frac{\partial P^i}{\partial x_j}
+\frac{\partial P^j}{\partial x_i}
-\frac{2}{3}\eta^{ij}
\frac{\partial P^l}{\partial x^l}
\right)
+\zeta
\frac{1}{\rho_0+p_0}
\frac{\partial}{\partial x_i}
\frac{\partial P^l}{\partial x^l}.
\end{align}
We note that there appear the relativistic corrections with
the time derivative. If those terms are dropped, the dissipation
terms reduce to the non-relativistic ones. Moreover we note that
the thermal conductivity $\lambda$ enters into the equation for
$P^i$ as a relativistic effect.

When moving to the Fourier space, we finally obtain
\begin{align}
\frac{\partial E_{\vec{q}}}{\partial t}
=&
-\text i q_j P_{\vec{q}}^j
+\lambda
\left(
-\frac{1}{C}q^2 E_{\vec{q}}
+\frac{T_0}{\rho_0+p_0}\text i q_j
\frac{\partial P_{\vec{q}}^j}{\partial t}
\right),
\\
\frac{\partial P_{\vec{q}}^i}{\partial t}
=&
-\text i q_j T_{0\vec{q}}^{ij}
+\lambda
\left(
\frac{1}{C}\text i q^i\frac{\partial E_{\vec{q}}}{\partial t}
+\frac{T_0}{\rho_0+p_0}
\frac{\partial^2 P_{\vec{q}}^i}{\partial t^2}
\right)
\nonumber\\
&+
\frac{\eta}{\rho_0+p_0}\text i q_j
\left(
\text i q^j P_{\vec{q}}^i
+\text i q^i P_{\vec{q}}^j
-\frac{2}{3}\eta^{ij}\text i q_l P_{\vec{q}}^l
\right)
+
\frac{\zeta}{\rho_0+p_0}\text i q^i\text i q_l P_{\vec{q}}^l,
\end{align}
from which we find the dissipation terms in Eqs.\
(\ref{kineticeqE}), (\ref{kineticeqP}).

\section{Scaling properties of the commutators of the slow variables}
\label{app:xjl}
\addcontentsline{toc}{subsection}{Appendix C}
In this appendix, we examine the scaling properties of the
commutators of the fluctuating amplitudes, $[A_j,A_l]$ for
$\{\phi, \phi^\dagger, Q, E, P^i\}$ in the chiral phase transition.
The results provide us with the exponents $x_{jl}$, which are
used to find the scaling properties of the commutators of the
reduced amplitudes, $[\tilde{A}_j, \tilde{A}_l]$.

\begin{itemize}
\item $[\phi_{\alpha\vec{q}}, (\phi_{\alpha\vec{q}'})^\dagger]$

The commutation relation is given in Eq.(\ref{phiphi}).
\begin{equation}
\left[\phi_{\alpha\vec{q}}, (\phi_{\alpha'\vec{q}'})^\dagger\right]
=
\frac{1}{2\omega_{\vec{q}}}\delta_{\alpha\alpha'}\delta_{\vec{q}\vec{q}'}
\sim \xi^{\frac{\gamma}{2\nu}},
\end{equation}
where we have used $\omega_{\vec{q}}=(2\chi_{\phi\vec{q}})^{-1/2}
\sim \xi^{-\gamma/2\nu}$.
\item $[\phi_{\alpha\vec{q}}, (\phi_{\alpha'-\vec{q}'}^{\dagger})^\dagger]$

This is zero.
\item $[\phi_{\alpha\vec{q}}, (Q_{\alpha'\beta'\vec{q}'})^\dagger]$

The commutation relation is given in Eq.\ (\ref{phiQ}).
Noting that
\begin{equation}
V^{-\frac{1}{2}}=\left(\frac{V}{\xi^d}\xi^d\right)^{-\frac{1}{2}}
\sim\xi^{-\frac{d}{2}},
\end{equation}
we find
\begin{align}
\left[
\phi_{\alpha\vec{q}}, \left(Q_{\alpha'\beta'\vec{q}'}\right)^\dagger
\right]
&=
\frac{-\text i}{2\sqrt{V}}\epsilon_{\alpha\alpha'\beta'\gamma}
\left\{
\phi_{\gamma\vec{q}-\vec{q}'}
+\phi_{\gamma\vec{q}'-\vec{q}}^{\dagger}
+\frac{\omega_{\vec{q}-\vec{q}'}}{\omega_{\vec{q}}}
\left(
\phi_{\gamma\vec{q}-\vec{q}'}
-\phi_{\gamma\vec{q}'-\vec{q}}^{\dagger}
\right)
\right\}
\nonumber\\
&\sim\xi^{-\frac{d}{2}+\frac{\gamma}{2\nu}}.
\end{align}
\item $[\phi_{\alpha\vec{q}}, (E_{\vec{q}'})^\dagger]$

Using the approximation
$[A_{\vec{k}}, B_{\vec{k}'}]\sim[A_{\vec{k}+\vec{l}}, B_{\vec{k}'-\vec{l}}]$
with $|\vec{k}|, |\vec{k}'|, |\vec{l}|\ll 1$, we find
\begin{equation}
\left[\phi_{\alpha\vec{q}}, E_{\vec{q}'}^\dagger\right]
=\left[\phi_{\alpha\vec{q}}, E_{-\vec{q}'}\right]
\sim
\left[\phi_{\alpha\vec{q}-\vec{q}'}, E_{\vec{0}}\right]
=\left[\phi_{\alpha\vec{q}-\vec{q}'}, \frac{1}{\sqrt{V}}
\mathcal{H}\right]
=\frac{\text i}{\sqrt{V}}\dot{\phi}_{\alpha\vec{q}-\vec{q}'},
\end{equation}
namely
\begin{equation}
\left[\phi_{\alpha\vec{q}}, E_{\vec{q}'}^\dagger\right]
=\frac{\text i}{\sqrt{V}}
\dot{\phi}_\alpha(\vec{q}-\vec{q}',\xi,V).
\end{equation}
The scaling property of $\dot{\phi}$ can be found by considering
\begin{align}
\left(
\text i\dot{\phi}_{\alpha\vec{k}},
\phi_{\beta-\vec{k}'}^{\dagger}Q_{\alpha'\beta'-\vec{k}-\vec{k}'}
\right)
&=
k_{\text B}T
\left\langle\left[
\phi_{\alpha\vec{k}},
\phi_{\beta-\vec{k}'}^{\dagger}Q_{\alpha'\beta'-\vec{k}-\vec{k}'}
\right]\right\rangle
\nonumber\\
&\hspace{-2.5cm}
=
\frac{-\text i}{2\sqrt{V}}k_{\text B}T
\left\langle
\phi_{\beta-\vec{k}}^{\dagger}
\epsilon_{\alpha\alpha'\beta'\gamma}
\left\{
\phi_{\gamma-\vec{k}'}
+\phi_{\gamma\vec{k}'}^{\dagger}
+\frac{\omega_{\vec{k}'}}{\omega_{\vec{k}}}
\left(
\phi_{\gamma-\vec{k}'}
-\phi_{\gamma\vec{k}'}^{\dagger}
\right)
\right\}
\right\rangle.
\end{align}
Denoting the exponent of $\dot{\phi}$ as $x_{\dot{\phi}}$, we find
\begin{equation}
L^{\frac{3}{2}d-x_{\dot{\phi}}-x_{\phi}-x_Q}
=L^{-\frac{d}{2}}L^{d-2x_{\phi}},
\end{equation}
which gives $x_{\dot{\phi}}=d/2+\beta/\nu$. Thus $\dot{\phi}$ scales as
\begin{equation}
\dot{\phi}(\vec{k},\xi,V)
=
L^{\frac{d}{2}-x_{\dot{\phi}}}
\dot{\phi}(L\vec{k},\xi/L,V/L^d)
\sim\xi^{-\frac{\beta}{\nu}},
\end{equation}
leading to
\begin{equation}
\left[\phi_{\alpha\vec{q}}, (E_{\vec{q}'})^\dagger\right]
\sim\xi^{-\frac{d}{2}-\frac{\beta}{\nu}}.
\end{equation}
\item $[\phi_{\alpha\vec{q}}, (P_{\vec{q}'}^i)^\dagger]$

From Eq.\ (\ref{phiP}),
\begin{equation}
\left[
\phi_{\alpha\vec{q}}, \left(P_{\vec{q}'}^i\right)^\dagger
\right]
=
\frac{1}{2\sqrt{V}}
\left\{
(q-q')^i
\left(
\phi_{\alpha\vec{q}-\vec{q}'}
+\phi_{\alpha\vec{q}'-\vec{q}}^{\dagger}
\right)
+\frac{\omega_{\vec{q}-\vec{q}'}}{\omega_{\vec{q}}}q^i
\left(
\phi_{\alpha\vec{q}-\vec{q}'}
-\phi_{\alpha\vec{q}'-\vec{q}}^{\dagger}
\right)
\right\}.
\end{equation}
This gives the scaling properties,
\begin{equation}
\left[
\phi_{\alpha\vec{q}}, (P_{\vec{q}'}^{\text Li})^\dagger
\right]
\sim
\left[
\phi_{\alpha\vec{q}}, (P_{\vec{q}'}^{\text Ti})^\dagger
\right]
\sim
\xi^{-\frac{d}{2}-1+\frac{\gamma}{2\nu}}
\end{equation}
for the longitudinal and transverse components of the momentum.
\item $[Q_{\alpha\beta\vec{q}}, (Q_{\alpha'\beta'\vec{q}'})^\dagger]$

The commutation relation Eq.\ (\ref{QQ}) immediately gives
\begin{equation}
\left[
Q_{a\vec{q}}^V, \left(Q_{b\vec{q}'}^V\right)^\dagger
\right]
=
\frac{\text i}{\sqrt{V}}\epsilon_{abc}Q_{c\vec{q}-\vec{q}'}^V
\sim\xi^{-\frac{d}{2}}.
\end{equation}

\item $[Q_{\alpha\beta\vec{q}}, (E_{\vec{q}'})^\dagger]$

As before, the commutator is approximated as
\begin{align}
\left[
Q_{\alpha\beta\vec{q}}, E_{-\vec{q}'}
\right]
&\sim
\left[
Q_{\alpha\beta\vec{q}-\vec{q}'}, E_{\vec{0}}
\right]
=
\left[
Q_{\alpha\beta\vec{q}-\vec{q}'}, \frac{1}{\sqrt{V}}\mathcal{H}
\right]
=
\frac{\text i}{\sqrt{V}}
\dot{Q}_{\alpha\beta\vec{q}-\vec{q}'},
\nonumber\\
\intertext{that is,}
\left[
Q_{\alpha\beta\vec{q}}, E_{-\vec{q}'}
\right]
&=
\frac{\text i}{\sqrt{V}}
\dot{Q}_{\alpha\beta}\left(\vec{q}-\vec{q}',\xi,V\right).
\end{align}
Consider
\begin{align}
\left(
\text i
\dot{Q}_{\alpha\beta\vec{k}},
Q_{\alpha'\beta'\vec{k}'}
Q_{\gamma'\delta'-\vec{k}-\vec{k}'}
\right)
&=
k_{\text B}T
\left\langle\left[
Q_{\alpha\beta\vec{k}},
Q_{\alpha'\beta'\vec{k}'}
Q_{\gamma'\beta'-\vec{k}-\vec{k}'}
\right]\right\rangle
\nonumber\\
&\sim
k_{\text B}T
\left\langle
\frac{i}{\sqrt{V}}
Q_{\vec{k}+\vec{k}'}Q_{-\vec{k}-\vec{k}'}
+\frac{i}{\sqrt{V}}
Q_{\vec{k}'}Q_{-\vec{k}'}
\right\rangle.
\end{align}
This gives
\begin{equation}
L^{\frac{3}{2}d-x_{\dot{Q}}-2x_Q}
=L^{-\frac{d}{2}}L^{d-2x_Q}
\end{equation}
to find $x_{\dot{Q}}=d$ where $x_{\dot{Q}}$ denotes the exponent
for $\dot{Q}$.
The scaling property of $\dot{Q}$ is thus
\begin{equation}
\dot{Q}_{\alpha\beta}\left(\vec{k},\xi,V\right)
=L^{\frac{d}{2}-x_{\dot{Q}}}
\dot{Q}_{\alpha\beta}\left(L\vec{k},\xi/L,V/L^d\right)
\sim\xi^{-\frac{d}{2}},
\end{equation}
from which we obtain
\begin{equation}
\left[
Q_{\alpha\beta\vec{q}}, E_{-\vec{q}'}
\right]
=
\frac{\text i}{\sqrt{V}}
\dot{Q}_{\alpha\beta}\left(\vec{q}-\vec{q}',\xi,V\right)
\sim
\xi^{-d}.
\end{equation}
\item $[Q_{\alpha\beta\vec{q}}, (P_{\vec{q}'}^i)^\dagger]$

The commutation relation Eq.\ (\ref{QP}) gives
\begin{equation}
\left[
Q_{\alpha\beta\vec{q}}, \left(P_{\vec{q}'}^{\text Li}\right)^\dagger
\right]
\sim
\left[
Q_{\alpha\beta\vec{q}}, \left(P_{\vec{q}'}^{\text Ti}\right)^\dagger
\right]
\sim\xi^{-\frac{d}{2}-1}.
\end{equation}
\item $[E_{\vec{q}}, (E_{\vec{q}'})^\dagger]$

The commutation relation Eq.\ (\ref{EE}) gives
\begin{equation}
\left[E_{\vec{q}}, (E_{\vec{q}'})^\dagger\right]
\sim\xi^{-\frac{d}{2}-1}.
\end{equation}
\item $[E_{\vec{q}}, (P_{\vec{q}'}^i)^\dagger]$

The same procedure as before leads us to
\begin{align}
\left[
E_{\vec{q}}, \left(P_{\vec{q}'}^i\right)^\dagger
\right]
&\sim
\left[
E_{\vec{0}}, P_{\vec{q}-\vec{q}'}^i
\right]
=\frac{-\text i}{\sqrt{V}}
\dot{P}_{\vec{q}-\vec{q}'}^i,
\nonumber\\
\intertext{that is,}
\left[
E_{\vec{q}}, P_{-\vec{q}'}^i
\right]
&=
\frac{-\text i}{\sqrt{V}}
\dot{P}^i\left(\vec{q}-\vec{q}',\xi,V\right).
\end{align}
Consider
\begin{align}
\left(
\text i
\dot{P}_{\vec{k}}^i,
P_{\vec{k}'}^j P_{-\vec{k}-\vec{k}'}^l
\right)
&=
k_{\text B}T
\left\langle\left[
P_{\vec{k}}^i,
P_{\vec{k}'}^j P_{-\vec{k}-\vec{k}'}^l
\right]\right\rangle
\nonumber\\
&=
\frac{k_{\text B}T}{\sqrt{V}}
\left\{
k^j
\left\langle
P_{\vec{k}+\vec{k}'}^i P_{-\vec{k}-\vec{k}'}^l
\right\rangle
-k^i
\left\langle
P_{\vec{k}+\vec{k}'}^j P_{-\vec{k}-\vec{k}'}^l
\right\rangle
\right.
\nonumber\\
&\hspace{1.5cm}+
\left.
k^l
\left\langle
P_{\vec{k}'}^j P_{-\vec{k}'}^i
\right\rangle
+(k+k')^i
\left\langle
P_{\vec{k}'}^j P_{-\vec{k}'}^l
\right\rangle
\right\},
\end{align}
which gives the exponent for $\dot{P}$ as $x_{\dot{P}}=d+1$. Thus
\begin{equation}
\dot{P}^i\left(\vec{k},\xi,V\right)
=
L^{\frac{d}{2}-x_{\dot{P}}}
\dot{P}^i\left(L\vec{k},\xi/L,V/L^d\right)
\sim\xi^{-\frac{d}{2}-1},
\end{equation}
and we find
\begin{equation}
\left[
E_{\vec{q}}, \left(P_{\vec{q}'}^i\right)^\dagger
\right]
=
\frac{-\text i}{\sqrt{V}}
\dot{P}^i\left(\vec{q}-\vec{q}',\xi,V\right)
\sim\xi^{-d-1}.
\end{equation}
\item $[P_{\vec{q}}^i, (P_{\vec{q}'}^j)^\dagger]$

The commutation relation Eq.\ (\ref{PP}) gives
\begin{equation}
\left[
P_{\vec{q}}^i, \left(P_{\vec{q}'}^j\right)^\dagger
\right]
\sim\xi^{-\frac{d}{2}-1}.
\end{equation}
\end{itemize}

\end{document}